\documentclass[journal]{IEEEtran}

\newcommand{\rev}[1]{#1}
\newcommand{\inlineRev}{}
\usepackage[dvipsnames]{xcolor}
\usepackage{changepage}
\usepackage{tikz}
\usetikzlibrary{calc}
\usepackage{color,hyperref,listings}
\usepackage[shortlabels]{enumitem}
\usepackage{algorithm}
\usepackage{algorithmic}
\usepackage{amssymb}
\usepackage{tkz-euclide}
\usepackage{siunitx}
\usepackage{physics}
\usepackage{multicol}
\usepackage{pgfplots}
\usepackage{adjustbox}
\usepackage{cancel}
\usepackage{commath}
\usepackage{graphicx}
\usepackage{mathdots}
\usepackage{tabularx}
\usepackage{mathtools}
\usepackage{booktabs}
\usepackage{environ}
\usepackage{cleveref}
\usepackage{mathrsfs}
\usepackage{fancyhdr}
\usepackage{centernot}
\usepackage[margin=0.75in]{geometry}
\usepackage[american,siunitx,arrowmos]{circuitikz}
\usepackage[final]{pdfpages}
\pgfplotsset{compat=1.18}
\usepackage[english]{babel}
\usepackage{fontawesome}
\usepackage{listings}
\usepackage{fixmath}
\usepackage{xspace}
\usepackage{listings}
\usepackage{amsthm}
\usepackage{upgreek}
\usetikzlibrary{trees}
\usepackage{amsmath,amsfonts}
\usepackage{array}
\usepackage{subcaption}
\usepackage{textcomp}
\usepackage{stfloats}
\usepackage{url}
\usepackage{verbatim}
\usepackage{cite}
\usepackage{graphicx}
\usepackage{prettyref}
\usepackage{wrapfig}
\usepackage{dsfont}
\newcommand{\Ebb}{\mathbb{E}}
\newcommand{\Hbb}{\mathbb{H}}

\newcommand{\xv}{\mathbf{x}}

\newcommand{\Bv}{\mathbf{B}}
\newcommand{\Xv}{\mathbf{X}}

\newcommand{\Acal}{\mathcal{A}}

\newcommand{\Ccal}{\mathcal{C}}

\newcommand{\Mcal}{\mathcal{M}}
\newcommand{\Scal}{\mathcal{S}}
\newcommand{\Zcal}{\mathcal{Z}}

\renewcommand{\Gamma}{\Upgamma}
\renewcommand{\Theta}{\Uptheta}
\renewcommand{\Omega}{\Upomega}

\newcommand{\convas}{\stackrel{\text{a.s.}}{\longrightarrow}}

\newcommand{\indic}[1]{\mathbf{1}\left\{#1\right\}}

\newcommand{\simiid}{\stackrel{\text{i.i.d.}}{\sim}}
\DeclareMathOperator{\supp}{supp}

\def\ie{\textit{i.e.}\@\xspace}
\def\eg{\textit{e.g.}\@\xspace}
\def\etc{\textit{etc.}\@\xspace}

\newtheorem{theorem}{Theorem}[section]

\newtheorem{lemma}[theorem]{Lemma}
\newtheorem{corollary}[theorem]{Corollary}
\newtheorem{fact}[theorem]{Fact}

\theoremstyle{definition}
\newtheorem{definition}[theorem]{Definition}
\newtheorem{construction}[theorem]{Construction}
\newtheorem{example}[theorem]{Example}

\newtheorem{exercise-easy}[theorem]{Exercise}
\newtheorem{exercise-med}[theorem]{Exercise}
\newtheorem{exercise-hard}[theorem]{Exercise$^\star$}
\newtheorem{claim}[theorem]{Claim}
\newtheorem*{claim*}{Claim}

\newtheorem{remark}[theorem]{Remark}
\newtheorem*{remark*}{Remark}

\newtheorem*{observation*}{Observation}

\newcommand{\savehyperref}[2]{\texorpdfstring{\hyperref[#1]{#2}}{#2}}

\newrefformat{eq}{\savehyperref{#1}{\textup{(\ref*{#1})}}}
\newrefformat{ineq}{\savehyperref{#1}{\textup{(\ref*{#1})}}}
\newrefformat{eqn}{\savehyperref{#1}{\textup{(\ref*{#1})}}}
\newrefformat{lem}{\savehyperref{#1}{Lemma~\ref*{#1}}}
\newrefformat{con}{\savehyperref{#1}{Construction~\ref*{#1}}}
\newrefformat{def}{\savehyperref{#1}{Definition~\ref*{#1}}}
\newrefformat{thm}{\savehyperref{#1}{Theorem~\ref*{#1}}}
\newrefformat{cor}{\savehyperref{#1}{Corollary~\ref*{#1}}}
\newrefformat{cha}{\savehyperref{#1}{Chapter~\ref*{#1}}}
\newrefformat{sec}{\savehyperref{#1}{Section~\ref*{#1}}}
\newrefformat{app}{\savehyperref{#1}{Appendix~\ref*{#1}}}
\newrefformat{tab}{\savehyperref{#1}{Table~\ref*{#1}}}
\newrefformat{fig}{\savehyperref{#1}{Figure~\ref*{#1}}}
\newrefformat{hyp}{\savehyperref{#1}{Hypothesis~\ref*{#1}}}
\newrefformat{alg}{\savehyperref{#1}{Algorithm~\ref*{#1}}}
\newrefformat{rem}{\savehyperref{#1}{Remark~\ref*{#1}}}
\newrefformat{item}{\savehyperref{#1}{Item~\ref*{#1}}}
\newrefformat{step}{\savehyperref{#1}{step~\ref*{#1}}}
\newrefformat{conj}{\savehyperref{#1}{Conjecture~\ref*{#1}}}
\newrefformat{fact}{\savehyperref{#1}{Fact~\ref*{#1}}}
\newrefformat{prop}{\savehyperref{#1}{Proposition~\ref*{#1}}}
\newrefformat{prob}{\savehyperref{#1}{Problem~\ref*{#1}}}
\newrefformat{claim}{\savehyperref{#1}{Claim~\ref*{#1}}}
\newrefformat{relax}{\savehyperref{#1}{Relaxation~\ref*{#1}}}
\newrefformat{rem}{\savehyperref{#1}{Remark~\ref*{#1}}}
\newrefformat{red}{\savehyperref{#1}{Reduction~\ref*{#1}}}
\newrefformat{part}{\savehyperref{#1}{Part~\ref*{#1}}}
\newrefformat{ex}{\savehyperref{#1}{Exercise~\ref*{#1}}}
\newrefformat{property}{\savehyperref{#1}{Property~\ref*{#1}}}
\newrefformat{type}{\savehyperref{#1}{Type~\ref*{#1}}}
\newrefformat{eg}{\savehyperref{#1}{Example~\ref*{#1}}}
\newrefformat{obs}{\savehyperref{#1}{Observation~\ref*{#1}}}

\definecolor{deepblue}{rgb}{0,0,0.5}
\definecolor{deepred}{rgb}{0.6,0,0}
\definecolor{deepgreen}{rgb}{0,0.5,0}

\begin{document}
\title{The LZ78 Source\\ \Large Definition, Entropic Properties, and Application to In-Context Learning}

\author{Naomi~Sagan,
        Amir~Dembo,
        Matthew~Ho,
        and~Tsachy~Weissman
\thanks{N.Sagan, M. Ho, and T. Weissman are with the Department of Electrical Engineering, Stanford University, Stanford, CA, 94305 USA (email:  nsagan@stanford.edu; matthho@stanford.edu; tsachy@stanford.edu).}% <-this % stops a space
\thanks{A. Dembo is with the Departments of Mathematics and Statistics, Stanford University, Stanford, CA, 94305 USA (email: adembo@stanford.edu).}% <-this % stops a space
\thanks{Manuscript received March 14, 2025; revised December 15, 2025.}}


\markboth{IEEE Transactions on Information Theory}%
{Sagan \MakeLowercase{\textit{et al.}}: The LZ78 Source}



\maketitle

\renewcommand{\Pi}{\Uppi}
\newcommand{\EE}{\mathbb{E}}

\newcommand{\opI}{I}
\providecommand{\MI}[2]{\ensuremath{\opI\left(#1 ; #2\right)}}
\providecommand{\CMI}[3]{\ensuremath{\opI \left( #1 ; #2 \,\middle| #3 \right)}}
\providecommand{\Ber}{\text{Ber}\,}

\begin{abstract}
We study a family of processes generated according to sequential probability assignments induced by the LZ78 universal compressor.
We characterize entropic and distributional properties such as their entropy and relative entropy rates,  finite-state compressibility and log loss of their realizations,  and the empirical distributions that they induce.
Though not quite stationary, these sources are ``almost stationary and ergodic'';  similar to stationary and ergodic processes, they satisfy a Shannon-McMillan-Breiman-type property: the normalized log probability of their realizations converges almost surely to their entropy rate.
Further, they are locally ``almost i.i.d.'' in the sense that the finite-dimensional empirical distributions of their realizations converge almost surely to a deterministic i.i.d. law.
However, unlike stationary ergodic sources, the  finite-state compressibility of their realizations is almost surely strictly larger than their entropy rate by a ``Jensen gap''.
We present simulations demonstrating the theoretical results.  
These sources allow to gauge the performance of sequential probability models, \rev{both classical and deep learning-based,} on non-Markovian  non-stationary data.
\rev{As such, we apply realizations of the LZ78 source to the study of in-context learning in transformer models.}
\end{abstract}


% \fi
\IEEEpeerreviewmaketitle

\section{Introduction}\label{sec:introduction}
\IEEEPARstart{T}he celebrated LZ78 compression algorithm \cite{originalLZ78paper} has been studied extensively and has proven an indispensable tool in developing universal schemes, \ie, those that work for arbitrary, deterministic input sequences.
In addition to universal compression, it has been used in settings such as gambling \cite{feder1991gambling}, sequence prediction \cite{federMerhavGutman1992}, noisy channel decoding \cite{ziv1984FSchannels}, filtering \cite{weissman2007Filtering}, and probability modeling \cite{sagan2024familylz78baseduniversalsequential}.
\textit{Cf.}\  \cite{merhav2024jacobzivsindividualsequenceapproach} for a review of LZ78-based universal schemes.

LZ78, however, has yet to be thoroughly investigated as a probability source.
\rev{Though \cite{merhav2022universalrandomcodingensemble} recently presented a similar probability source in the context of lossy compression (specifically, universal lossy compression can be performed via a random codebook in which codewords are drawn with log probability proportional to their LZ78 codelength), the study of this source did not extend beyond the lossy compression use case.}
In this paper, we start to fill this gap by characterizing the basic entropic and probabilistic properties of a family of sources generated from the respective family of LZ78 Sequential Probability Assignments (SPAs) introduced recently in  \cite{sagan2024familylz78baseduniversalsequential}.
Specifically,  we determine the entropy rate of such sources and establish the almost sure convergence of a realization's normalized log probability to this entropy rate (a Shannon-McMillan-Breiman type result).
In addition, we show that this source is highly non-Markovian: the log loss of any finite-context probability model on the realized source sequence---and therefore also its finite-state compressibility---almost surely approach another, higher value. 

Key to our results on the entropic properties are results of independent interest pertaining to the  empirical distribution of realizations from these sources.
Despite their non-trivial dependence structure, they are locally ``almost i.i.d.'' in the sense that the finite-dimensional empirical distributions of their realizations converge almost surely to a deterministic i.i.d. law.
This property is reminiscent of that of good channel and source codes (\textit{cf.}\   \cite{ShamaiVerdu1997, weissman_empirical_2005}, and references therein).     

\textbf{In-Context Learning.}
These LZ78 sources provide us with new non-Markovian non-stationary laws for which the entropic properties---and thus the fundamental limits of compression and sequential probability modeling---are now known.
As such, \rev{we apply them} to studying performance tradeoffs in sequential probability and online models, \rev{in particular transformer-based \cite{vaswani2017attentionneed} deep learning models.}

\rev{Transformer models have been found to have the capability for \textit{in-context learning (ICL)}, or the ability to apply sequential prediction algorithms on the data seen in-context at test time.
In particular, beyond learning the statistics of their training data, transformers can learn to apply an algorithm, e.g., something akin to a Markov model or Context Tree Weighting \cite{willems1995CTW}, over the data in its context.
In the past years, there have been efforts to use structured random training data (e.g., generated from Markov or Variable-order Markov sources) to study the ICL properties of small transformer models.
\cite{edelman2024evolutionstatisticalinductionheads,makkuva2025attentionmarkovframeworkprincipled} trains transformers on second-and third-order Markov data, showing (both empirically and theoretically) that two-layer transformers can learn a first-order Markov model via a phenomenon known as induction heads \cite{olsson2022context}.
\cite{edelman2024evolutionstatisticalinductionheads} states that a network with $k$ attention heads can learn a $k$-order Markov model in-context.
\cite{rajaraman2024transformersmarkovdataconstant} provides tighter bounds, showing that a transformer with $\log_2(k+1)$ layers can learn a $k$-order Markov model.
\cite{zhou2024transformerslearnvariableordermarkov} applies the study of ICL to variable-order Markov data generated from context trees \cite{willems1995CTW}, showing that transformer models can learn CTW in-context.
}

\rev{In this paper, we perform a similar increment as \cite{zhou2024transformerslearnvariableordermarkov}: instead of studying constant-order Markov models or variable-order Markov models with a fixed depth, the LZ78 Source provides a non-Markovian model where the context needed to make predictions grows with the sequence length.
The non-Markovian nature of the source, especially the ``Jensen gap'' between the entropy rate and finite-state predictability, allow for a nuanced empirical study of ICL.
This Jensen gap means that, for any finite sequence length, different order Markov SPAs will have different cumulative log losses, with a lower potential log loss for higher-order models.
It also means that this source is much more difficult to learn than even variable-order Markov sources.
Whereas, in \cite{zhou2024transformerslearnvariableordermarkov}, the benchmarked transformers learn the context tree data to global optimum, transformers have a more difficult time learning the LZ78 source, though they do better than many classical algorithms (see \prettyref{sec:experiments}).
This allows us to benchmark transformers against, e.g., the best Markov models of varying depths, as well as practical algorithms like CTW and the LZ78 SPA from \cite{sagan2024familylz78baseduniversalsequential}.
Thus, the LZ78 source constitutes a harder and more discriminative benchmark for probing ICL than any previously studied Markov or variable-order Markov family.
}

\section{An LZ78-Based Probability Source}\label{sec:source-definition}
\subsection{Preliminaries and Notation}
\begin{definition}[Alphabets and Sequences]
    Let $\Acal$ be a finite alphabet, and $\Mcal(\Acal)$ be the simplex of probability mass functions (PMFs) over that alphabet.
    In the case of the binary alphabet $\Acal = \{0, 1\}$,  $\Mcal(\Acal)$ is the set of all Bernoulli distributions.
    % which we identify  
    %  with the unit interval via the Bernoulli parameter.

    For $\theta \in \Mcal(\Acal)$, we write $X \sim \theta$ to denote that $X$ has the PMF $\theta$. 

    Let an infinite stochastic sequence (process) with components in $\Acal$ be denoted, \eg, as $\Xv = (X_1, X_2, \cdots)$.
    $X^n$ represents the first $n$ components of $\Xv$.
    $X_t^\ell = (X_t, X_{t+1}, \dots, X_\ell)$ for $t \leq \ell$ and is equal to the empty sequence for $t > \ell$.
    If $P$ is the probability law of $\Xv$, the law of any subsequence is denoted via a subscript on $P$, \eg, $X_t^\ell \sim P_{X_t^\ell}$, where 
     $P_{X_t^\ell} \in \Mcal(\Acal^{\ell - t + 1})$.
An infinite deterministic sequence (also referred to as an \textit{individual sequence}) is denoted by $\xv$, with analogous notation for subsequences.
\end{definition}
\begin{remark}
    In the case of a binary alphabet, for conciseness of notation, $\Theta \in \Mcal(\Acal)$ is identified with the unit interval via the scalar Bernoulli parameter.
    % The proofs carry over seamlessly to accommodate a general finite alphabet (only the notation would be  more burdensome).
\end{remark}
\begin{definition}[Order of Growth]
    If sequence $a_k$ and function $f(k)$ satisfy $a_k \cong f(k)$, $a_k$ grows on the order of $f(k)$.
    \rev{Likewise, $a_k \lesssim f(k)$ means that $a_k$ grows at most on the order of $f(k)$.}
\end{definition}
\begin{remark}[Default Logarithmic Base]
    $\log(\cdot)$ and $\exp(\cdot)$ refer to the base-2 logarithm and exponentiation, respectively.
\end{remark}

This paper relies heavily on LZ78 incremental parsing and the corresponding prefix tree interpretation.
Refer to \cite{originalLZ78paper} for details on incremental parsing, and to the appendix of \cite{sagan2024familylz78baseduniversalsequential} for an example of constructing an LZ78 prefix tree.

\subsection{Definition of the LZ78 Probability Source}
%  Generating a sequence via a model from the LZ78 family of SPAs\cite{sagan2024familylz78baseduniversalsequential} is equivalent to the following: 
% \begin{construction}[LZ78 Probability Source]\label{con:lz78-probability-source}
%     Given a probability distribution $\Pi$ over the simplex $\Mcal(\Acal)$, 
%     %denote the LZ78 probability source as $Q^{\text{LZ},\Pi}$.
%     we produce two sequences: $\Xv$, comprising components in alphabet $\Acal$, and $\Bv$, with components in $\Mcal(\Acal)$,  representing the distributions from which   the corresponding $\Xv$ values are drawn.
%     The sequences are generated as follows:
% We refer to the law of $\Xv$ as the ``LZ78 source'', and denote it by $Q^{\text{LZ}, \Pi}$.
% The properties of $Q^{\text{LZ}, \Pi}$ as a model and a SPA were thoroughly investigated in \cite{sagan2024familylz78baseduniversalsequential}. The present paper is dedicated to its properties as a data generating source. 
% \end{construction}
\begin{construction}[LZ78 Probability Source]\label{con:lz78-probability-source}
    We produce a stochastic process over alphabet $\Acal$, denoted $\Xv = (X_1, X_2, \dots)$.
    $\Xv$ is closely tied to an auxiliary process $\Bv$, with components in $\Mcal(\Acal)$.
    As we describe below,  conditioned on $B_t$,  $X_t \sim B_t$.

    First, fix some probability distribution $\Pi$ over the simplex $\Mcal(\Acal)$. 
    Then, the components of the processes $\Xv$ and $\Bv$  are generated as follows:
    \begin{enumerate}
        \item Generate\footnote{Computationally, this step occurs lazily: values are only generated as they are needed during subsequent steps.}  $(\Theta_0, \Theta_1, \Theta_2, \dots) \simiid \Pi$.
        \item Grow an LZ78 prefix tree as follows. For $t \geq 1$:
        \begin{enumerate}
            \item If the current ($t$\textsuperscript{th}) node of the tree lacks an associated $\Theta$ value, assign the next $\Theta$ generated in step 1 to the current node.
            Let $B_t$ denote the $\Theta$ value corresponding to the current node. 
            \item Given all that was generated thus far, namely $X^{t-1}$ and $ B^t$,  generate  $X_t \sim B_t$.
            \item Traverse, and potentially grow, the LZ78 tree for the newly-drawn symbol $X_t$.
            If a new leaf is added, return to the root of the tree.
        \end{enumerate}
    \end{enumerate}
    
    This formulation is exactly equivalent to drawing from the LZ78-based family of SPAs from \cite{sagan2024familylz78baseduniversalsequential}.
    \textit{I.e.}, let $\left\{q^\text{LZ}(x_t|x^{t-1})\right\}_t$ represent one SPA from that family.
    At a specific time $t$, $q^\text{LZ}(\cdot|x^{t-1}) \triangleq \left(q^\text{LZ}(a|x^{t-1})\,:\, a \in \Acal \right)$ can be thought of as a probability mass function over $\mathcal{A}$ corresponding to $\Pr{X = a} = q^\text{LZ}(a|x^{t-1})$.
    Using this perspective, we produce process $\Xv$ by drawing random variable $X_1$ according to $q(\cdot)$, then $X_2 \sim q(\cdot|X_1)$, $\dots$, $X_t \sim q(\cdot|X^{t-1})$, \textit{etc}.

    % \begin{enumerate}
    %     \item Create an ``empty'' LZ78 prefix tree, with only a root node.
    %     Draw $\Theta_0 \sim \Pi$, and ``assign'' it to the root node.
    %     Every node in the tree will be assigned such a random variable.
    %     \item Grow the LZ78 prefix tree as follows. For $t \geq 1$, starting from the root of the tree:
    %     \begin{enumerate}
    %         \item Set $B_t$ to be the value of the ``$\Theta$'' random variable assigned to the current node of the tree.
    %         \item Draw $X_t \sim B_t$.
    %         \item Traverse the LZ78 tree according to the newly-drawn $X_t$.
    %         If adding a new leaf, draw $\Theta \sim \Pi$ and assign it to the new leaf (and then return to the root of the tree).
    %     \end{enumerate}
    % \end{enumerate}
We refer to the law of $\Xv$ as the ``LZ78 source'', and denote it by $Q^{\text{LZ}, \Pi}$.
The properties of $Q^{\text{LZ}, \Pi}$ as a model and a SPA were thoroughly investigated in \cite{sagan2024familylz78baseduniversalsequential}, which may include useful context for understanding the LZ78 source.
The present paper is dedicated to its properties as a data generating source. 
\end{construction}

\begin{figure*}[t]
\centering
    \begin{minipage}[t]{0.48\linewidth}
        \textbf{Initial step}: Draw $\Theta_0 \sim \text{Unif}([0, 1])$.
        For this realization, say that we get $\Theta_0 = 0.05$.
        
        \vspace{0.5em}
        
       \begin{center}
            \begin{tikzpicture}[
              level distance=1cm,
              level 1/.style={sibling distance=6cm}
            ]
            \node[rectangle, fill=yellow!25, rounded corners, draw]
            {\parbox[t]{4em}{\centering\small root\\$\Theta_0=0.05$}};
            \end{tikzpicture}
        \end{center}
    
        \vspace{0.8em}
        
        \textbf{Step 1} ($t=1$): We are at the root, so $B_1 = \Theta_0 = 0.05$.
        We draw $X_1 \sim \text{Ber}(0.05)$, which results in $X_1 = 0$ for this realization.
        So, we add a new leaf corresponding to the LZ78 phrase \texttt{0}, assign $(\Theta_1 \sim \text{Unif}([0, 1])) = 0.8$ to the new leaf, and return to the root.
        \begin{center}
            \begin{tikzpicture}[
              level distance=1cm,
              level 1/.style={sibling distance=6cm}
            ]
            \node[rectangle, fill=yellow!25, rounded corners, draw]
            {\parbox[t]{4em}{\centering\small root\\$\Theta_0=0.05$}}
            child {
              node[rectangle, fill=gray!5, rounded corners, draw]
              {\parbox[t]{4em}{\centering\small node \texttt{0}\\$\Theta_1=0.8$}}
            };
            
            \node[xshift=2.2cm, yshift=-0.6cm]
            {\parbox[t]{6em}{\small $X_1=0$\\$B_1=0.05$}};
            \end{tikzpicture}
        \end{center}
        
        \vspace{0.8em}
        
        \textbf{Step 2} ($t=2$): We are again at the root, so $B_2 = \Theta_0 = 0.05$.
        $(X_2 \sim \text{Ber}(0.05)) = 0$, so we traverse to the leaf we added in the previous step.

        \begin{center}
            \begin{tikzpicture}[
              level distance=1cm,
              level 1/.style={sibling distance=6cm}
            ]
            \node[rectangle, fill=yellow!25, rounded corners, draw]
            {\parbox[t]{4em}{\centering\small root\\$\Theta_0=0.05$}}
            child {
              node[rectangle, fill=green!15, rounded corners, draw]
              {\parbox[t]{4em}{\centering\small node \texttt{0}\\$\Theta_1=0.8$}}
            };
            
            \node[xshift=2.5cm, yshift=-0.6cm]
            {\parbox[t]{8em}{\small $X_1=0,X_2=0$\\$B_1=B_2=0.05$}};
            \end{tikzpicture}
        \end{center}
    
        \vspace{0.8em}
    \end{minipage}
    \hfill
    \begin{minipage}[t]{0.48\linewidth}
    % ---- Step 3 ----
    \textbf{Step 3} ($t=3$): The current node is assigned $\Theta_1$, so $B_3 = \Theta_1 = 0.8$.
        We draw $(X_3 \sim \text{Ber}(0.8)) = 0$, and add a new leaf corresponding to the phrase \texttt{00}.
        This leaf gets assigned $(\Theta_2 \sim \text{Unif}([0, 1])) = 0.44$, and we return to the root.
        
    \begin{center}
        \begin{tikzpicture}[
          level distance=1cm,
          level 1/.style={sibling distance=6cm},
          level 2/.style={sibling distance=2cm}
        ]
        \node[rectangle, fill=green!15, rounded corners, draw]
        {\parbox[t]{4em}{\centering\small root\\$\Theta_0=0.05$}}
        child {
          node[rectangle, fill=yellow!25, rounded corners, draw]
          {\parbox[t]{4em}{\centering\small node \texttt{0}\\$\Theta_1=0.8$}}
          child {
            node[rectangle, fill=gray!5, rounded corners, draw]
            {\parbox[t]{4em}{\centering\small node \texttt{00}\\$\Theta_2=0.44$}}
          }
        };
        
        \node[xshift=3.1cm, yshift=-0.6cm]
        {\parbox[t]{9em}{\small $X_1=X_2=X_3=0$\\$B_3=0.8$}};
        \end{tikzpicture}
    \end{center}
    
    \vspace{0.5em}

    \textbf{Later steps}: After a few more steps, the tree and corresponding generated sequences might look like the following, with $X_{1:8} = 0,0,0,0,1,0,1,0$ and $B_{1:8} = 0.05,0.05,0.8,0.05,0.8,0.05,0.8,0.05$:
    \begin{center}
        \begin{tikzpicture}[
          level distance=1cm,
          level 1/.style={sibling distance=5cm},
          level 2/.style={sibling distance=2cm},
          level 3/.style={sibling distance=1.5cm}
        ]
        \node[rectangle, fill=green!15, rounded corners, draw]
        {\parbox[t]{4em}{\centering\small root\\$\Theta_0=0.05$}}
        child {
          node[rectangle, fill=gray!5, rounded corners, draw]
          {\parbox[t]{4em}{\centering\small node \texttt{0}\\$\Theta_1=0.8$}}
          child {
            node[rectangle, fill=gray!5, rounded corners, draw]
            {\parbox[t]{4em}{\centering\small node \texttt{00}\\$\Theta_2=0.44$}}
          }
          child {
            node[rectangle, fill=yellow!25, rounded corners, draw]
            {\parbox[t]{4em}{\centering\small node \texttt{01}\\$\Theta_3=0.05$}}
            child {
              node[rectangle, fill=gray!5, rounded corners, draw]
              {\parbox[t]{4em}{\centering\small node \texttt{010}\\$\Theta_4=0.74$}}
            }
          }
        };
        \end{tikzpicture}
    \end{center}
    
    \end{minipage}

\caption{Illustration of the tree-based generation process for a realization of the LZ78 source with random numbers drawn via \texttt{numpy} for random seed $78$.
After each step, the current LZ tree is drawn, with the node at which the step was performed colored yellow, and the node traversed to, if different than the current node, colored green.}
\label{fig:tree_generation_example}
\end{figure*}

\begin{example}[Realization from the LZ78 Source]
    Let the alphabet be binary, and let $\Pi$ be the uniform distribution over the unit interval $[0, 1]$.
    Recall that we identify a probability mass function in the simplex $\Mcal(\{0, 1\})$ with the scalar Bernoulli parameter (probability of drawing a $1$).
    
    We provide an example of the first few components of $\Xv$ and $\Bv$ drawn from $Q^{\text{LZ}, \Pi}$ in \prettyref{fig:tree_generation_example}.
    
\end{example}

\begin{remark}[Non-Stationarity of the LZ78 Source]
    This probability law is not stationary. \textit{E.g.}, consider the case of a binary  alphabet and  an arbitrary distribution $\Pi$ over $[0, 1]$.
    The distributions of $B_1$, $B_2$, and $B_3$ generated from the LZ78 source are:
    \begin{enumerate}
        \item $t=1$: We start at the root, so $B_1 = \Theta_0$.
        A new leaf is generated based on $X_1 \sim B_1$.
        \item $t=2$: We are again at the root, so $B_2 = \Theta_0$.
        A new leaf is added if $X_1 \neq X_2$, which happens, conditioned on $\Theta_0$,  with probability $\alpha(\Theta_0) \triangleq 2\Theta_0 (1-\Theta_0)$.
        \item $t=3$: If a new leaf was added at $t=2$, we are at the root and $B_3 = \Theta_0$.
        Otherwise, $B_3 = \Theta_1$.
    \end{enumerate}
    Thus, $B_1$ and $B_2$ are both $\sim \Pi$.
    However, even though $\Theta_0$ and $\Theta_1$ were generated i.i.d. $\Pi$, $B_3$ does not (for general $\Pi$) have distribution $\Pi$.
    Indeed, for an arbitrary function $f: [0, 1] \to \mathbb{R}$, as $\Theta_0$ and $\Theta_1$ are independent,
    \begin{align*}
        \Ebb[f(B_3)] &= \Ebb[f(\Theta_0) \alpha(\Theta_0)] + \Ebb[f(\Theta_1) (1-\alpha(\Theta_0))] \\
        &=  \Ebb[f(\Theta_0) \alpha(\Theta_0)] + \Ebb[f(\Theta_1)](1 - \Ebb[\alpha(\Theta_0)]),
    \end{align*}
    which is equal to $\Ebb_{\Theta\sim\Pi}[f(\Theta)]$ iff $\mathrm{cov}(f(\Theta_0), \alpha(\Theta_0)) = 0$, \ie, $f(\Theta_0)$ and $\alpha(\Theta_0)$ are uncorrelated.
    The covariance is not zero for general $f$ and $\Pi$ (for instance, set $f(\Theta) = \Theta$ and let $\Pi$ be any non-trivial law over $[0,1]$).
This non-stationarity of the $\Bv$ process is readily seen to translate to non-stationarity of the $\Xv$ process.
\end{remark}

\section{Main Results: Entropic Properties of the LZ78 Source}\label{sec:main-results}
The LZ78 source of \prettyref{con:lz78-probability-source}, under benign conditions on $\Pi$, has the following properties:
\begin{enumerate}[1.]
    \item Its entropy rate is $\Ebb [H(\Theta)]$, where $\Theta \sim \Pi$ and $H(\cdot)$ is the Shannon entropy.
    \item The normalized log probability of $X^n$ drawn from the LZ78 source converges almost surely to its entropy rate.
    \item The finite-state compressibility and finite-state SPA log-loss of $\Xv$ is almost surely equal to $H(\Ebb [\Theta])$.
\end{enumerate}
%One condition on $\Pi$ is a moment condition, for which, to %streamline the results, we define notation.
Evidently, on one hand, similarly to a stationary and ergodic process, the normalized log probability of a realization from the LZ78 source converges almost surely to its entropy rate.
On the other hand, it is unlike a stationary ergodic source, whose finite-state compressibility almost surely coincides with its entropy rate  \cite{originalLZ78paper}.    
We now proceed toward a formal statement of these and related properties.

We can now state the main results, deferring further discussion and proof sketches to \prettyref{sec:entropic-properties}.
\newcommand{\smbtheoremtitle}{A Shannon-McMillan-Breiman-Type Result}
\newcommand{\smbtheoremcontents}{
    Let $\Xv$ be generated from the LZ78 source $Q^{\text{LZ}, \Pi}$ with $\supp(\Pi) = \Mcal(\Acal)$.
    Then, almost surely,
    \[ \lim_{n \rightarrow \infty} \frac{1}{n}\log \frac{1}{Q^{\text{LZ}, \Pi}_{X^n}(X^n)} =  \Ebb_{\Theta\sim\Pi}[H(\Theta)]. \]
}
\begin{theorem}[\smbtheoremtitle]\label{thm:shannon-mcmillan-breiman}
    \smbtheoremcontents
\end{theorem}

\newcommand{\entropyratetitle}{Entropy Rate}
\newcommand{\entropyratecontents}{
    Let $\Xv$ be generated from the LZ78 source.
    Then
    \[\lim_{n\to\infty} \frac{1}{n} H(X^n) = \Ebb_{\Theta \sim \Pi}[H(\Theta)].\]
}
\begin{theorem}
    [\entropyratetitle]\label{thm:entropy-rate}
    \entropyratecontents
\end{theorem}
\begin{definition}[Markov Sequential Probability Assignment Log Loss]\label{def:mu-x}
    For any infinite sequence $\xv$, the optimal log-loss of a Markovian probability model is
    \begin{align*}
        \mu(\xv) &\triangleq \lim_{k\to\infty} \mu_k(\xv) \triangleq \lim_{k\to\infty} \limsup_{n\to\infty} \mu_k(x^n) \\
        &\triangleq \lim_{k\to\infty} \limsup_{n\to\infty} \hat{H}_{X_{k+1}|X^k}(x^n),
    \end{align*}
    where $\hat{H}_{X_{k+1}|X^k}(x^n)$ is the $k$\textsuperscript{th}-order empirical conditional entropy of $x^n$.
\end{definition}
Note that, for any individual sequence $\xv$, $\mu(\xv)$ is also equal to the optimal finite-state SPA log loss which, in turn, is also equal to its  finite-state compressibility \cite{originalLZ78paper, sagan2024familylz78baseduniversalsequential}.  
\newcommand{\markovloglosstitle}{Optimal Finite-State and Markov Model Log Loss on $\Xv$}
\newcommand{\markovlogloscontents}{
    Let $\Xv$ be generated from the LZ78 source.
    Then, almost surely, 
    \[\mu(\Xv) = H(\Ebb_{\Theta\sim\Pi}[\Theta]).\]
}
\begin{theorem}[\markovloglosstitle]\label{thm:markov-log-loss}
    \markovlogloscontents
\end{theorem}
Evidently, the finite-state compressibility of $\Xv$ is a ``Jensen gap'' larger than its entropy rate. 
\begin{remark}[Mutual Information Interpretation of the Jensen Gap] Consider $(\Theta, Y)$ jointly distributed as follows:  $\Theta\sim\Pi$ and, conditioned on $\Theta$, $Y \sim \Theta$. Then 
\begin{align*}
    H(Y | \Theta) &= \Ebb_{\Theta \sim \Pi}[H(\Theta)], \\
    H(Y) &= H (\Ebb_{\Theta \sim \Pi}[\Theta]),
\end{align*}
and, therefore, the ``Jensen gap'' is 
\begin{align*}
    H (\Ebb_{\Theta \sim \Pi}[\Theta]) -  \Ebb_{\Theta \sim \Pi}[H(\Theta)] &=  H(Y) - H(Y|\Theta) \\
    &= I (\Theta ; Y)  . 
\end{align*}
\label{rem:jensen-gap-as-mutual-info}
\end{remark}

To motivate our next result, we recall that, under benign conditions on $\Pi$, among the results of     
\cite{sagan2024familylz78baseduniversalsequential}
is the universality of $Q^{\text{LZ}, \Pi}$ in the sense that   
 \[ \lim_{n \rightarrow \infty} \frac{1}{n} D\left(P_{X^n} \big\lVert Q_{X^n}^{\text{LZ}, \Pi} \right) = 0 \ \ \ \forall P \mbox{ stationary}.  \]
\rev{For the sake of completeness, we also remark on the} behavior of the discrepancy between these laws when measured under the relative entropy in the other direction.  
The framework for proving \prettyref{thm:markov-log-loss} lends itself to answering that question for the family of  stationary $k$th-order Markov sources, which we denote by $\Mcal_k$. 

\newcommand{\pointwisedivergencetitle}{Divergence with Markovian Law: Pointwise Result}
\begin{theorem}[\pointwisedivergencetitle]\label{thm:relative-entropy-result-pointwise}
    Let $\Xv$ be generated from the LZ78 source $Q^{\text{LZ}, \Pi}$ with
    $\supp(\Pi) = \Mcal(\Acal)$. For any $k$\textsuperscript{th}-order Markov law
    $P \in \Mcal_k$, almost surely
    \begin{equation}
    \label{eq: pointwise divergence result}
    \begin{aligned}
    &\lim_{n \to \infty}
    \frac{1}{n}
    \log
    \frac{Q_{X^n}^{\text{LZ}, \Pi}(X^n)}{P_{X^n}(X^n)} = H(\Ebb[\Theta])
    - \Ebb[H(\Theta)]
    \\
    &\hspace{5em} + \Ebb \Bigl[
    D\Bigl(
    \Ebb[\Theta]
    \,\Big\lVert\,
    P_{X_0 \mid X_{-k}^{-1}}
    (\cdot \mid Y_{-k}^{-1})
    \Bigr)
    \Bigr],
    \end{aligned}
    \end{equation}
    where on the right-hand side
    $\Theta \sim \Pi$ and
    $Y_{-k}^{-1} \simiid \Ebb[\Theta]$.
\end{theorem}

\newcommand{\relativeentropytitle}{Relative Entropy with Markovian Law}
\begin{corollary}[\relativeentropytitle]\label{thm:relative-entropy-result}
    For $Q^{\text{LZ}, \Pi}$ with $\supp(\Pi) = \Mcal(\Acal)$, and any $k$\textsuperscript{th}-order Markov law $P \in \Mcal_k$,
    \begin{equation} 
    \begin{aligned}
        &\lim_{n\to\infty} \frac{1}{n} D\left(Q_{X^n}^{\text{LZ}, \Pi} \big\lVert P_{X^n}\right) = H(\Ebb[\Theta]) - \Ebb[H(\Theta)] \\
        &\hspace{5em} + \Ebb \left[D\left(\Ebb[\Theta] \,\Big\lVert\, P_{X_0|X_{-k}^{-1}}(\cdot|Y_{-k}^{-1})\right)\right],
    \end{aligned}
    \label{eq: relative entropy non pointwise result}
    \end{equation} 
    where on the right side $\Theta \sim \Pi$ and $Y_{-k}^{-1} \simiid \Ebb[\Theta]$.
    \rev{Note that, whereas \prettyref{thm:relative-entropy-result-pointwise} is a an almost-sure result, this result is in expectation.}
\end{corollary}
\begin{remark}
As in \prettyref{rem:jensen-gap-as-mutual-info}, the expression in the right sides of (\ref{eq: pointwise divergence result}) and (\ref{eq: relative entropy non pointwise result}) can be equivalently written as 
\[ D\left(P_Y  \,\Big\lVert\, P_{X_0|X_{-k}^{-1}} \Big| P_{Y_{-k}^{-1}} \right) +  I (\Theta ; Y), \]
where: $\Theta\sim\Pi$, 
%conditioned on $\Theta$, 
$Y \sim \Theta$, $Y_{-k}^{-1} \simiid Y$, $P_{X_0|X_{-k}^{-1}}$ is that associated with the $P \in \Mcal_k$, and here we assume the notation of  
\cite{Csiszár_Körner_2011} for relative entropy between conditional distributions (kernels) averaged with respect to another distribution.  
\label{remark: alternative limiting expression for the relative entropy} 
\end{remark}
\begin{remark}
It is straightforward to extend Theorem \ref{thm:relative-entropy-result-pointwise} (and its corollary) to stationary laws beyond Markov that are sufficiently mixing, \eg,  in the sense that  
\begin{align*}
\lim_{k \to \infty}
\sup_{x_{-\infty}^{-k-1},\,\tilde{x}_{-\infty}^{-k-1}}
\max_{x_{-k}^0}
\;
\Delta_k\!\left(
x_{-k}^0,\,
x_{-\infty}^{-k-1},\,
\tilde{x}_{-\infty}^{-k-1}
\right)
= 0 ,
\end{align*}
where
\begin{align*}
    \Delta_k(\cdot)
    &\triangleq
    \bigg|
    P_{X_0 \mid X_{-\infty}^{-1}}
    \!\left(
    x_0 \mid x_{-k}^{-1}, x_{-\infty}^{-k-1}
    \right) \\
    &\qquad-
    P_{X_0 \mid X_{-\infty}^{-1}}
    \!\left(
    x_0 \mid x_{-k}^{-1}, \tilde{x}_{-\infty}^{-k-1}
    \right)
    \bigg| .
\end{align*}
% \[ \lim_{k \rightarrow \infty} \sup_{x_{-\infty}^{-k-1} , \tilde{x}_{-\infty}^{-k-1}} \max_{x_{-k}^0} \left| P_{X_0|X_{-\infty}^{-1}} \left( x_0 | x_{-k}^{-1} , x_{-\infty}^{-k-1}  \right) - P_{X_0|X_{-\infty}^{-1}} \left( x_0 | x_{-k}^{-1} , \tilde{x}_{-\infty}^{-k-1}  \right) \right|  = 0 ,  \]
as is the case, e.g., for hidden Markov processes under benign conditions on their transition kernels \cite{EphraimMerhav2002}. In that case, Theorem \ref{thm:relative-entropy-result-pointwise} (and its corollary) remain valid, with $k = \infty$ in  the right sides of (\ref{eq: pointwise divergence result}) and (\ref{eq: relative entropy non pointwise result}). 
\label{remark: extension beyond markov to limiting expression for the relative entropy} 
\end{remark}

\section{Roadmap of Proofs}\label{sec:roadmap}
In this section, we introduce some additional  notation and then describe the structure and order of the main proofs.
We first prove results (of independent interest) on the empirical measures of $r$-tuples drawn from the LZ78 source and the empirical distribution of the $\Bv$ sequence, and then harness them for establishing  the main results characterizing the entropic properties of the source.

\subsection{Additional Notation and Definitions}
\subsubsection{Empirical Measures}
In a sense we make precise shortly, as the number of samples goes to $\infty$, the empirical distribution of $r$-tuples drawn from the LZ78 source of \prettyref{con:lz78-probability-source} converges to a law defined as follows. 
\newcommand{\etaStarRA}{\eta_r^*(A)}
\begin{definition}[Asymptotic Measure of $r$-Tuples]\label{def:eta-r-star}
    Let $\eta_r^*$ denote the probability law governing the pair of $r$-tuples $(\Theta^r, Y^r)$ where the components of $\Theta^r$ are drawn $\simiid \Pi$ and, conditioned on $\Theta^r$, those of $Y^r$ are independently generated with $Y_i \sim\Theta_i$.
    Note that marginally, under $\eta_r^*$, the components of $Y^r$ are $\simiid E[\Theta]$, where $\Theta \sim \Pi$. 
\end{definition}
We additionally need the following notation to state the empirical distribution results:
\begin{definition}[Single-Sequence Event]\label{def:single-sequence-event}
    Let event $A \subseteq \Mcal(\Acal)^r \times \Acal^r$ (\ie, corresponding to $r$-tuples of $\Theta$ values and corresponding symbols from the LZ78 source) be a \textit{single-sequence event} if there exists sequences $X_*^r \in \Acal^r$ and $\{A_i \subseteq \Mcal(\Acal)\}_{i=1}^r$ such that: 
    \[(X^r, \Theta^r) \in A \iff X^r = X_*^r, \,\,\text{and}\]
    \[\Theta_i \in A_i,\, \forall 1 \leq i \leq r,\]
    We denote the set of all single-sequence events by $\Scal_r$.
\end{definition}
\begin{definition}[Empirical Measure of $r$-Tuples]\label{def:empirical-measure-of-r-tuples}
    Let $\delta_\alpha$, where $\alpha$ is a point in some set $\Scal$, be the law that puts full mass on $\alpha$.
    For event $A \subseteq \Scal$, $\delta_\alpha(A) = \indic{\alpha \in A}$.
     Define the empirical measure of pairs of $r$-tuples from the LZ78 source as
    \[L_n^{(r)} = \frac{1}{n} \sum_{i=1}^n \delta_{B_i^{i+r-1}, X_i^{i+r-1}}.\]
\end{definition}
\begin{definition}[Zero-Order Empirical Measure of $\Bv$]
    The empirical measure of the $\Theta$ values encountered at each time point is denoted
    \[M_n = \frac{1}{n} \sum_{i=1}^n \delta_{B_i} \]
Note that $M_n$ can be obtained from $L_n^{(r)}$ by marginalizing out all but its  $\Theta_1$ component. 
\end{definition}
While the $\Theta_i$ values are $\simiid \Pi$ and, therefore, their empirical measure converges to $\Pi$ by the law of large numbers, the $B_i$ values have a non-trivial dependence structure.
Nevertheless, in a sense stated and proven formally below, their empirical measure converges to $\Pi$ as well.

\subsubsection{Further Definitions and Notation from \cite{sagan2024familylz78baseduniversalsequential}}
%To discuss the entropic properties of the LZ78 source, we  notation from %\cite{sagan2024familylz78baseduniversalsequential}.
\begin{definition}[Root Visits]\label{def:root-visits}
    For any sequence $x^n$, let $T_n = T_n (x^n)$ be the number of phrases in its LZ78 parsing, \ie, the number of times the root has been visited.\footnote{When the argument of $T(\cdot)$ is not specified, it is taken to be a realization, $X^n$, of the LZ78 source.}
    Note that this quantity was denoted $C(x^n)$ in \cite{sagan2024familylz78baseduniversalsequential}.
\end{definition}
\begin{definition}[LZ78 Tree Properties]\label{def:lz78-tree-properties}
    For individual sequence $x^n$, let $\Zcal(x^n)$ be the set of nodes in the LZ78 prefix tree representation of $x^n$.
    Note that $T_n(x^n)$ from \prettyref{def:root-visits} is equal to $|\Zcal(x^n)|$.
\end{definition}
\begin{definition}[Node Subsequences]
    For sequence $x^n$ and $z \in \mathcal{Z}(x^n)$, $y_z^{m_z}$ is the subsequence of $x^n$ seen while traversing node $z$ of the LZ78 tree, and $m_z$ is the length of that subsequence.
\end{definition}
\begin{definition}[Bayesian Mixture Distribution]\label{def:bayesian-mixture-dist}
    For probability measure $\Pi$ over $\Mcal(\Acal)$, we let  $q^\Pi$ denote the associated mixture of i.i.d. laws, \ie, for any individual sequence $x^n$, 
    \[q^\Pi(x^n) = \int_{\Mcal(\Acal)} \left[ \prod_{a\in\Acal}\Theta[a]^{\Ccal(a|x^n)} \right] d\Pi(\Theta),\]
    where $\Ccal(a|x^n)$ is the number of occurrences of $a$ along  $x^n$.
\end{definition}
\begin{remark}\label{rem:probability-of-lz78-sequence}
    For $X^n$ generated from the LZ78 probability source and any node $z \in \Zcal(X^n)$, the distribution of the symbols generated from that node satisfies
    \[Q^{\text{LZ},\Pi}_{Y_z^{m_z}}(y_z^{m_z}) = q^\Pi(y_z^{m_z}).\]
    As the current node of the LZ78 tree is a deterministic function of the preceding symbols generated and, conditioned on the current node, the next symbol depends only on the corresponding $\Theta$ value,
    \begin{align*}
        Q^{\text{LZ},\Pi}_{X^n}(x^n) = \prod_{i=1}^n Q^{\text{LZ},\Pi}_{X_i|X^{i-1}}(x_i|x^{i-1}) = \prod_{z\in \Zcal(x^n)} q^\Pi(y_z^{m_z}).
    \end{align*}
\end{remark}
% \begin{definition}[Root Visits]\label{def:root-visits}
%     Let $T_n$ be the number of times the root has been visited in the first $n$ steps, \ie, 
%     \[T_n = \max \{\ell \geq 1 \,:\, N(\ell) \leq n\}.\]
% \end{definition}

\subsection{Proof Roadmap}
The key empirical measure result is as follows:
\newcommand{\rtupletitle}{Empirical Distribution of $r$-Tuples}
\newcommand{\rtuplecontents}{
    Let $A \in \Scal_r$ be any single-sequence event.
    Then, for any LZ78 source,
    \[L_n^{(r)}(A) \convas \eta^*_r(A)\quad\text{as }n\to\infty.\]
}
\begin{theorem}[\rtupletitle]\label{thm:emp-dist-of-r-tuples}
    \rtuplecontents
\end{theorem}
\begin{corollary}\label{cor:emp-dist-of-r-tuples}
    Applying \prettyref{thm:emp-dist-of-r-tuples} with single-sequence event $A$ such that $A_1 = \cdots = A_r = \Mcal(\Acal)$, we obtain 
        \[\frac{1}{n}\sum_{i=1}^n \delta_{X_i^{i+r-1}}(a^{r}) \convas \prod_{t\in[r]}\Ebb[\Theta[a_t]]  \]
        as $n\to\infty,  \quad \forall a^r \in \Acal^r$.
        
        In words, the $r$th-order empirical distribution of the realized sequence from the LZ78 source converges,  almost surely, to the law of $r$ i.i.d. drawings from the PMF $\Ebb[\Theta]$. 
\end{corollary}
Although this theorem can be proven directly, it is more insightful to prove a similar result on the first-order empirical distribution of $\Bv$ and then extend the proof to $r$-tuples from $(\Bv, \Xv)$.
The proof of the first-order result contains the main ingredients of that of the $r$-tuple result, and is useful for establishing the entropic properties of the LZ78 source.

\newcommand{\zeroordertitle}{Zero-Order Empirical Distribution}
\newcommand{\zeroordercontents}{
    For any LZ78 source, $M_n$ converges to $\Pi$ weakly, almost surely.
}
\begin{theorem}[\zeroordertitle]\label{thm:zero-order-empirical-distribution}
    \zeroordercontents
\end{theorem}
Specifically, we prove that $M_n(A) \convas \Pi(A)$ for any measurable $A \subseteq \Mcal(\Acal)$.
The proof of this theorem has two main components.
First, we prove that the ratio of steps with $B_t \in A$ in the first $\ell$ LZ78 phrases approaches $\Pi(A)$ as $\ell\to\infty$, using the tree-recursive structure inherent in the LZ78 source, along with Chebyshev's inequality and the Borel-Cantelli lemma.
We also prove that, loosely, the number of symbols in the first $\ell$ phrases does not grow too quickly with $\ell$.
This is achieved by showing that the expected phrase length is on the order of $\log \ell$, and then showing that the deviation of the actual number of symbols about the expectation is small.
These two results, along with the monotonicity of the number of symbols in the first $\ell$ phrases, suffice to prove the theorem.

The extension to $r$-tuples is straightforward, as $\Theta$ values in any given path from the root to a leaf are drawn $\simiid \Pi$, and returns to the root are increasingly infrequent (so a vanishing number of $r$-tuples will contain a return to the root).

Subsequently, we prove the main results from \prettyref{sec:main-results}: namely, the entropic properties of the LZ78 source, fundamental limits of finite-state models on $\Xv$, and the relative entropy between the LZ78 source and a Markov law.
Given Theorem III.11 of\cite{sagan2024familylz78baseduniversalsequential}, the results are reasonably direct extensions of the study on empirical measures of $(\Xv, \Bv)$.
% \begin{lemma}[Extended Theorem III.10 from \cite{sagan2024familylz78baseduniversalsequential}]\label{lem:log-probability-converges-to-lz-compression-ratio}
%     For Bayesian mixture distribution $q^\Pi$ such that $\supp(\Pi) = \Mcal(\Acal)$,
%     \[\lim_{n\to\infty} \max_{x^n\in\Acal^n} \left| \frac{1}{n} \sum_{z\in\Zcal(x^n)} \log \frac{1}{q^\Pi(y_z^{m_z})} - \frac{T_n(x^n) \log T_n (x^n)}{n}\right| = 0.\]
% \end{lemma}
% This follows from some loose lower bounds on $q^{\Pi}(\cdot)$, the $\chi^2$ bound on relative entropy, and the fact that most steps occur at nodes with long corresponding subsequences. 
% Applying this lemma, we can leverage components of the proof of \prettyref{thm:zero-order-empirical-distribution} to prove that the normalized log probability of $\Xv$ almost surely converges to $\Ebb[H(\Theta)]$.
\rev{For priors with support over the full simplex, the entropy rate follows from~\thmref{thm:shannon-mcmillan-breiman} by means of the dominated convergence theorem.
However, the entropy rate result holds with no conditions on the prior. In the case where the prior does not have support over the full simplex, we are able to use~\lemref{lem:smb-in-prob-no-support} as a substitute for~\thmref{thm:shannon-mcmillan-breiman} to show convergence in probability and then use Vitali's theorem to find the entropy rate.}
Finally, the finite-state SPA log loss and relative entropy results follow from \prettyref{cor:emp-dist-of-r-tuples}. 
%\prettyref{thm:emp-dist-of-r-tuples}.

\section{Asymptotics of the  Empirical Distribution}\label{sec:asymp-dist}
\subsection{Zero-Order Empirical Distribution of $\Bv$ Sequence}\label{sec:zero-order-empirical-distribution}
\newcommand{\NAratioclaimcontents}{
    Suppose $\Ebb_{\Theta\sim\Pi} H(\Theta) > 0$.
    Then, for any strictly increasing sequence $\ell_k$ such that
    \[\ell_k = f(k) \exp{\sqrt{k}(\log k)^\alpha},\quad \frac{1}{2} < \alpha < 1,\]
    where $f(k)> 0$, $f(k+1) = f(k) + o(1)$, and $0 < \lim_{k\to\infty} f(k) < \infty$,
    \[\lim_{k\to\infty} \tfrac{N^A(\ell_k)}{N(\ell_k)} = \Pi(A)\quad(\text{a.s.}).\]
}
\newcommand{\Nratioclaimcontents}{
    If $\Ebb_{\Theta\sim\Pi} H(\Theta) > 0$, then, for the same $\ell_k$ as in \prettyref{claim:NA-to-N-ratio-converges-to-measure},
    \[\lim_{k\to\infty} \tfrac{N(\ell_{k+1})}{N(\ell_k)} = 1\quad(\text{a.s.}).\]
}
\textbf{Theorem \ref{thm:zero-order-empirical-distribution}} (\zeroordertitle)\textbf{.}
    \zeroordercontents
    
    \textit{Proof Sketch}.
        As stated above, we prove that $M_n(A) \convas \Pi(A)$, for any measurable $A \subseteq \Mcal(\Acal)$.
        
        First, note that distributions $\Pi$ satisfying $\Ebb_{\Theta\sim\Pi} H(\Theta) = 0$ can be considered a special case, for which this result is proven in \prettyref{app:proofs-for-supp-pi-01}.
        The proof is relatively direct, leveraging the Chernoff bound on the binomial distribution and the Borel-Cantelli lemma.
        For the remainder of the proof, we assume that $\Ebb_{\Theta\sim\Pi} H(\Theta) \triangleq h > 0$.

        For this proof and subsequent ones, it is helpful to consider empirical distributions over a certain number of fully-parsed phrases, \eg, evaluating $M_n$ for $n$ such that the currently-traversed node is the root of the tree.
        To this extent, we define the following notation:
        \begin{definition}[Step Counts]\label{def:step-counts}
            Let $N(\ell)$ be the number of symbols generated for the first $\ell$ visits to the root.
            $\forall A \subseteq \Mcal(\Acal)$, $N^A(\ell)$ is the number of such steps where $B_i \in A$.
        \end{definition}

        \begin{remark}
            $T_n (X^n)$ from \prettyref{def:root-visits} is equivalently
            \[T_n = T_n(X^n) = \max \{\ell \geq 1 \,:\, N(\ell) \leq n\}.\]
        \end{remark}

        \begin{remark}[Extreme Values of $N(\ell)$]\label{rem:extreme-values-of-N-ell}
            The tree grows no faster than one consisting of a single branch, so $N(\ell) \leq \frac{\ell(\ell+1)}{2}$.
            It also grows no slower than a full $|\Acal|$-ary tree, so $N(\ell) \geq C \ell \log \ell$, for some constant $C$ that solely depends on the alphabet.
            Due to the upper bound, $T_n \to \infty$ deterministically (\ie, on all individual sequences $\xv$) as $n\to\infty$.
        \end{remark}
        The proof of this theorem follows from the subsequent two claims, which are proven in Appendices \ref{app:NA-to-N-ratio-converges-to-measure-proof} and \ref{app:N-ratio-converges-to-one-proof}.
        
        \begin{claim}\label{claim:NA-to-N-ratio-converges-to-measure}
            \NAratioclaimcontents
            \textit{Proof Sketch}.
            Using the tree-structured incremental parsing of the LZ78 source, we develop a recursive formula for $N^A(\ell)$.
            This allows us to prove via induction that the second moment of $N^A(\ell)$ is bounded by a constant times $\ell^2$.
            Then, the almost sure convergence of $\frac{N^A(\ell_k)}{N(\ell_k)}$ follows from Chebyshev's inequality and the Borel-Cantelli lemma.
        \end{claim}
        \begin{claim}\label{claim:N-ratio-converges-to-one}
            \Nratioclaimcontents
            \textit{Proof Sketch}.
            First, we modify the problem statement: it suffices to show that the deviation of $(N(\ell_{k+1}) - N(\ell_k))$ about its first moment grows slower than $\ell_k\log \ell_k$.
            This requires us to show that the first moment of $N(\ell)$ also grows slower than $\ell_k \log \ell_k$, which follows from proving, via induction, that the expected length of the $\ell$\textsuperscript{th} LZ78 phrase is on the order of $\log \ell$.
            Then, via Chebyshev's inequality and the Borel-Cantelli lemma, the result of this claim reduces to proving a second-moment bound on $(N(\ell_{k+1}) - N(\ell_k))$.
            This moment bound is achieved using components of the proof for \prettyref{claim:NA-to-N-ratio-converges-to-measure} and the result on the order of the $\ell$\textsuperscript{th} phrase length.
        \end{claim}

        From these claims, \prettyref{thm:zero-order-empirical-distribution} can be proven via direct computation, leveraging the monotonicity of $N$ and $N^A$.
        Refer to \prettyref{app:end-of-zero-order-proof} for the remainder of the proof.

\subsection{Extension to $r$-Tuples}\label{sec:extension-to-r-tuples}
Given \prettyref{thm:zero-order-empirical-distribution}, we now prove the main result:

\textbf{Theorem \ref{thm:emp-dist-of-r-tuples}} (\rtupletitle)\textbf{.}
    \rtuplecontents
\textit{Proof Sketch}.
    The proof is similar to that of  \prettyref{thm:zero-order-empirical-distribution} and can re-use many results thereof, with some additional subtleties.
    As before, $\Ebb_{\Theta\sim\Pi} H(\Theta) = 0$ is considered a special case and covered in \prettyref{app:proofs-for-supp-pi-01}; for the rest of the proof, $\Ebb_{\Theta\sim\Pi} H(\Theta)$ is assumed to be $> 0$.
    
    First, some terms of $L_n^{(r)}$ involve a return to the root in the middle of an $r$-tuple.
    This would complicate the proof by causing dependence between disjoint branches of the tree, so we define the following two variants of $L_n^{(r)}$:
    \begin{align*}
        \underline{L}_n^{(r)}(A) &= \frac{1}{n}\sum_{i=1}^n (1-R_i)\delta_{B_i^{i+r-1}, X_i^{i+r-1}}(A); \\
        \overline{L}_n^{(r)}(A) &= \frac{1}{n} \sum_{i=1}^n \max\left(R_i, \delta_{B_i^{i+r-1}, X_i^{i+r-1}}(A)\right),
    \end{align*}
    where $R_i$ is an indicator for a return to the root in the $i$\textsuperscript{th} $r$-tuple.
    In other words, a term of $\underline{L}_n^{(r)}(A)$ is always $0$ if there is a return to the root between timepoints $i$ and $i+r-1$, and the corresponding terms of $\overline{L}_n^{(r)}(A)$ are always $1$.
    
    $\underline{L}_n^{(r)}(A) \leq L_n^{(r)}(A) \leq \overline{L}_n^{(r)}(A)$, so
    %, via the squeeze theorem, 
    the following two results suffice:
    \begin{lemma}\label{lem:LnrA-lower-bound}
        Assume $\Ebb_{\Theta\sim\Pi} H(\Theta) > 0$. Then, for any $A \in \Scal_r$, $\underline{L}_n^{(r)}(A) \convas \etaStarRA$ as $n\to\infty$.
    \end{lemma}
    \begin{lemma}\label{lem:LnrA-upper-bound}
        Assume $\Ebb_{\Theta\sim\Pi} H(\Theta) > 0$. Then, for any $A \in \Scal_r$, $\overline{L}_n^{(r)}(A) \convas \etaStarRA$ as $n\to\infty$.

        \textit{Proof Sketch}.
    \textup{Almost the entire proof of \prettyref{thm:zero-order-empirical-distribution} carries over to $\underline{L}_n^{(r)}(A)$ and $\overline{L}_n^{(r)}(A)$, except the equivalent of \prettyref{claim:NA-to-N-ratio-converges-to-measure}.
    Because of the returns to the root, the expectation of $N^{(r),A}(\ell)$ (the $r$-tuple equivalent of $N^A(\ell)$) is not quite $\etaStarRA \Ebb[N(\ell)]$.
    It, however, differs by an additive term of smaller order than $\Ebb[N(\ell)]$, so it is still possible to bound $\Ebb\left[(N^{A,(r)}(\ell) - \eta_r^*(\ell) \Ebb[N(\ell)])^2\right]$ by a constant times $\ell^2$, and then conclude that $\frac{N^{A,(r)}(\ell)}{N(\ell)} \convas \etaStarRA$ by Chebyshev's inequality and the Borel-Cantelli lemma.
    Then, the convergence results of $\underline{L}_n^{(r)}(A)$ and $\overline{L}_n^{(r)}(A)$ follow directly from work done in \prettyref{thm:zero-order-empirical-distribution}.}

    \textup{Refer to \prettyref{app:r-tuple-proofs} for the full proofs of these two lemmas.}
    \end{lemma}

\section{Entropic Properties, Finite-State Compressibility, and Relative Entropy Rates}\label{sec:entropic-properties}
Given components of \prettyref{thm:zero-order-empirical-distribution}, we can prove that the entropy rate of the $\Xv$ sequence from the LZ78 probability source is $\Ebb_{\Theta \sim \Pi}[H(\Theta)]$ (where $H(\cdot)$ is the Shannon entropy), and that its normalized log probability almost surely converges to the entropy rate.
% For both tasks, it is helpful to extend Theorem III.10 of \cite{sagan2024familylz78baseduniversalsequential} to priors with full support of $\Mcal(\Acal)$, rather than only those admitting densities bounded away from zero.

% \subsection{Log Probability and LZ78 Compression Ratio}
\rev{
\begin{remark}[Theorem III.11 from \cite{sagan2024familylz78baseduniversalsequential}]\label{rem:lz78-codelength-convergence}
    For convenience of reference, we restate Theorem III.11 of \cite{sagan2024familylz78baseduniversalsequential} using our notation.
    
    For $q^\Pi$ as defined in \prettyref{def:bayesian-mixture-dist} for prior with $\supp(\Pi) = \Mcal(\Acal)$,
    \[\lim_{n\to\infty} \max_{x^n\in\Acal^n} \left| \frac{1}{n} \sum_{z\in\Zcal(x^n)} \log \frac{1}{q^\Pi(y_z^{m_z})} - \ell_\text{LZ}(x^n)\right| = 0,\]
    where $\ell_\text{LZ}(x^n) \triangleq \frac{T_n (x^n) \log T_n (x^n) }{n}$ is the normalized asymptotic LZ78 codelegth for $x^n$.
\end{remark}}

\subsection{Entropy Rate; Shannon-McMillan-Breiman-Type Result}
We first prove the Shannon-McMillan-Breiman-style \cite{breiman1957Ergodic} result that the normalized log probability of $\Xv$ drawn from the LZ78 probability source almost surely converges to $\Ebb_{\Theta \sim \Pi}[H(\Theta)]$ given $\supp(\Pi) = \Mcal(\Acal)$.
We also show that the entropy rate, under no support assumption, follows from a modified version of the Shannon-McMillan-Breiman-style result.

\textbf{Theorem \ref{thm:shannon-mcmillan-breiman}} (\smbtheoremtitle)\textbf{.}
    \smbtheoremcontents

    \textit{Proof Sketch}.
    \textup{
    Applying \prettyref{rem:lz78-codelength-convergence}, it remains to show that $\frac{T_n \log T_n}{n} \convas \Ebb[H(\Theta)]$.
    Equivalently, we show $\frac{\ell \log \ell}{N(\ell)}$ almost surely converges to $\Ebb[H(\Theta)]$.
    By \prettyref{cor:m1ell-bound}, this holds if the denominator is replaced by $\Ebb[N(\ell)]$.
    So, we can complete the proof by showing that the ratio between $N(\ell)$ and its expectation converges almost surely to $1$, which follows directly from results in \prettyref{sec:asymp-dist}.}

    \textup{The full proof can be found in \prettyref{app:smb-proof}}.

\rev{\textbf{Theorem \ref{thm:entropy-rate}} (\entropyratetitle)\textbf{.}
    \entropyratecontents

\begin{proof}
    From~\lemref{lem:smb-in-prob-no-support} in the appendix, we know that if $\Ebb[H(\Theta)] > 0$, then
    \[ 
        L_n: =\frac{1}{n} \log \frac{1}{P_{X^n}(X^n)} \overset{p}{\to} \Ebb[H(\Theta)]
    \]
    as $n \to \infty$. By Vitali's theorem, it suffices to show uniform integrability of $\set{L_n}$, which is shown in~\lemref{lem:ui_for_L}. If we have that $\Ebb[H(\Theta)] = 0$, then we note that the proof method of Lemma D.1 in~\cite{sagan2024familylz78baseduniversalsequential} can be used to show the entropy rate is $0$ by arguing that the entropy added per phrase is $1$.
\end{proof}}

\subsection{Finite-State Modeling of the LZ78 Source}
The optimal log loss of a Markov sequential probability assignment (from \prettyref{def:mu-x}) for any $\Xv$ emitted by the LZ78 source almost surely converges to $H(\Ebb[\Theta])$, as we now show using \prettyref{cor:emp-dist-of-r-tuples}.

By \cite{sagan2024familylz78baseduniversalsequential}, this is also equal to the optimal finite-state sequential probability assignment log loss and the finite-state compressibility of \cite{originalLZ78paper} (scaled by $\log |\Acal|$).

\textbf{Theorem \ref{thm:markov-log-loss}} (\markovloglosstitle) \textbf{.}
    \markovlogloscontents

    \begin{proof}
        By \prettyref{cor:emp-dist-of-r-tuples}, $\forall x^{k+1} \in \Acal^{k+1}$, almost surely
        \begin{align*}
            \lim_{n\to\infty} \frac{\Ccal(x^{k+1}|X^{n+k})}{n} &\triangleq \lim_{n\to\infty} \frac{1}{n}\sum_{i=1}^n \delta_{X_i^{i+k}}(x^{k+1})\\
            &= \prod_{t\in[k+1]}\Ebb[\Theta[x_t]],
        \end{align*}
        which is the law of i.i.d. $(k+1)$-tuples with distribution $\Ebb[\Theta]$.
        
        $\mu_k(X^n)$ is a continuous function of
        \[\left\{ \frac{\Ccal(x^{k+1}|X^{n+k})}{n} : x^{k+1} \in \Acal^{k+1} \right\},\]
        so, by the continuous mapping theorem, almost surely
        \[\lim_{n\to\infty} \mu_k(X^n) = \lim_{n\to\infty} \hat{H}_{X_{k+1}|X^k}(X^n) = \Hbb(\Ebb[\Theta]) \]
        since the $(k+1)$-order empirical distribution converges to an i.i.d. law with distribution $\Ebb[\Theta]$.

        As this holds for any finite $k$, it also holds in the limit $k\to\infty$, so $\mu(\Xv) = H(\Ebb[\Theta])$ almost surely.
    \end{proof}

$H(\Ebb[\Theta])$ differs from the entropy rate of $\Xv$ by a ``Jensen gap,'' as entropy is concave.
Thus, given any finite amount of memory, the per-symbol log loss of any sequential probability model will be bounded away from the entropy rate by at least the Jensen gap, on almost all realizations of the LZ78 source.

\rev{Note that this is an asymptotic result; in the finite sample regime, finite-state probability models can and sometimes do get close to the log probability of the realized sequences.}

\subsection{Relative Entropy between the LZ78 Source and a Markovian Law}
With the framework developed thus far, it is also possible to explore the relative entropy between the LZ78 source and a $k$-order Markov probability law.
By \cite{sagan2024familylz78baseduniversalsequential}, the normalized relative entropy between $P_{X^n}$ and $Q_{X^n}^{\text{LZ}, \Pi}$, where $P$ is any stationary law, is asymptotically zero.
We now show that this is not case for the normalized relative entropy between $Q_{X^n}^{\text{LZ}, \Pi}$ and $P_{X^n}$.

\textbf{Theorem \ref{thm:relative-entropy-result-pointwise}} (\pointwisedivergencetitle)\textbf{.}
     Let $\Xv$ be generated from the LZ78 source $Q^{\text{LZ}, \Pi}$ with
    $\supp(\Pi) = \Mcal(\Acal)$. For any $k$\textsuperscript{th}-order Markov law
    $P \in \Mcal_k$, almost surely
    \begin{align*}
    &\lim_{n \to \infty}
    \frac{1}{n}
    \log
    \frac{Q_{X^n}^{\text{LZ}, \Pi}(X^n)}{P_{X^n}(X^n)} = H(\Ebb[\Theta])
    - \Ebb[H(\Theta)]
    \\
    &\hspace{5em} + \Ebb \Bigl[
    D\Bigl(
    \Ebb[\Theta]
    \,\Big\lVert\,
    P_{X_0 \mid X_{-k}^{-1}}
    (\cdot \mid Y_{-k}^{-1})
    \Bigr)
    \Bigr],
    \end{align*}
    where on the right-hand side
    $\Theta \sim \Pi$ and
    $Y_{-k}^{-1} \simiid \Ebb[\Theta]$.

\begin{proof}
    For this proof, define shorthand notation
    \begin{align*}
        D_{QP}(X^n) &\triangleq \frac{1}{n} \log \frac{Q_{X^n}^{\text{LZ}, \Pi}(X^n)}{P_{X^n}(X^n)}, \\
        \ell_n(X^n) &\triangleq \frac{1}{n} \log \frac{1}{P_{X^n}(X^n)}.
    \end{align*}
    
    Rearranging terms in $D_{QP}(X^n)$,
    \begin{align*}
        \lim_{n\to\infty} D_{QP}(X^n) &= \lim_{n\to\infty} \ell_n(X^n) - \frac{1}{n}\log \frac{1}{Q^{\text{LZ}, \Pi}_{X^n}(X^n)} \\
        &= \lim_{n\to\infty} \ell_n(X^n) - \Ebb[H(\Theta)]\quad(\text{a.s.}),
    \end{align*}
    by \prettyref{thm:shannon-mcmillan-breiman}.
    We now must evaluate the first term of the right-hand side.
    As $P$ is $k$-order Markov,
    \begin{align*}
        \ell_n(X^n) &= \frac{1}{n} \sum_{t=k+1}^n \log \frac{1}{P_{X_0|X_{-k}^{-1}}(X_{t}|X^{t-1}_{t-k})} \\
        &\qquad+ \frac{1}{n}\underbrace{ \sum_{t=1}^k \log \frac{1}{P_{X_0}|X_{-t+1}^{-1}(X_t|X^{t-1})}}_{\triangleq f(X^k)} \\
        &=\!\!\sum_{x^{k+1} \in \Acal^{k+1}} \frac{\Ccal(x^{k+1}|X^{n+k})}{n} \log \frac{1}{P_{X_0|X_{-k}^{-1}}(x_{k+1}|x^k)} \\
        &\quad+ \frac{1}{n}f(X^k) - \frac{1}{n}\underbrace{ \sum_{t=n+1}^{n+k} \log \frac{1}{P_{X_0|X_{-k}^{-1}}(X_t|X^{t-1}_{t-k})}}_{\triangleq g(X_{n+1}^{n+k})}.
    \end{align*}

   By \prettyref{cor:emp-dist-of-r-tuples}, $\frac{\Ccal(x^{k+1}|X^{n+k})}{n}$ almost surely converges to the law of i.i.d. $(k+1)$-tuples with distribution $\Ebb[\Theta]$.
    
    We now consider two cases on $P$.
    First, assume $\exists x^{k+1} \in \Acal^{k+1}$ such that $P_{X^{k+1}}(x^{k+1}) = 0$.
    Then, almost surely, there will be a term in $\frac{1}{n} \log \frac{1}{P(X^n)}$ that is infinite: as $\supp(\Pi) = \Mcal(\Acal)$, $\Ebb[\Theta]$ is bounded away from the edge of the simplex, so $\frac{\Ccal(x^{k+1}|X^{n+k})}{n}$ almost surely converges to a positive value.
    In this case, however, the right-hand side in the theorem statement is also infinite, so this theorem still holds.

    Now, assume that $P_{X^{k+1}}(x^{k+1}) > 0$ for all $x^{k+1} \in \Acal^{k+1}$.
    Then, $f(X^k)$ and $g(X_{n+1}^{n+k})$ from above are uniformly bounded for all $X^k$, $X_{n+1}^{n+k}$, so $\frac{f(X^k)}{n}$ and $\frac{g(X_{n+1}^{n+k})}{n}$ are deterministically $o(1)$.
    We can apply this fact and Slutsky's theorem to get
    \begin{align*}
        &\frac{1}{n} \log \frac{1}{P_{X^n}(X^n)} \convas \Ebb_{Y_{-k}^0 \simiid \Ebb[\Theta]} \left[\log \frac{1}{P_{X_0|X_{-k}^{-1}}(Y_0|Y_{-k}^{-1})}\right],
    \end{align*}
    which evaluates to
    \begin{align*}
        \Ebb_{Y_{-k}^{-1} \simiid \Ebb[\Theta]} \left[D\left(\Ebb[\Theta] \,\Big\lVert\, P_{X_0|X_{-k}^{-1}}(\cdot|Y_{-k}^{-1})\right)\right] + H(\Ebb[\Theta]).
    \end{align*}
    Putting everything together,
    \begin{align*}
        \frac{1}{n} \log \frac{Q_{X^n}^{\text{LZ}, \Pi}(X^n)}{P_{X^n}(X^n)} &\convas  H(\Ebb[\Theta]) - \Ebb[H(\Theta)] \\
        &\hspace{-1em}+\Ebb_{Y_{-k}^{-1} \simiid \Ebb[\Theta]} \left[D\left(\Ebb[\Theta] \,\Big\lVert\, P(\cdot|Y_{-k}^{-1})\right)\right].
    \end{align*}
\end{proof}

\textbf{Corollary \ref{thm:relative-entropy-result}} (\relativeentropytitle)\textbf{.}
    For $Q^{\text{LZ}, \Pi}$ and any $k$\textsuperscript{th}-order Markov law $P \in \Mcal_k$,
    \begin{align*}
        &\lim_{n\to\infty} \frac{1}{n} D\left(Q_{X^n}^{\text{LZ}, \Pi} \big\lVert P_{X^n}\right) = H(\Ebb[\Theta]) - \Ebb[H(\Theta)] \\
        &\hspace{5em} + \Ebb \left[D\left(\Ebb[\Theta] \,\Big\lVert\, P_{X_0|X_{-k}^{-1}}(\cdot|Y_{-k}^{-1})\right)\right],
    \end{align*}
    where on the right side $\Theta \sim \Pi$ and $Y_{-k}^{-1} \simiid \Ebb[\Theta]$.
    \rev{Note that, whereas \prettyref{thm:relative-entropy-result-pointwise} is a an almost-sure result, this result is in expectation.}
\begin{proof}
    The proof is nearly identical to that of \prettyref{thm:relative-entropy-result-pointwise}, and thus we only state where it differs: we only need to make two modifications.
    First, we can apply \prettyref{thm:entropy-rate} (the entropy rate of the LZ78 source) in lieu of \prettyref{thm:shannon-mcmillan-breiman} (the almost-sure convergence of the normalized log probability).
    In addition, recognizing that $\frac{\Ccal(x^{k+1}|X^{n+k})}{n}$ is bounded by $1$, its almost-sure convergence to the law of i.i.d. $(k+1)$-tuples also occurs in expectation by dominated convergence.
\end{proof}

\rev{Note that \prettyref{thm:relative-entropy-result-pointwise} and \prettyref{thm:relative-entropy-result} imply that the Markovian law that minimizes the asymptotic relative entropy with the LZ78 source is the constant law $\Ebb[\Theta]$.
\textit{I.e.}, higher-order Markov models do not \textit{asymptotically} better approximate the LZ78 source.
This is also implied by \prettyref{thm:markov-log-loss}, the proof which states that the $k$-order Markov predictability for a realization from the LZ78 source, for any $k$, is $H(\Ebb_{\Theta \sim \Pi}[\Theta]$).
This is due to the fact that the LZ78 tree grows infinitely deep, so, for any fixed $k$, the ratio of symbols corresponding to nodes at depth $\leq k$ vanishes as $n\to\infty$.
However, if we consider finite-length realizations from the LZ78 source, it is true that higher-order Markov models can better learn the realizations.
In other words, the larger the Markov context $k$, the slower the convergence of $\mu_k(X^n)$ to $H(\Ebb[\Theta])$, \textit{i.e.}, the model can maintain a low log loss for longer sequences.
This is corroborated by empirical results from \prettyref{sec:experiments}.
}

\section{Experiments}\label{sec:experiments}
\begin{figure}[tbp]
    \includegraphics[width=\linewidth]{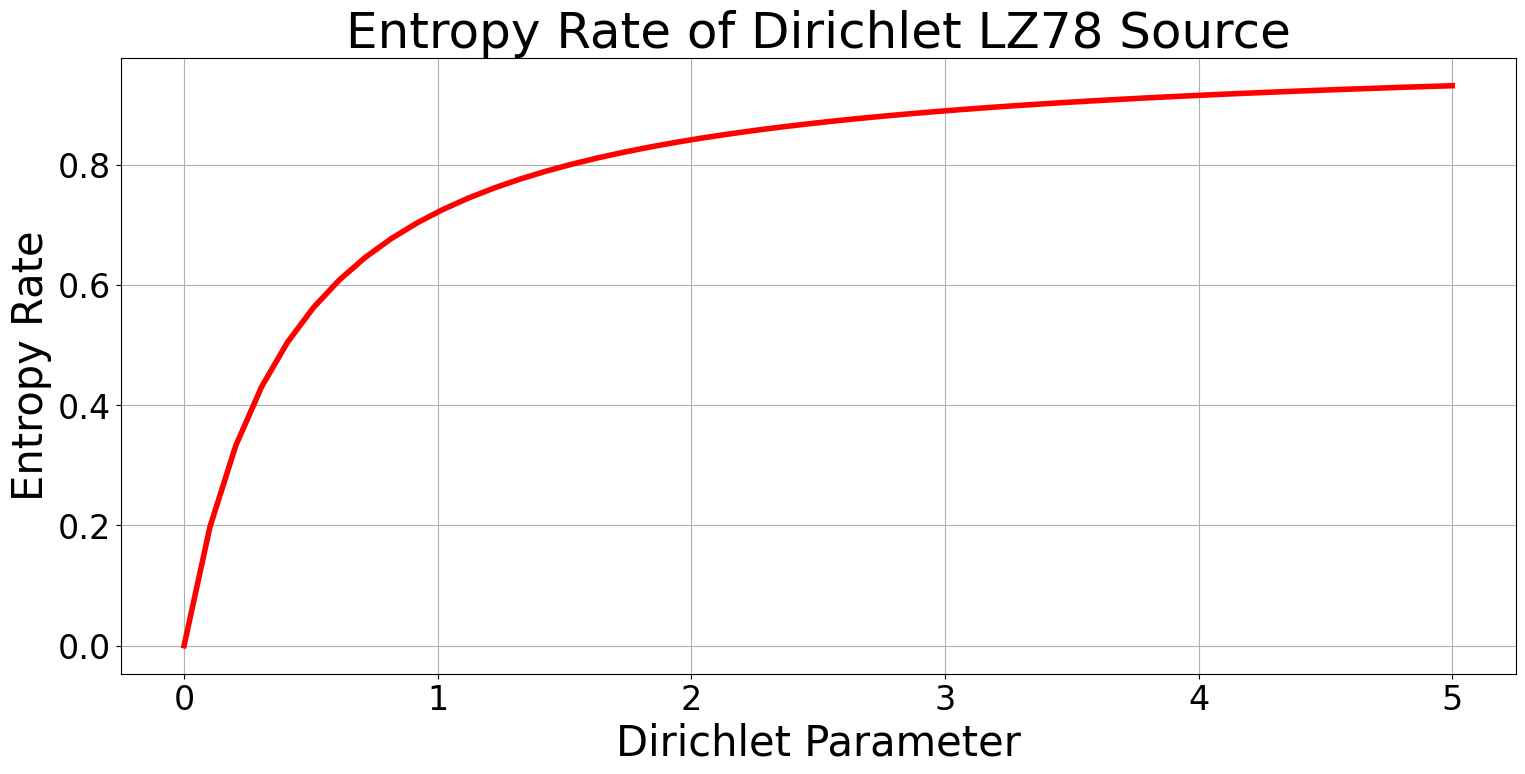}
    \caption{Entropy Rate of the LZ78 Source for a Dirichlet($\gamma \dots \gamma$) prior.}
    \label{fig:entropy-rates}
\end{figure}

\rev{First, in \prettyref{sec:simulations}, we present simulation results to empirically verify the paper's key theoretical results.
Next, in \prettyref{sec:icl_study}, we apply the LZ78 source to the study of in-context learning in transformer models.}

\subsection{Simulation Results}\label{sec:simulations}

We present simulations demonstrating  some of the key theoretical results: \prettyref{thm:shannon-mcmillan-breiman} (that the normalized log probability of the LZ78 source almost surely converges to $\Ebb[H(\Theta)]$) and \prettyref{thm:markov-log-loss} (that the log loss of the best Markov probability model, applied to the LZ78 source, will almost surely approach $H(\Ebb[\Theta])$).
\rev{These simulation results not only provide empirical validation of the key theoretical results, but also provide some insight into the finite-sample behavior of LZ78 source realizations' log probability and finite-context Markov predictability.
}

We study the LZ78 source under a Dirichlet prior, which is a canonical prior for Bayesian mixture distributions, as the mixture evaluates to a perturbation of the empirical distribution \cite{cover1972admissibility,krichevskiyTrofimov1981}.
We also consider a distribution that is mixture of point masses and a Dirichlet distribution.
This serves as an approximation of a point mass distribution that satisfies the condition $\supp(\Pi) = \Mcal(\Acal)$ (required in our proofs of the entropic properties).

For each prior, $\Pi$, we perform two experiments.
First, we plot the normalized log probability (with respect to the sequence length thus far) for five realizations of the LZ78 source and compare it to the entropy rate and $\mu(\Xv)$.
Then, for one realization, we compute the Markov probability model log losses $\mu_0$, $\mu_5$, and $\mu_{10}$, again comparing to the entropy rate and $\mu(\Xv)$.
\rev{Each data point for $\mu_k$ is computed by computing the $k$-tuple and $(k+1)$-tuple counts of the sequence up to (and including) that point, and then computing
\[\mu_k(X^\ell) = \frac{1}{\ell} \!\!\!\! \sum_{\substack{m^{k+1} \in \mathcal{A}^{k+1} :\\ \texttt{counts}_{k+1}(m^{k+1}) > 0}} \!\!\!\!\log\left(\frac{\texttt{counts}_k(m^k)}{\texttt{counts}_{k+1}(m^{k+1})}\right).\]
}

\subsubsection{The Dirichlet Prior}\label{sec:dirichlet-prior}
We consider three binary and one ternary LZ78 source with $\Pi$ of the form Dirichlet($\gamma, \dots, \gamma$).
For a binary LZ78 source with this prior, the entropy rate (and normalized asymptotic log probability) is\footnote{Derived via Mathematica}
\[\Ebb[H(\Theta)] = -\frac{\sqrt{\pi} 2\Gamma(2\gamma) 4^{-\gamma} (H_\gamma - H_{2\gamma})}{\Gamma(\gamma + 1/2) \Gamma(\gamma)^2 \ln 2} \triangleq H_\text{Dirichlet}(\gamma),\]
where $\Gamma$ is the Gamma function and $H_x$ is the harmonic number function (as the sum of the digamma function and Euler-Mascheroni constant).
This entropy rate function is plotted in \prettyref{fig:entropy-rates}.
Due to the symmetry of the Dirichlet prior we use, $\Ebb[\Theta[a]] = 1/|\Acal|$, $\forall a \in \Acal$, resulting in $\mu(\Xv) = \log |\Acal|$.

\begin{figure*}[tbp]
    \centering
    \begin{subfigure}{0.45\linewidth}
        \centering
        \includegraphics[width=0.99\linewidth]{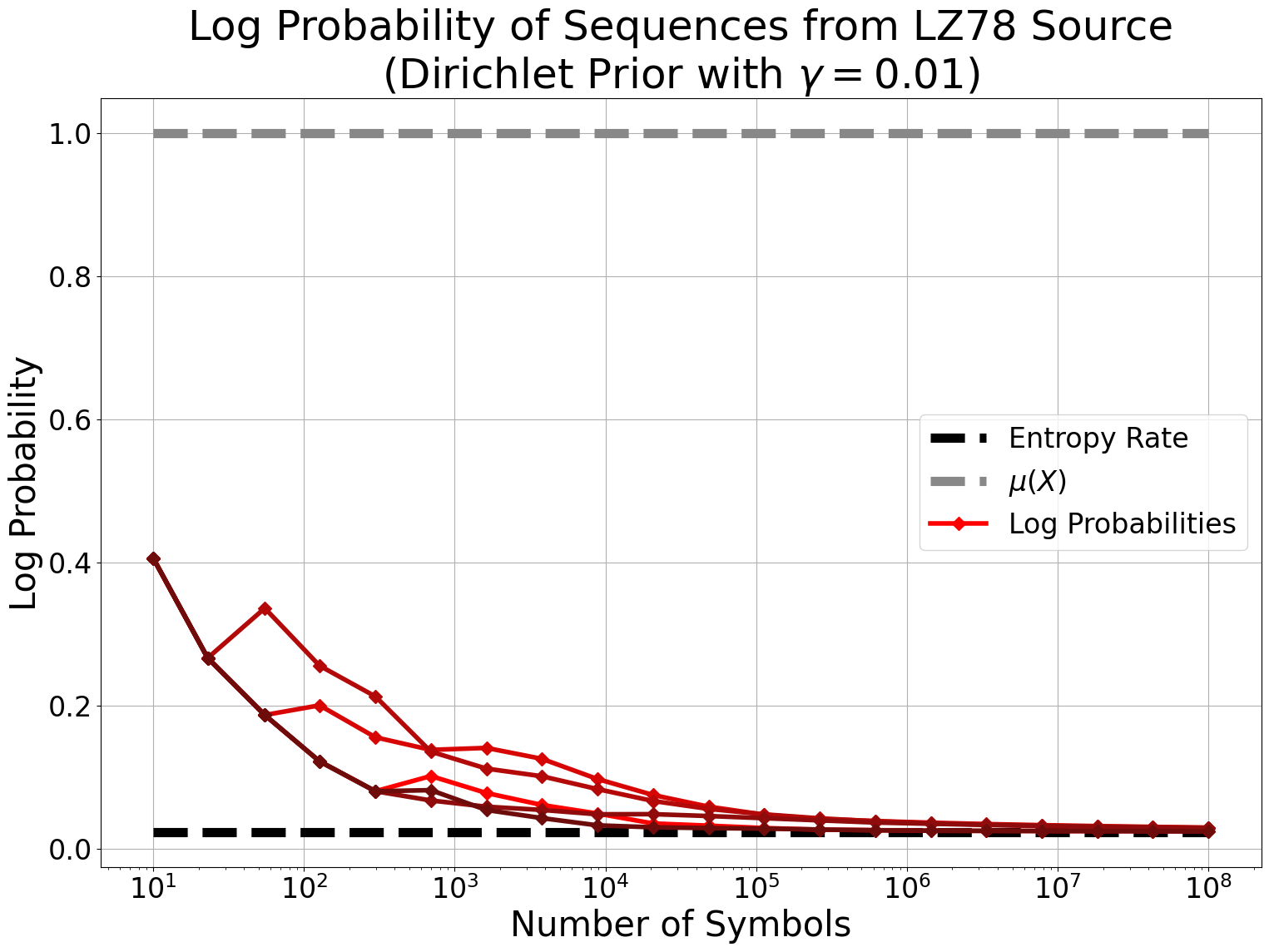}
        \caption{\inlineRev Normalized log probability of five realizations of the LZ78 source, with the entropy rate and $\mu(\Xv)$ labeled.}
    \end{subfigure}
    \begin{subfigure}{0.45\linewidth}
        \centering
        \includegraphics[width=0.99\linewidth]{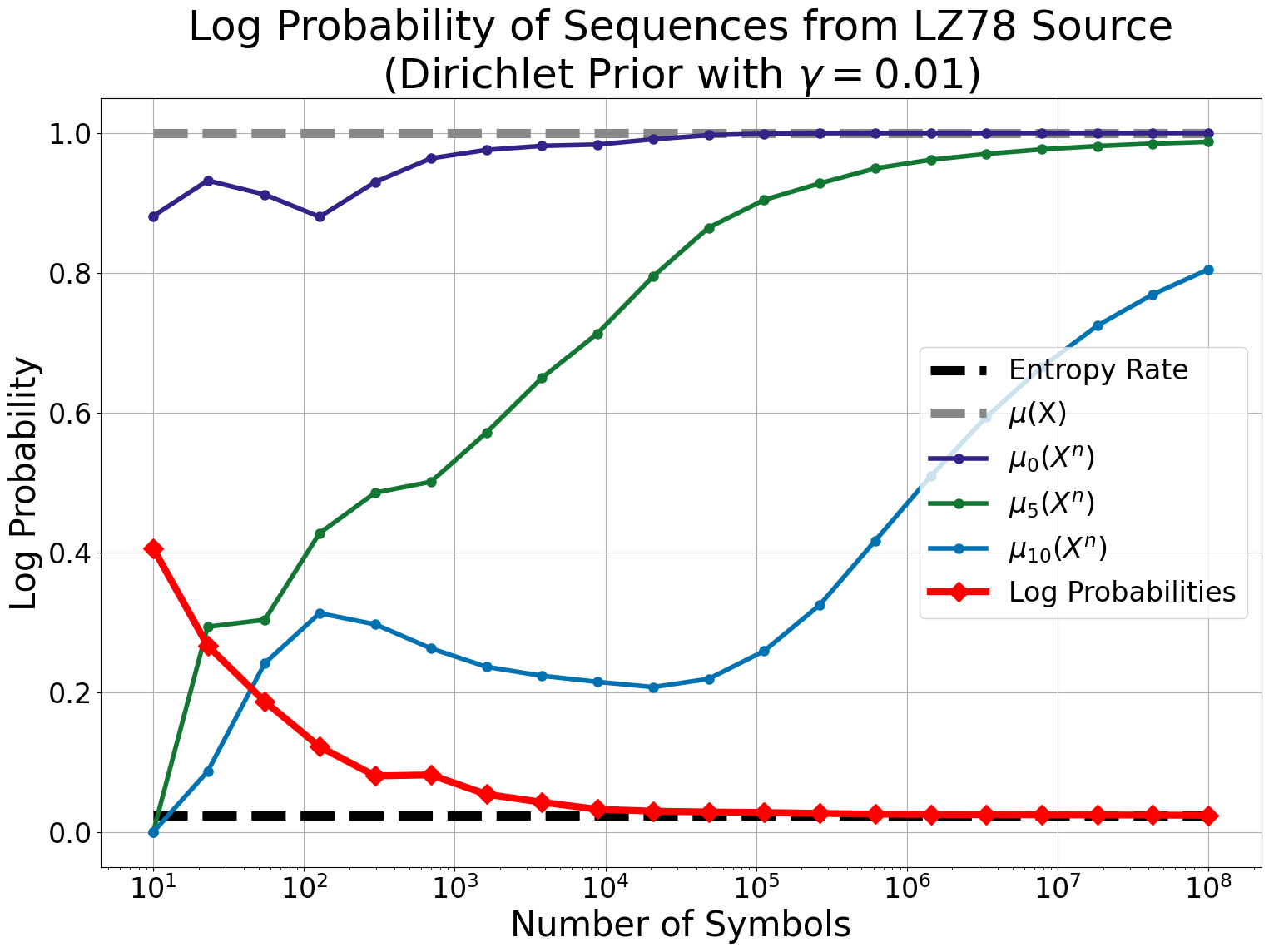}
        \caption{\inlineRev Normalized log probability of one realization of the LZ78 source, and log loss of Markov sequential probability models.}
    \end{subfigure}
    \caption{\inlineRev Simulation results, where $\Pi$ is the Dirichlet(0.01, 0.01) distribution.}
    \label{fig:dirichlet-gamma-0-01}
\end{figure*}

\begin{figure*}[tbp]
    \centering
    \begin{subfigure}{0.45\linewidth}
        \centering
        \includegraphics[width=0.99\linewidth]{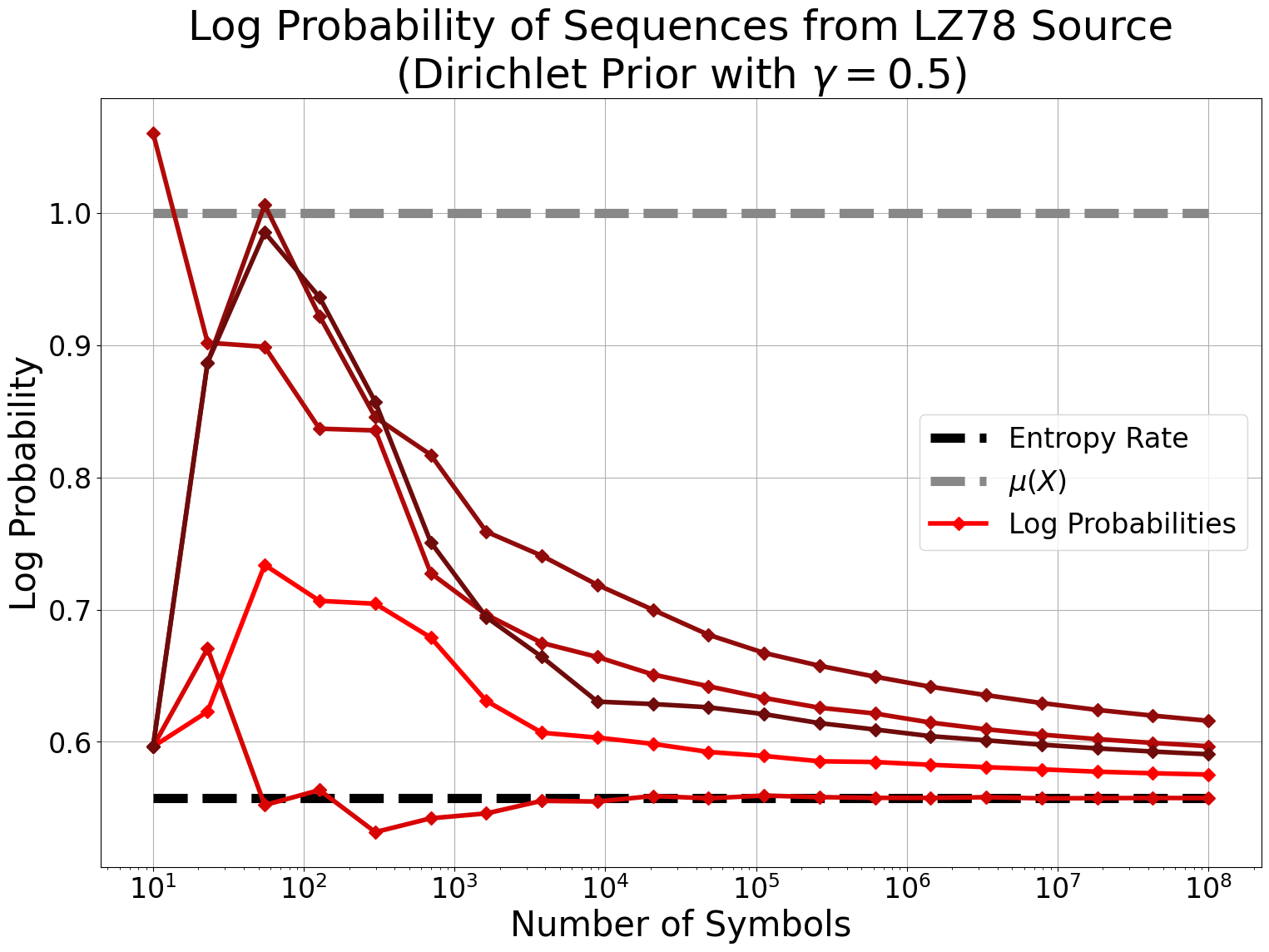}
        \caption{Normalized log probability of five realizations of the LZ78 source, with the entropy rate and $\mu(\Xv)$ labeled.}
    \end{subfigure}
    \begin{subfigure}{0.45\linewidth}
        \centering
        \includegraphics[width=0.99\linewidth]{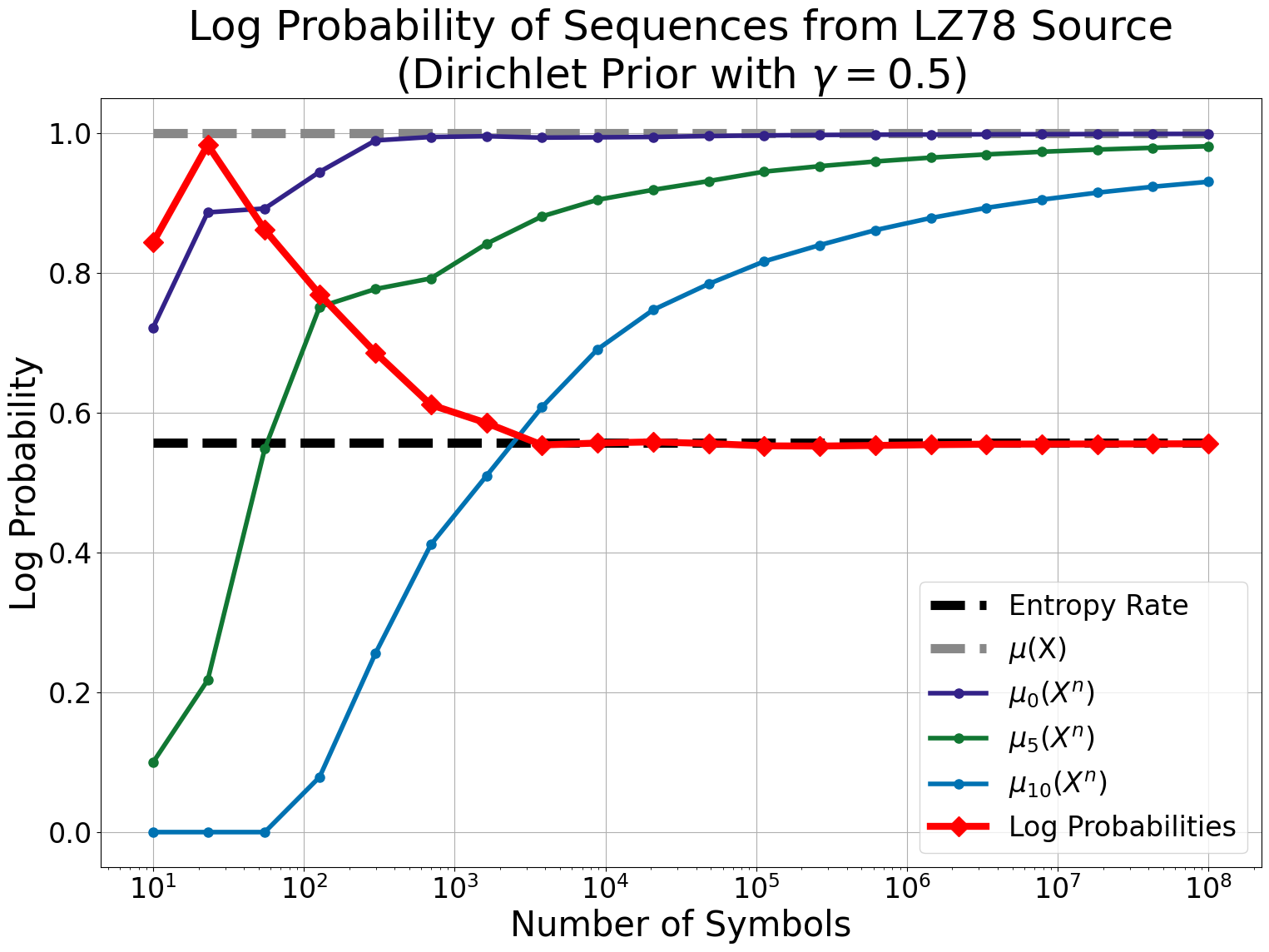}
        \caption{Normalized log probability of one realization of the LZ78 source, and log loss of Markov sequential probability models.}
    \end{subfigure}
    \caption{Simulation results, where $\Pi$ is the Jeffreys prior (Dirichlet(0.5, 0.5)).}
    \label{fig:dirichlet-gamma-0-5}
\end{figure*}

\begin{figure*}[tbp]
    \centering
    \begin{subfigure}{0.45\linewidth}
        \centering
        \includegraphics[width=0.99\linewidth]{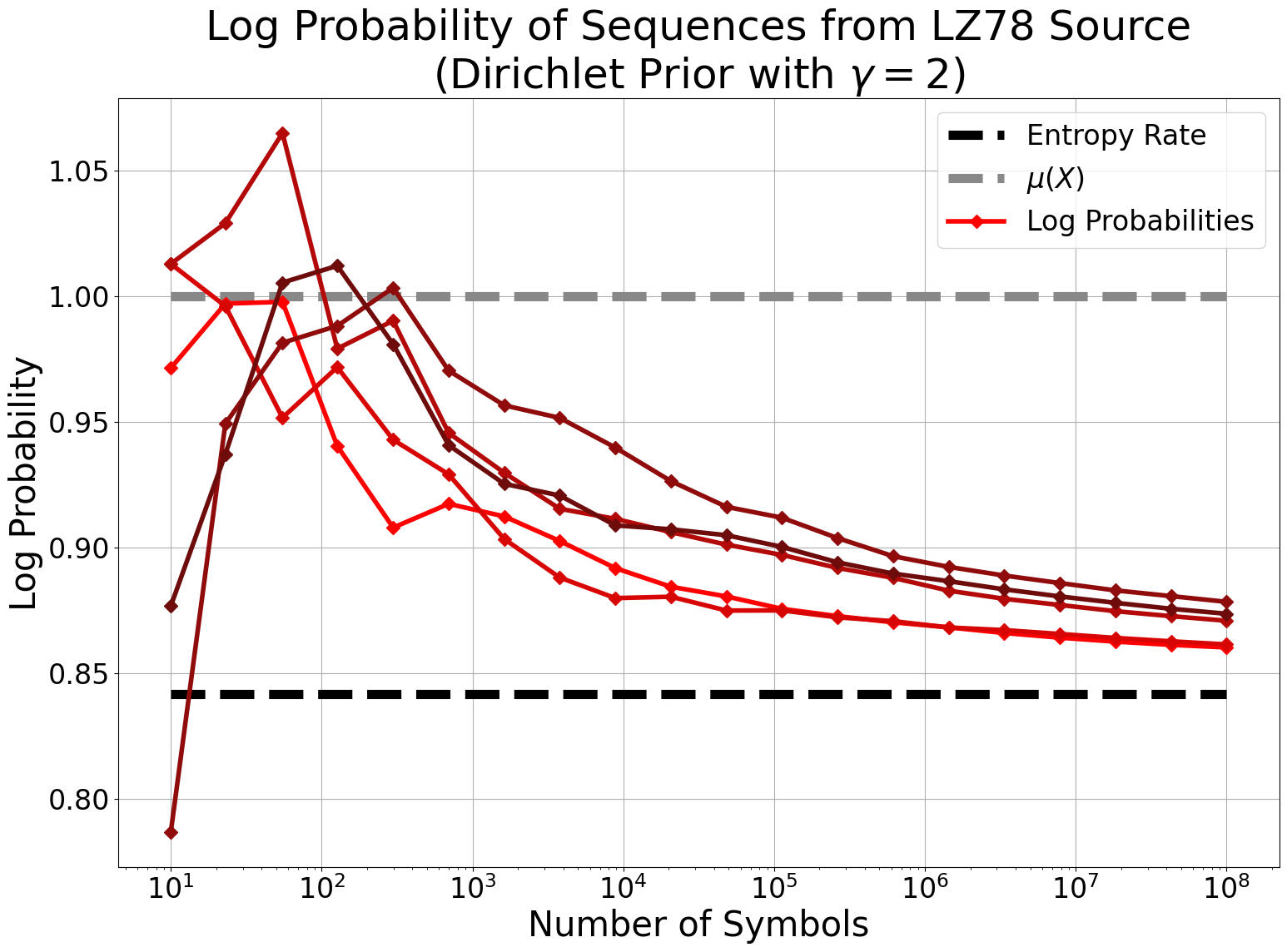}
        \caption{Normalized log probability of five realizations of the LZ78 source, with the entropy rate and $\mu(\Xv)$ labeled.}
    \end{subfigure}
    \begin{subfigure}{0.45\linewidth}
        \centering
        \includegraphics[width=0.99\linewidth]{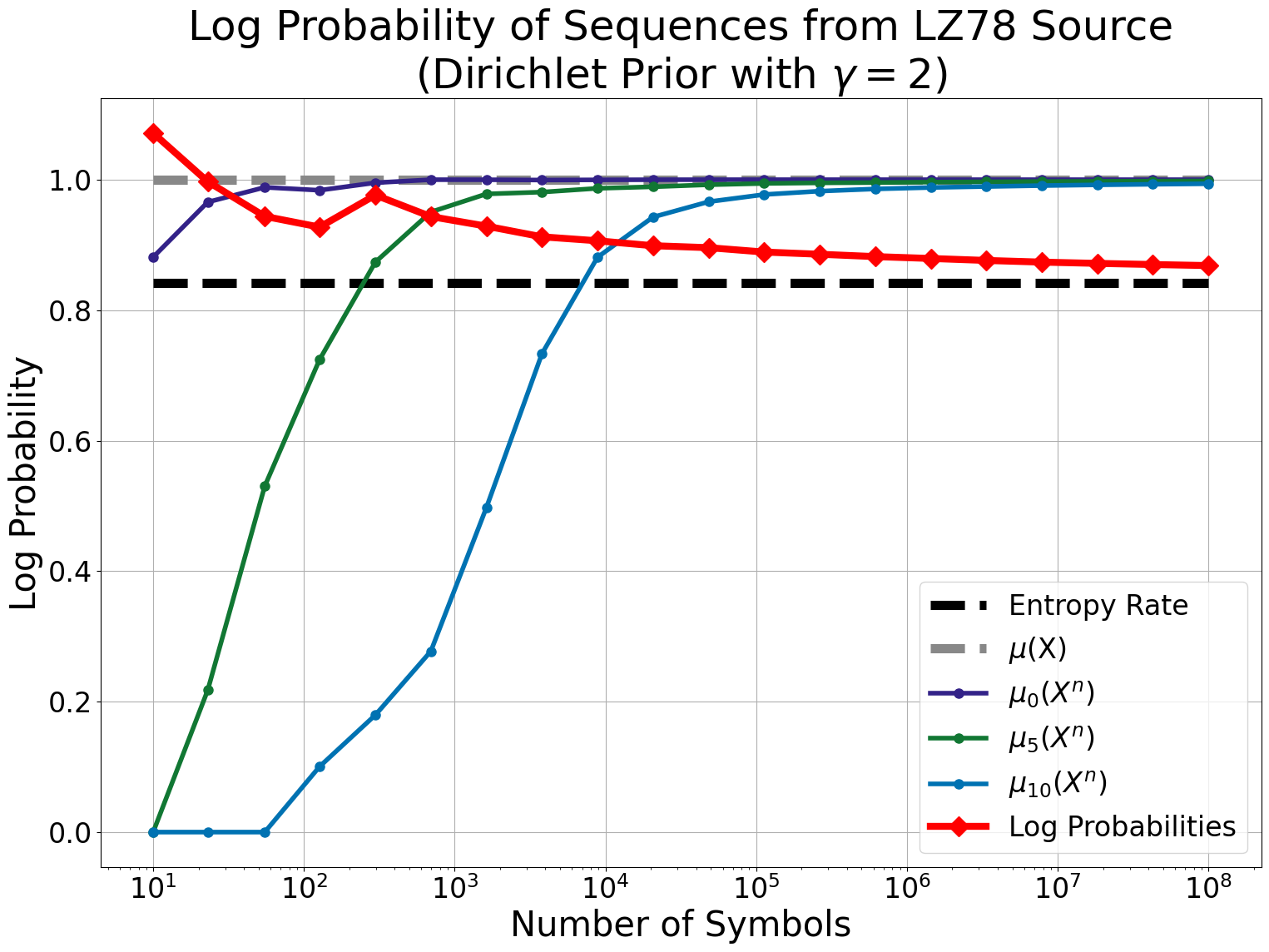}
        \caption{Normalized log probability of one realization of the LZ78 source, and log loss of Markov sequential probability models.}
    \end{subfigure}
    \caption{Simulation results, where $\Pi$ is the Dirichlet($2, 2$) distribution.}
    \label{fig:dirichlet-gamma-2}
\end{figure*}

\begin{figure*}[tbp]
    \centering
    \begin{subfigure}{0.45\linewidth}
        \centering
        \includegraphics[width=0.99\linewidth]{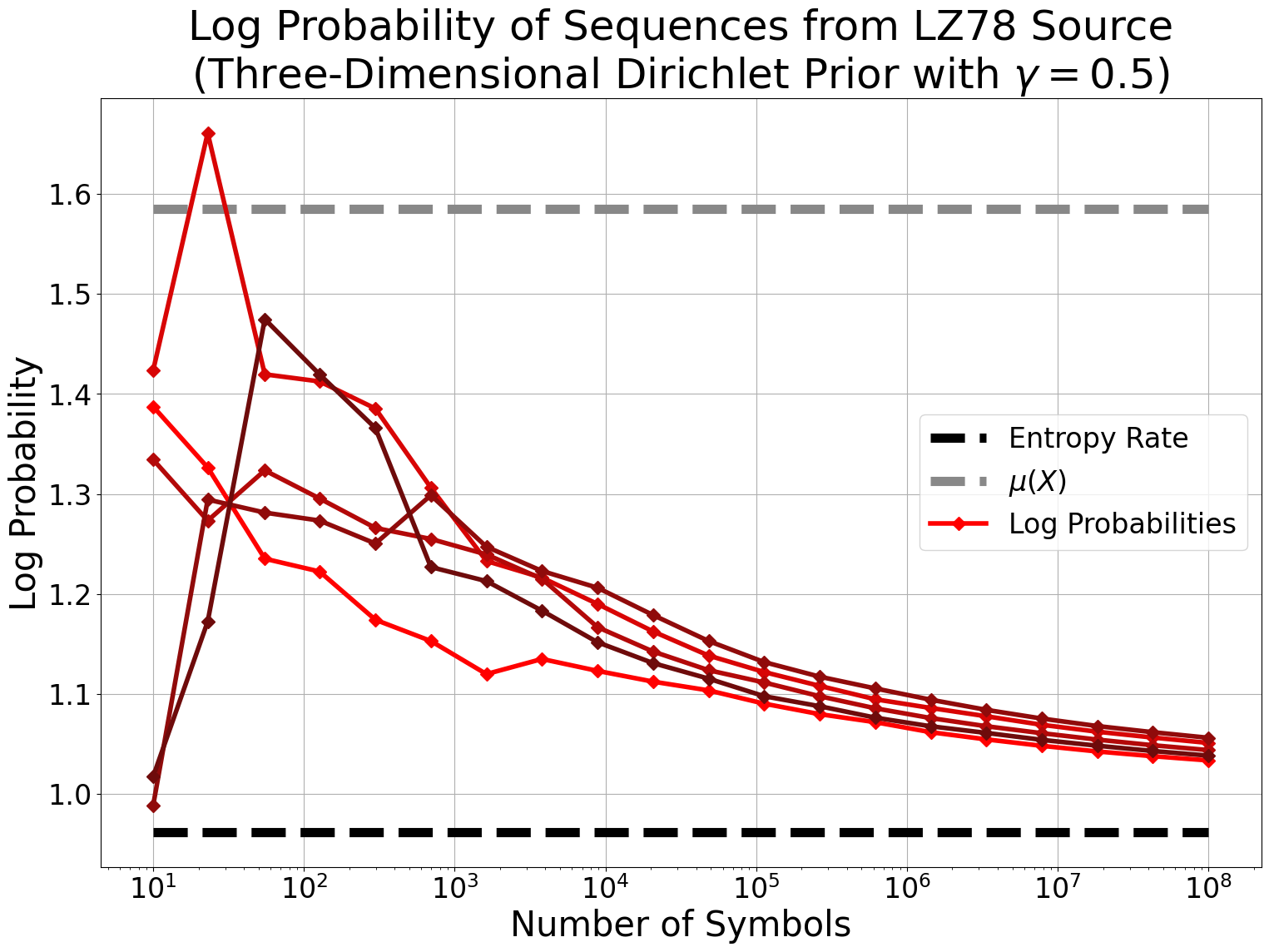}
        \caption{Normalized log probability of five realizations of the LZ78 source, with the entropy rate and $\mu(\Xv)$ labeled.}
    \end{subfigure}
    \begin{subfigure}{0.45\linewidth}
        \centering
        \includegraphics[width=0.99\linewidth]{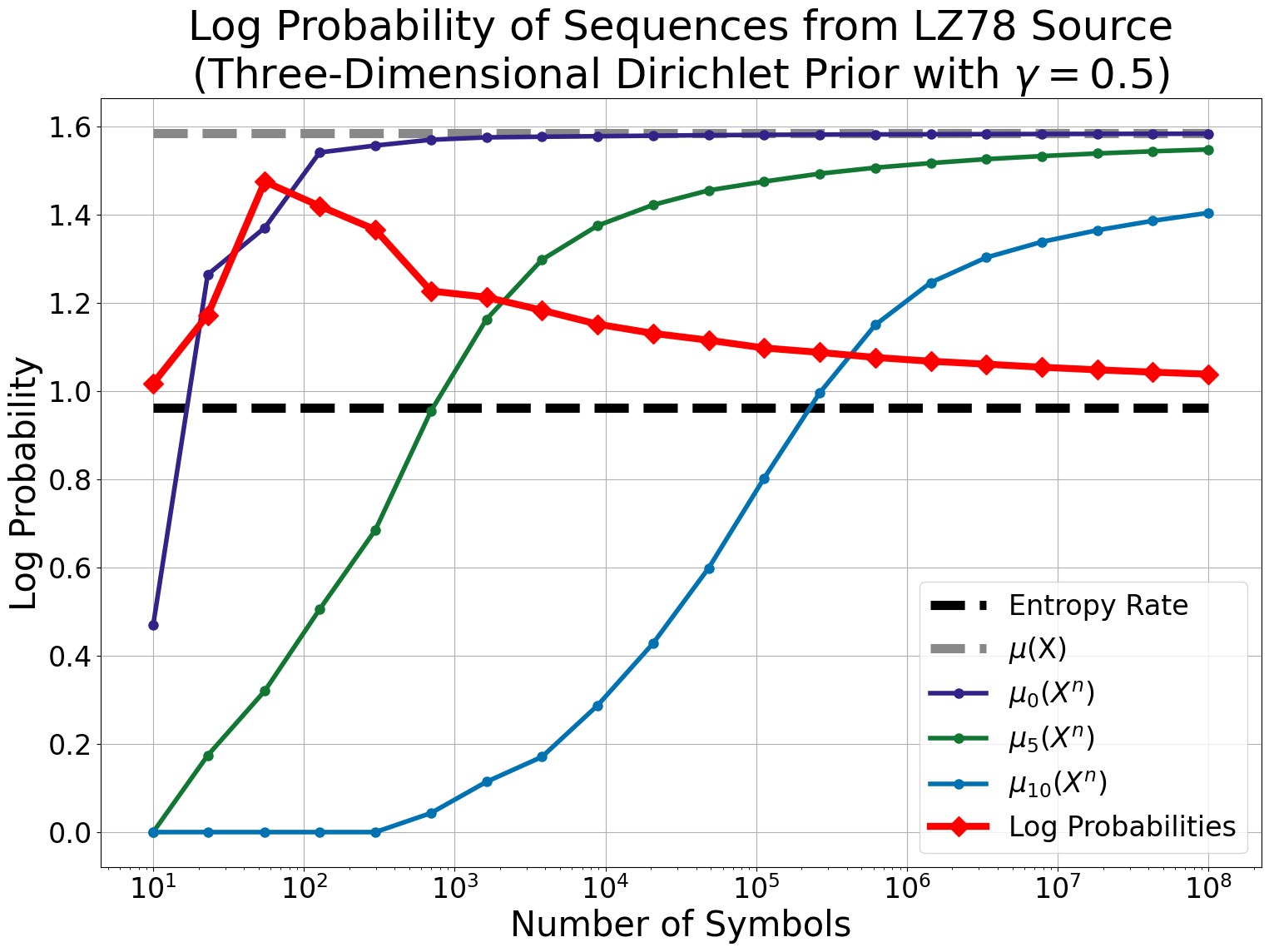}
        \caption{Normalized log probability of one realization of the LZ78 source, and log loss of Markov sequential probability models.}
    \end{subfigure}
    \caption{Simulation results, where $\Pi$ is the Dirichlet($0.5, 0.5, 0.5$) distribution.}
    \label{fig:dirichlet-gamma-1-1-A-3}
\end{figure*}

For a binary alphabet, we use \rev{$\gamma=0.01$ (to study the source under a very low entropy rate)}, $\gamma = 0.5$, and $\gamma = 2$.
For the ternary alphabet experiment, we use $\gamma = 0.5$.
The log probabilities of source realizations, and corresponding values of $\mu_k$, are plotted in \Cref{fig:dirichlet-gamma-0-01,fig:dirichlet-gamma-0-5,fig:dirichlet-gamma-2,fig:dirichlet-gamma-1-1-A-3}.
All priors tested exhibit the same behavior: over a logarithmic timeframe, the normalized log probabilities of all realizations approach the entropy rate.
\rev{In addition, the lower-entropy-rate source ($\gamma=0.01$), appears to have faster convergence to the entropy rate than the other two sources.}
In addition, all values of $\mu_k$ plotted approach $\mu(\Xv)$, at a slower rate for larger values of $k$.
\rev{For the low-entropy-rate $\gamma=0.01$ prior, $\mu_{10}$ shows an interesting phenomenon: the log loss initially increases, then dips for several orders of magnitude, and then starts increasing towards $\mu(\Xv)$.
From our experimentation, this behavior is consistent across realizations (random seeds).
This is likely due to two opposing forces: the decreasing log probability (for this particular realization $X^n$) as more symbols are drawn, and the eventual convergence of $\mu_k(X^n)$ to $\mu(\Xv)$}

% \begin{figure}[htbp]
%     \centering
%     \begin{subfigure}{0.45\linewidth}
%         \centering
%         \includegraphics[width=0.99\linewidth]{figures/entropy_rate_gamma-1.png}
%         \caption{Normalized log probability of five realizations of the LZ78 source, with the entropy rate and $\mu(\Xv)$ labeled.}
%     \end{subfigure}
%     \begin{subfigure}{0.45\linewidth}
%         \centering
%         \includegraphics[width=0.99\linewidth]{figures/mu_k_gamma-1.png}
%         \caption{Normalized log probability of one realization of the LZ78 source, and log loss of Markov sequential probability models.}
%     \end{subfigure}
%     \caption{Simulation results, where $\Pi$ is the Dirichlet($1, 1$) distribution.}
%     \label{fig:dirichlet-gamma-1-1}
% \end{figure}


\subsubsection{A Dirac-Dirichlet Mixture}\label{sec:dirac-dirichlet-mixture}
We also consider a prior over a binary alphabet that is the mixture of a Dirichlet distribution with $\gamma > 1$ and a point mass distribution.
$\Pi$ is defined as the following law governing $\Theta$ (which we refer to as a ``Dirac-Dirichlet mixture''):
\begin{enumerate}
    \item With probability $\zeta$, draw $\Theta\sim$ Dirichlet$(\gamma, \gamma)$, where $\gamma > 1$.
    \item Otherwise, draw $\Theta$ from the distribution with equal-height point masses at $\xi$ and $1-\xi$.
\end{enumerate}

As $\Theta$ is symmetric, $\mu(\Xv)=1$.
The entropy rate of the LZ78 source under this law is
\begin{align*}
    \Ebb[H(\Theta)] &= \zeta H_\text{Dirichlet}(\gamma) + (1-\zeta) h_2(\xi) \\
    &= -\zeta\frac{\sqrt{\pi} 2\Gamma(2\gamma) 4^{-\gamma} (H_\gamma - H_{2\gamma})}{\Gamma(\gamma + 1/2) \Gamma(\gamma)^2 \ln 2} \\
    &\qquad- (1-\zeta)\left(\xi \log \xi + (1-\xi)\log(1-\xi)\right).
\end{align*}
For this formulation, small $\xi$ and $\zeta$ correspond to lower entropy rates.

\begin{figure*}[!tbp]
    \centering
    \begin{subfigure}{0.45\linewidth}
        \centering
        \includegraphics[width=0.99\linewidth]{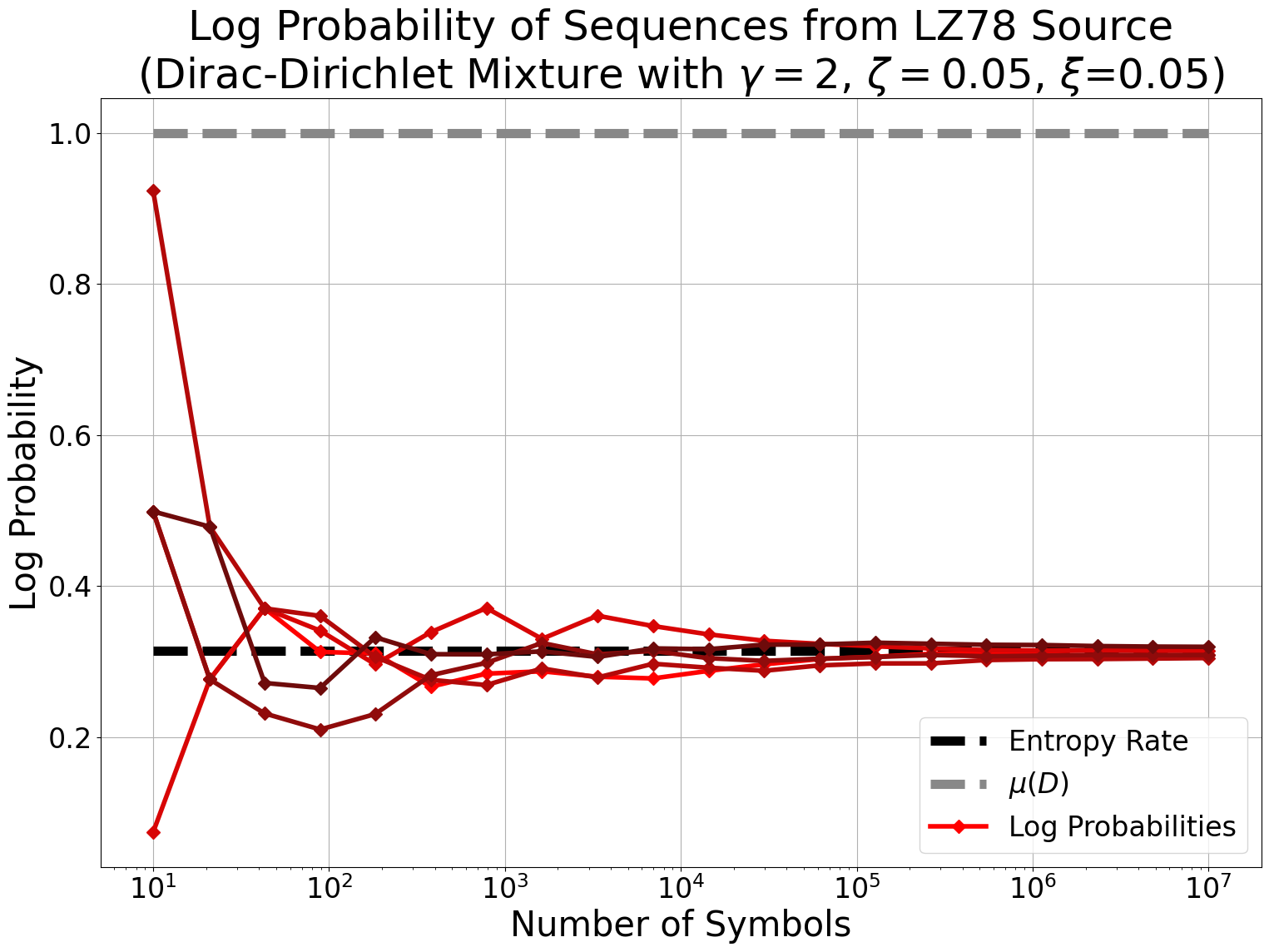}
        \caption{Normalized log probability of five realizations of the LZ78 source, with the entropy rate and $\mu(\Xv)$ labeled.}
    \end{subfigure}
    \begin{subfigure}{0.45\linewidth}
        \centering
        \includegraphics[width=0.99\linewidth]{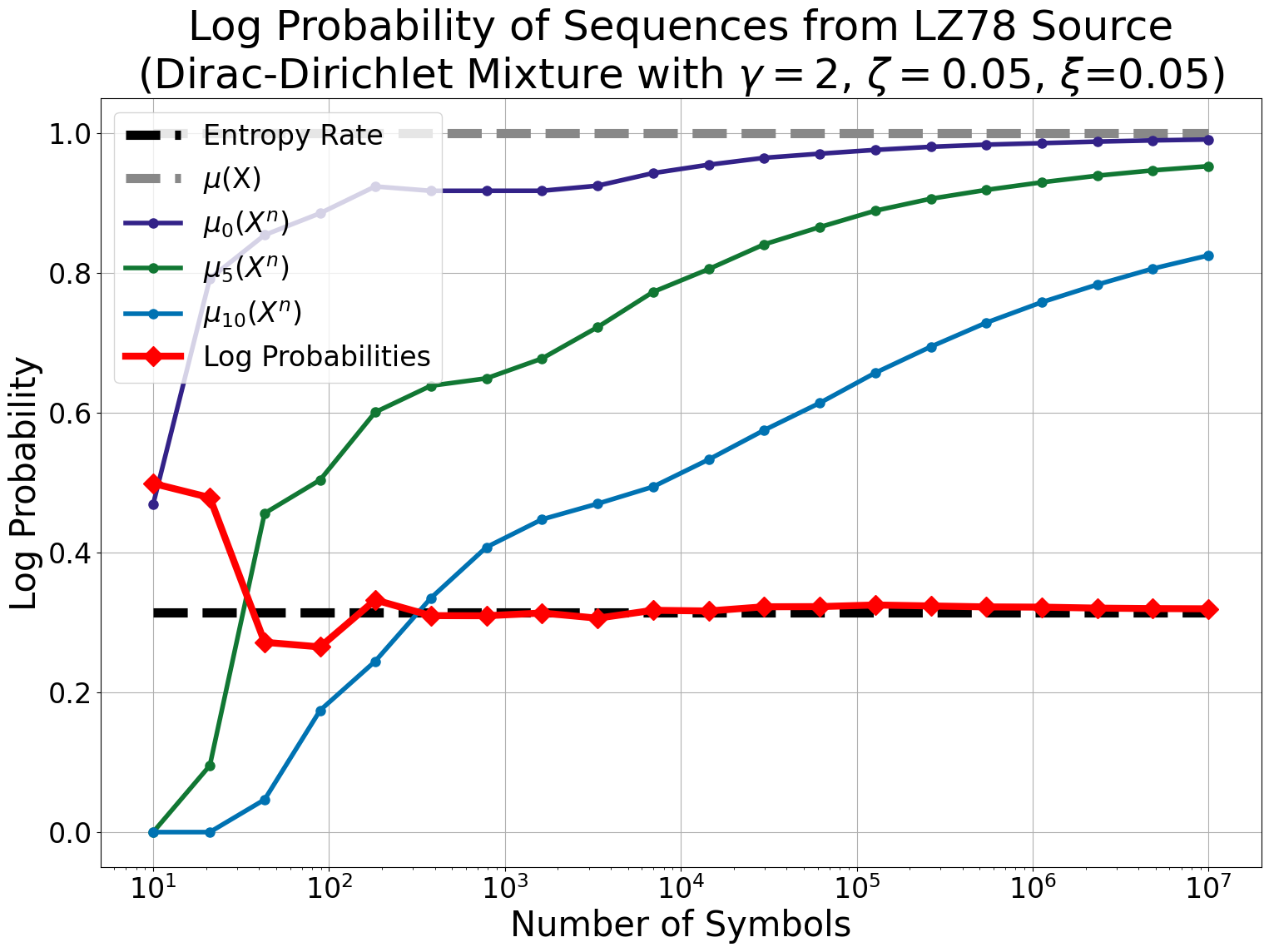}
        \caption{Normalized log probability of one realization of the LZ78 source, and log loss of Markov sequential probability models.}
    \end{subfigure}
    \caption{Simulation results, where $\Pi$ is the Dirac-Dirichlet Mixture with Dirichlet parameter $2$, point masses at $0.05$ and $0.95$, and weight $0.95$ placed on the point masses.}
    \label{fig:dirac-dirichlet-mixture}
\end{figure*}
\begin{figure*}[!tbp]
    \centering
    \begin{subfigure}{0.45\linewidth}
        \centering
        \includegraphics[width=0.99\linewidth]{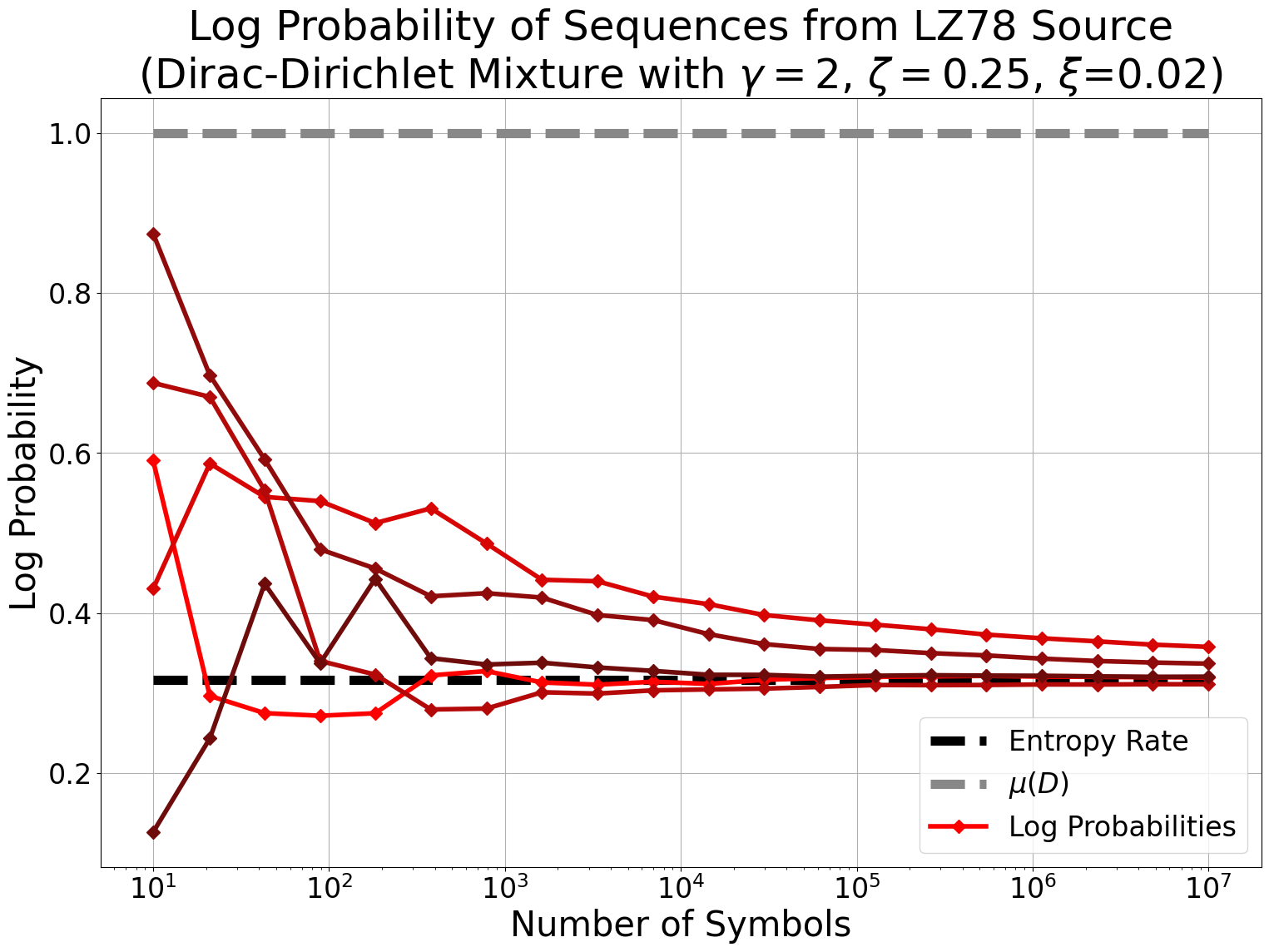}
        \caption{Normalized log probability of five realizations of the LZ78 source, with the entropy rate and $\mu(\Xv)$ labeled.}
    \end{subfigure}
    \begin{subfigure}{0.45\linewidth}
        \centering
        \includegraphics[width=0.99\linewidth]{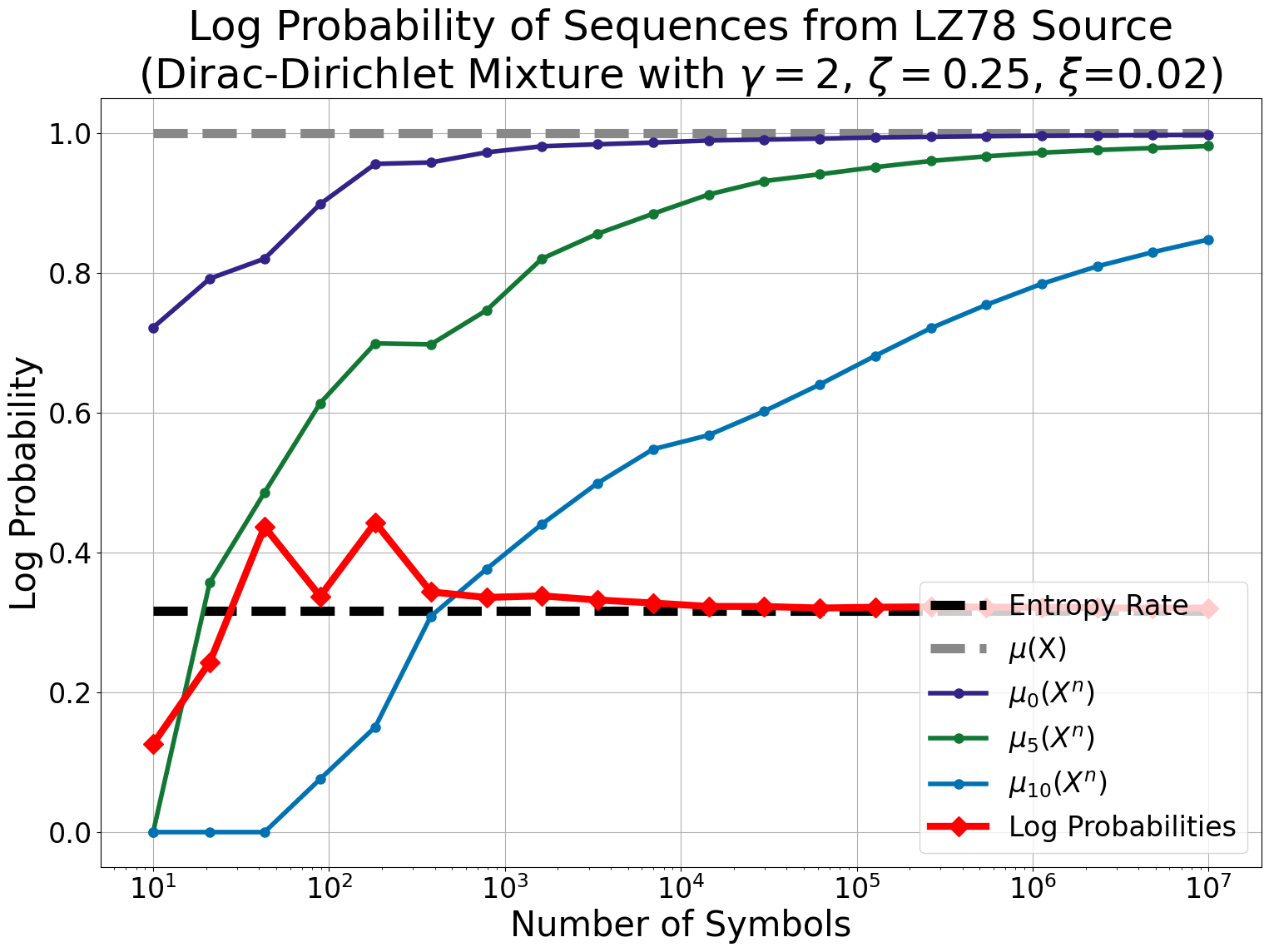}
        \caption{Normalized log probability of one realization of the LZ78 source, and log loss of Markov sequential probability models.}
    \end{subfigure}
    \caption{Simulation results, where $\Pi$ is the Dirac-Dirichlet Mixture with Dirichlet parameter $2$, point masses at $0.02$ and $0.98$, and weight $0.75$ placed on the point masses.}
    \label{fig:dirac-dirichlet-mixture-2}
\end{figure*}

Normalized log probabilities and values of $\mu_k$ are plotted in Figures \ref{fig:dirac-dirichlet-mixture} and \ref{fig:dirac-dirichlet-mixture-2}, which have different values of $\zeta$ and $\xi$ but similar entropy rates.
As opposed to in \prettyref{sec:dirichlet-prior}, where 100 million symbols were drawn, these plots include 10 million symbols, as the convergence of normalized log probabilities to the entropy rate is clear within that timeframe.
Convergence of the log probabilities to the entropy rate is much faster than in \prettyref{sec:dirichlet-prior}.
This is especially apparent in \prettyref{fig:dirac-dirichlet-mixture}, where very little weight is put on the Dirichlet component.

\rev{\subsection{Studying In-Context Learning of Transformers}\label{sec:icl_study}

In this section, we use realizations of the LZ78 source to study ICL abilities of transformer models with depths (\ie, number of layers) ranging from 1 to 5.
We systematically vary the priors used to generate the LZ78 source, producing training and test sets that differ in entropy rate and contextual structure.
This enables us to examine how transformer ICL adapts to increasingly challenging non-Markovian regimes.
Finally, by evaluating these same models on real genomic data, we observe qualitative patterns reminiscent of CTW, suggesting that the learned ICL heuristics transfer beyond synthetic sources.

As discussed in \prettyref{sec:introduction}, the LZ78 source’s expanding context and Jensen gap make it harder to model than fixed- or variable-order Markov processes.
This source thus allows us to test how well transformers can perform ICL in regimes where classical finite-state predictors are fundamentally mismatched to the data.}

\rev{\subsubsection{Experimental Setup}
For the transformer implementation, we used the NanoGPT GitHub repository \cite{Karpathy2025}, which implements a transformer with multi-headed group query attention.
We used all default settings except for the number of layers (\eg, embedding dimension and number of attention heads as a function of number of layers, optimizer, learning rate, \etc).

For sampling training and evaluation data, we fix a prior (either Dirichlet or a Dirac-Dirichlet mixture, as described in \prettyref{sec:simulations}), and generate each training or evaluation sequence i.i.d. from this prior.
All data is binary, with no additional tokenization.
Each realization is of length $n=2048$.
This length is chosen to allow the realizations' log probabilities to start converging to the entropy rate, while taking into consideration the fact that transformers become significantly more compute-intensive to train with longer context lengths due to quadratic scaling of the self-attention mechanism.
The training set has 100,000 realizations, and the evaluation set has 2048 realizations.
For each transformer, we perform 5 epochs of training with a batch size of 32, which takes approximately 5 hours on an NVIDIA A6000 GPU.\footnote{\inlineRev Due to limited computational resources, we could not train the models for longer.
Empirically, we saw reasonable convergence of the training and test loss curves over this duration, and see ICL properties emerge far earlier than 5 epochs.}

We also evaluate generalization performance selected transformer models trained on LZ78 source data by applying them to real-world genomic data.
This evaluation is performed by sampling 2048 length-1024 sequences from promoter and enhancer regions of Human Chromosome 1, and then mapping each of the 4 nucleotide symbols to a two-bit sequence (\eg, \texttt{A} becomes $00$, \texttt{C} becomes $01$, \etc).
We use the same sequences for all such experiments.

In our results, we plot the cumulative normalized log loss of the transformer model on prefixes of the length-2048 sequences, \ie, the same quantity plotted in \prettyref{sec:simulations}, averaged over the evaluation set.
As baselines, we plot $\mu_k$ for several values of $k$, CTW as implemented in \cite{Jiao_2013}, the realizations' log probability, $\mu(\Xv)$, and the entropy rate, all averaged over the evaluation set.

For more details, refer to our GitHub repository \href{https://github.com/NSagan271/LZ78-Source-ICL}{here}.}

\subsubsection{Experimental Results}

\begin{figure*}[tbp]
    \centering
    \begin{subfigure}{0.48\linewidth}
        \centering
        \includegraphics[width=\linewidth]{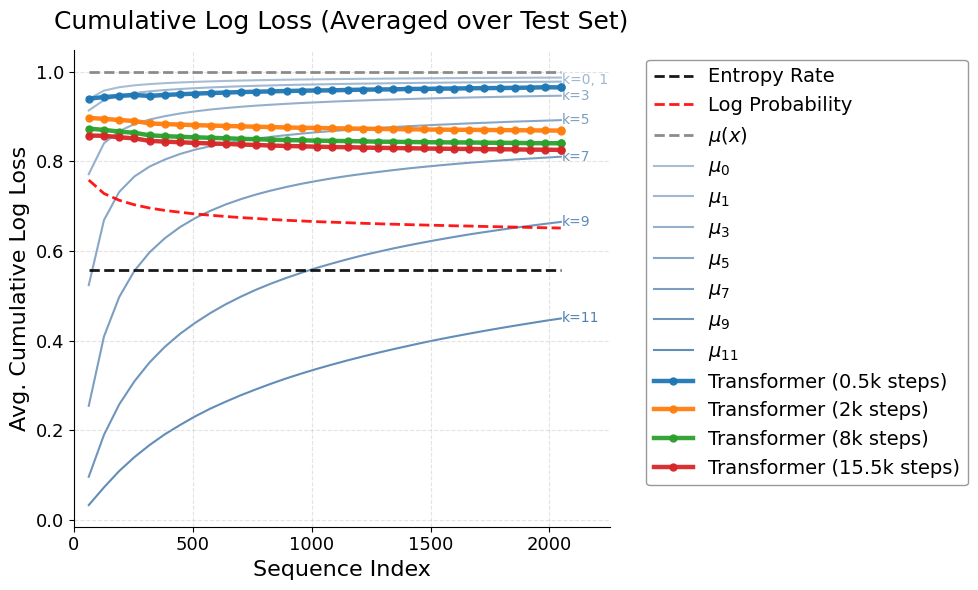}
        \subcaption{\inlineRev $\mu_k$ comparison}
        \label{fig:jeffreys_mu_k}
    \end{subfigure}
    \hfill
    \begin{subfigure}{0.48\linewidth}
        \centering
        \includegraphics[width=\linewidth]{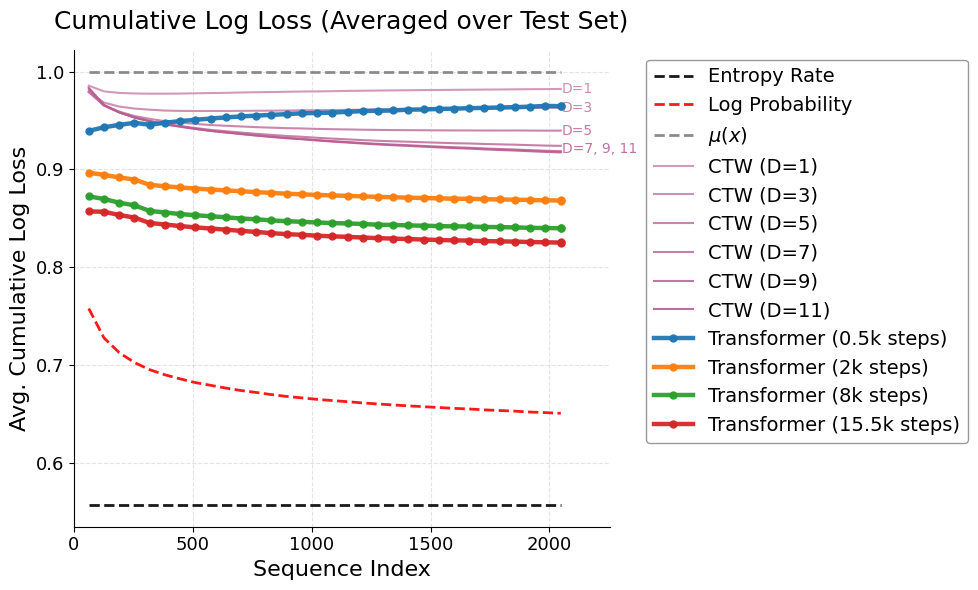}
        \subcaption{\inlineRev CTW comparison}
        \label{fig:jeffreys_ctw}
    \end{subfigure}

    \caption{\inlineRev Four-layer transformer compared to $\mu_k$ and CTW, trained on the Jeffreys prior LZ78 source.
    Training time is ablated, with ``steps'' referring to the number of 32-sequence batches used to train the transformer.
    All curves are for a single transformer over the course of training.}
    \label{fig:jeffreys_comparison}
\end{figure*}

\begin{figure*}[tbp]
    \centering
    \begin{subfigure}{0.48\linewidth}
        \centering
        \includegraphics[width=\linewidth]{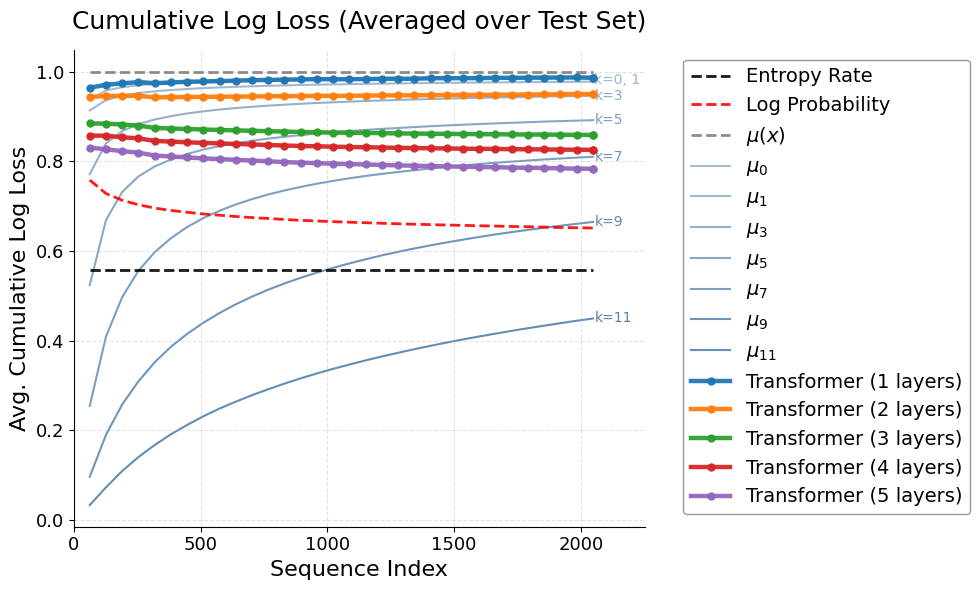}
        \subcaption{\inlineRev$\mu_k$ comparison}
        \label{fig:jeffreys_mu_k_depth}
    \end{subfigure}
    \hfill
    \begin{subfigure}{0.48\linewidth}
        \centering
        \includegraphics[width=\linewidth]{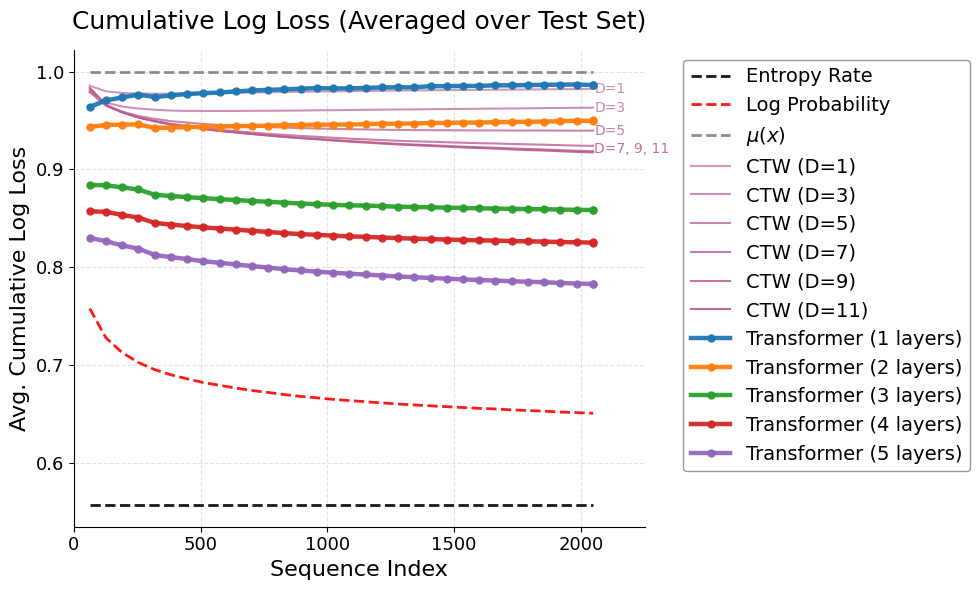}
        \subcaption{\inlineRev CTW comparison}
        \label{fig:jeffreys_ctw_depth}
    \end{subfigure}

    \caption{\inlineRev Transformers of different depths compared to $\mu_k$ and CTW, trained on the Jeffreys prior LZ78 source.}
    \label{fig:jeffreys_comparison_depth}
\end{figure*}

\paragraph{Jeffreys prior}
First, we consider the LZ78 source under the Jeffreys prior, ablating transformer depth and training time.
Comparisons against $\mu_k$ for different values of $k$ and CTW with different depths are shown in Figures \ref{fig:jeffreys_comparison} and \ref{fig:jeffreys_comparison_depth}, respectively.
For the ablation of training time, a four-layer transformer is used.

In general, transformers with enough layers and enough training examples display a log loss that decreases over the context, which \cite{olsson2022context} describes as a key empirical feature of in-context learning.
The four-layer transformer early on in training (500 steps) and the one- or two-layer transformers do not display such properties, with the log loss instead increasing over the context.
Over the full sequence, the one-layer transformer only does as well as CTW with depth $D=1$, indicating that it learns only local patterns (or perhaps global sequence statistics).
The two-layer transformer performs on par with $\mu_2$, possibly indicating ICL properties that take longer contexts into account.
Transformers with 3-5 layers achieve better log loss than all depths of CTW tested, and are on par with $\mu_k$ for $k \approx 7$.
As the training data comes from the LZ78 source under the Jeffreys prior (the same prior used for CTW), it is not the case that the transformer simply ``learns the prior better.''
Instead, it is likely adapting to the LZ-based dependency structure of the input data.

Each added transformer layer presents a clear improvement in performance.
For all transformer models tested, however, there is still a large gap between the model's log loss and the true log probability of the evaluation data.
More experimentation is needed to determine whether longer training or larger transformers will close the gap.
Based on the model depth ablation in \prettyref{fig:jeffreys_comparison_depth}, it is possible that a large enough transformer will be able to achieve a log loss close to the log probability (over these short sequences---based on our theoretical results, this would be impossible asymptotically).

\begin{figure*}[tbp]
    \centering
    \begin{subfigure}{0.48\linewidth}
        \centering
        \includegraphics[width=\linewidth]{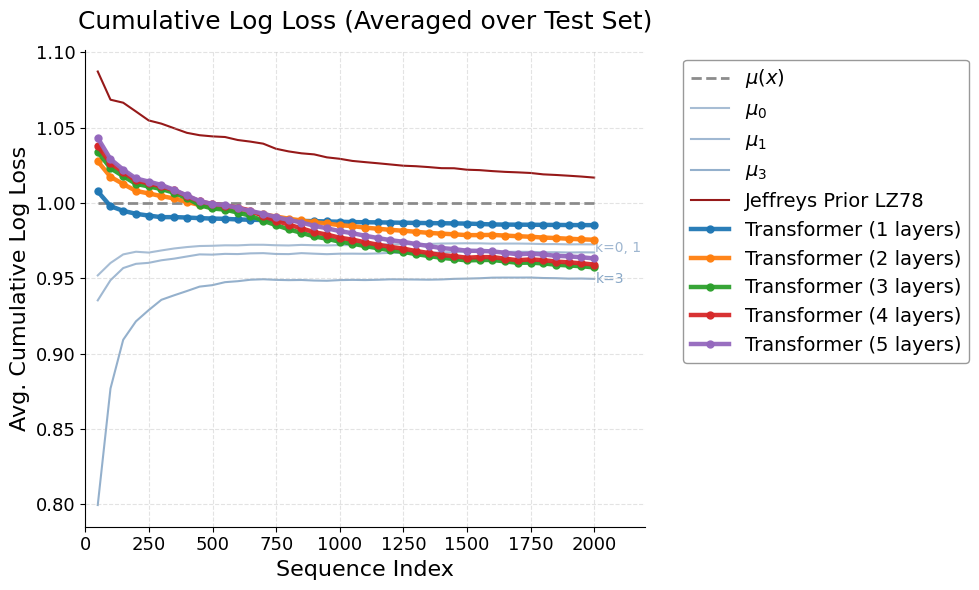}
        \subcaption{\inlineRev $\mu_k$ comparison}
        \label{fig:jeffreys_prom_enh_mu_k_depth}
    \end{subfigure}
    \hfill
    \begin{subfigure}{0.48\linewidth}
        \centering
        \includegraphics[width=\linewidth]{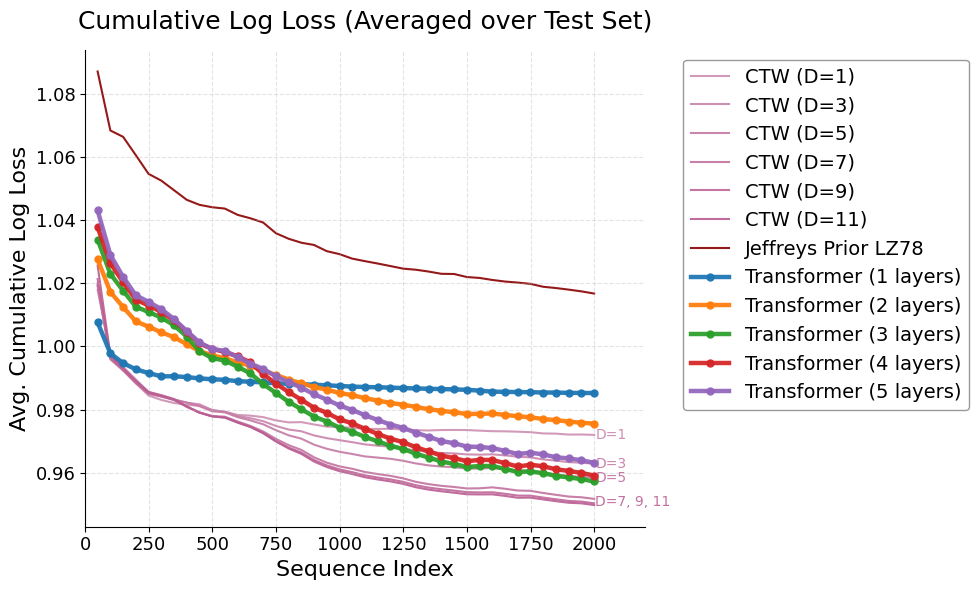}
        \subcaption{\inlineRev CTW comparison}
        \label{fig:jeffreys_prom_enh_ctw_depth}
    \end{subfigure}

    \caption{\inlineRev Transformers of varying depths compared to $\mu_k$ and CTW, trained on the Jeffreys prior LZ78 source and tested on human genomic data.}
    \label{fig:jeffreys_comparison_prom_enh_depth}
\end{figure*}

\paragraph{Jeffreys prior, tested on genomic data}
We apply the transformers from \prettyref{fig:jeffreys_comparison_depth} to human genomic data, to see how the ICL properties generalize to real-world data.
Results are plotted in \prettyref{fig:jeffreys_comparison_prom_enh_depth}.

Notably, all models have a log loss that decreases over the context, except for the one-layer transformer, which has a fairly constant log loss.
This indicates that the transformers are indeed learning some prediction algorithm in-context that generalizes to out-of-distribution data.
The transformers (except the one-layer model) exhibit decreasing log loss over the length of the sequences, with the three- to five-layer transformers doing on par with CTW of depths 3-5 (over the full length of the sequence).
The three-layer transformer has the best performance; potentially, the larger transformers fit the LZ78 training data better and thus do not generalize as well to this out-of-distribution data.
In this sense, the model size may act as a form of regularization.

% Except for the one-layer transformer, where the loss curve is mostly flat, the shape of the transformer loss curves resembles that of LZ78 (except with lower overall loss).
% This indicates the 
% This could indicate that the algorithm learned by the transformers in-context has some relation to LZ78.
% Given that the transformers are trained on LZ78 source data, this may be reasonable to expect.

\begin{figure*}[tbp]
    \centering
    \begin{subfigure}{0.48\linewidth}
        \centering
        \includegraphics[width=\linewidth]{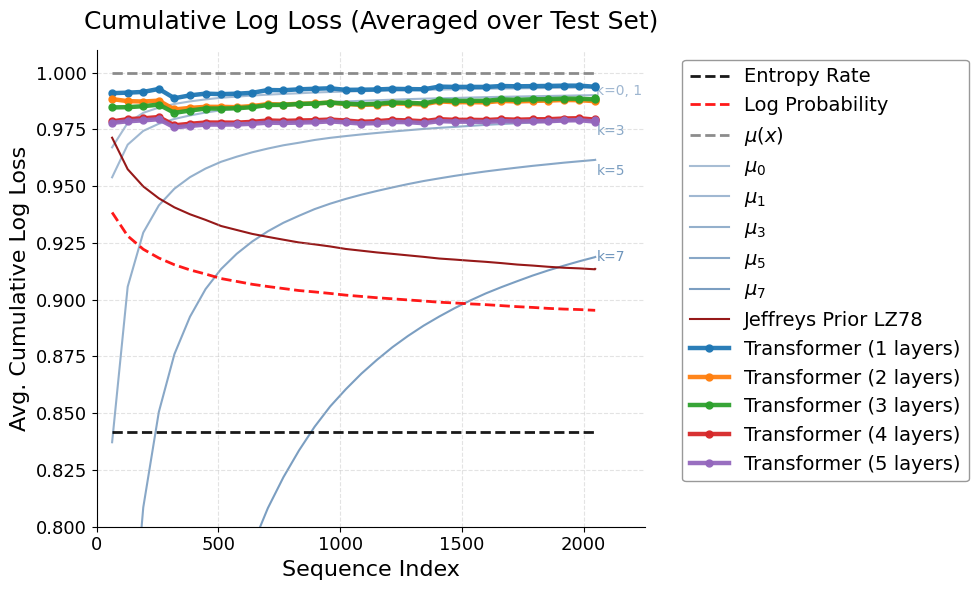}
        \subcaption{$\mu_k$ comparison}
        \label{fig:dirichlet_2_mu_k}
    \end{subfigure}
    \hfill
    \begin{subfigure}{0.48\linewidth}
        \centering
        \includegraphics[width=\linewidth]{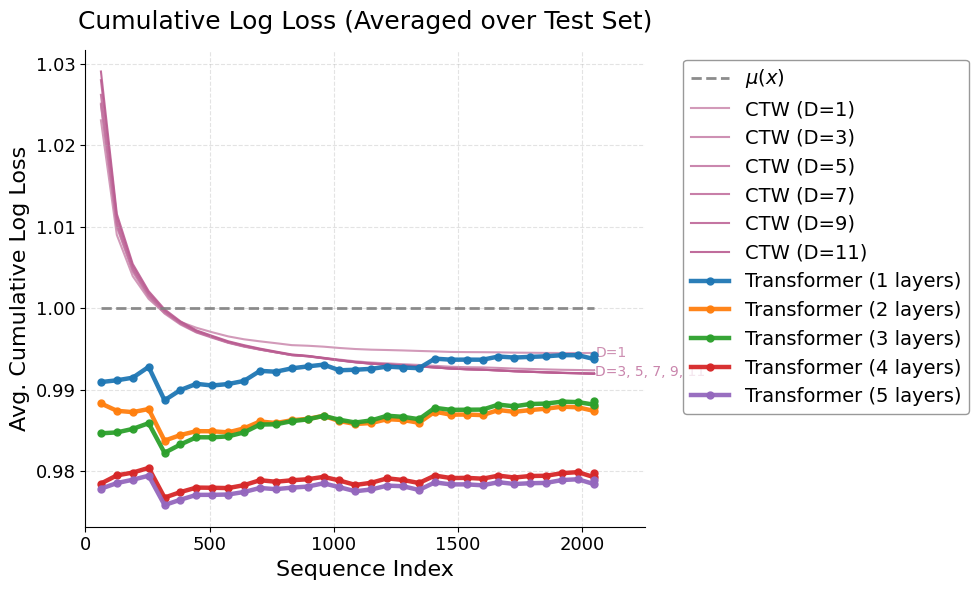}
        \subcaption{CTW comparison}
        \label{fig:dirichlet_2_ctw}
    \end{subfigure}

    \caption{Transformers of varying depths compared to $\mu_k$ and CTW, trained on the Dirichlet(2, 2) prior LZ78 source.}
    \label{fig:dirichlet_2_ctw_comparison}
\end{figure*}

\paragraph{Dirichlet(2, 2) prior}
Next, we train transformers on the LZ78 source under a Dirichlet(2, 2) prior, which has a significantly higher entropy rate.
Comparisons to $\mu_k$ and CTW are shown in \prettyref{fig:dirichlet_2_ctw_comparison}.
For the CTW plots, the log probability and LZ78 curves are not shown to better showcase the shape of the loss curves.

\begin{figure*}[tbp]
    \centering
    \begin{subfigure}{0.48\linewidth}
        \centering
        \includegraphics[width=\linewidth]{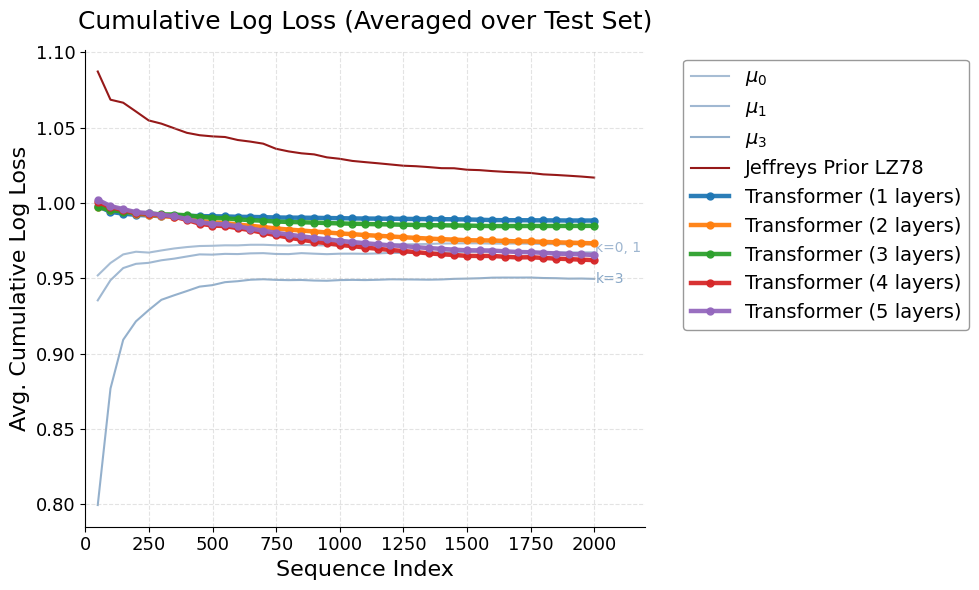}
        \subcaption{\inlineRev $\mu_k$ comparison}
        \label{fig:dirichlet_2_prom_enh_mu_k}
    \end{subfigure}
    \hfill
    \begin{subfigure}{0.48\linewidth}
        \centering
        \includegraphics[width=\linewidth]{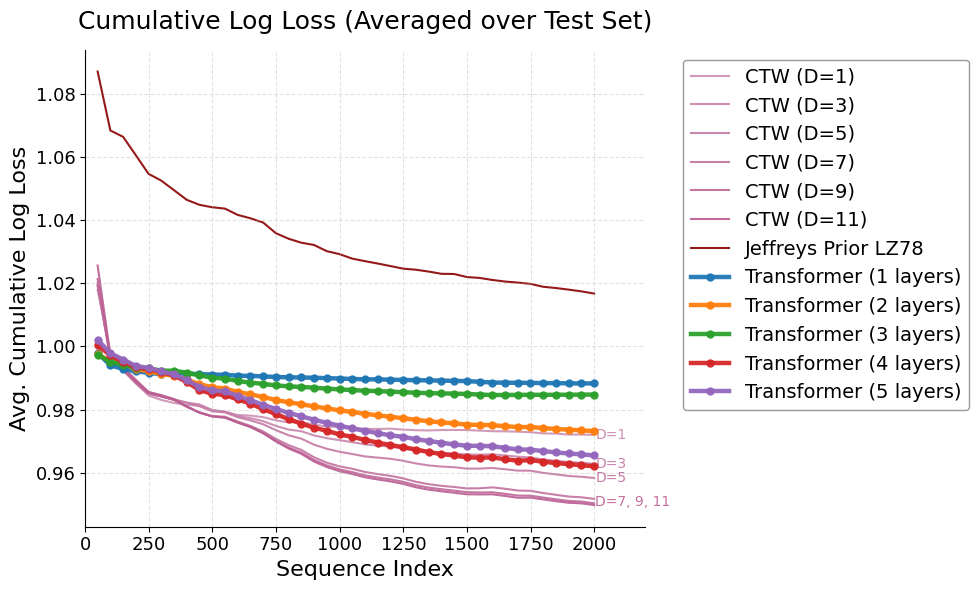}
        \subcaption{\inlineRev CTW comparison}
        \label{fig:dirichlet_2_prom_enh_ctw}
    \end{subfigure}

    \caption{\inlineRev Transformers of varying depths compared to $\mu_k$ and CTW, trained on the Dirichlet(2, 2) prior LZ78 source and tested on genomic data.}
    \label{fig:dirichlet_2_prom_enh}
\end{figure*}

Unlike with the Jeffreys prior, the loss curves do not smoothly decrease during the course of learning, even for the five-layer transformer. 
This may indicate that higher-entropy versions of the LZ78 source are harder to learn.
If this is the case, it appears to also hold for CTW: the transformers (except for the one-layer one) still perform better than CTW.
This could indicate that ICL is still present, but the sequences are too ``noisy'' for a smooth loss curve to appear.
Overall, the two largest transformers perform on par with $\mu_3$ and there is a significant gap between the transformers and a mismatched LZ78 SPA.
This indicates some ICL occurs (which is corroborated by evaluation on genomic data in \prettyref{fig:dirichlet_2_prom_enh}), but not to the extent seen with the Jeffreys prior.

\begin{figure}[tbp]
    \centering
    \includegraphics[width=1\linewidth]{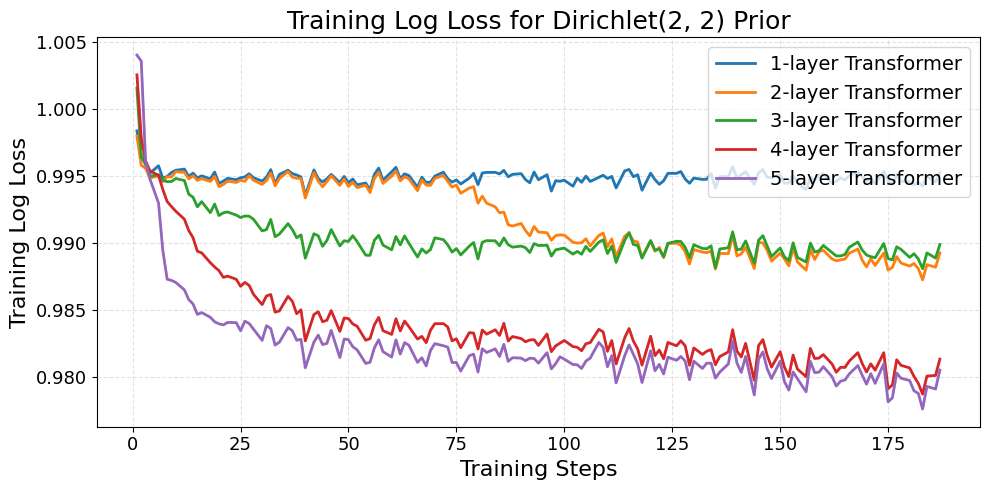}
    \caption{\inlineRev Training log loss curves for transformers of different sizes trained on the Dirichlet(2, 2) LZ78 source.}
    \label{fig:dirichlet_2_training_loss}
\end{figure}

An interesting phenomenon is that the two-layer transformer achieves slightly lower log loss than the three-layer transformer.
To understand this better, we plot the training log loss curves for each transformer in \prettyref{fig:dirichlet_2_training_loss}.
We hypothesize that this performance is due to the three-layer transformer getting stuck in a local minimum.
Similar behavior is apparent for the two-layer transformer, which has training loss comparable to the one-layer transformer until mid-training, when it breaks out of the local minimum.
It is possible that smaller models have more ICL capabilities than the results here show, but they are more likely to get stuck in local minima.

\paragraph{Dirichlet(2, 2) prior, tested on genomic data}
We also apply these transformers to genomic data, as shown in \prettyref{fig:dirichlet_2_prom_enh}.

The four- and five-layer transformers recover CTW-like performance on this data, so, despite the inconclusive results of \prettyref{fig:dirichlet_2_prom_enh}, the transformers do appear to be learning in-context.

\begin{figure*}[tbp]
    \centering
    \begin{subfigure}{0.48\linewidth}
        \centering
        \includegraphics[width=\linewidth]{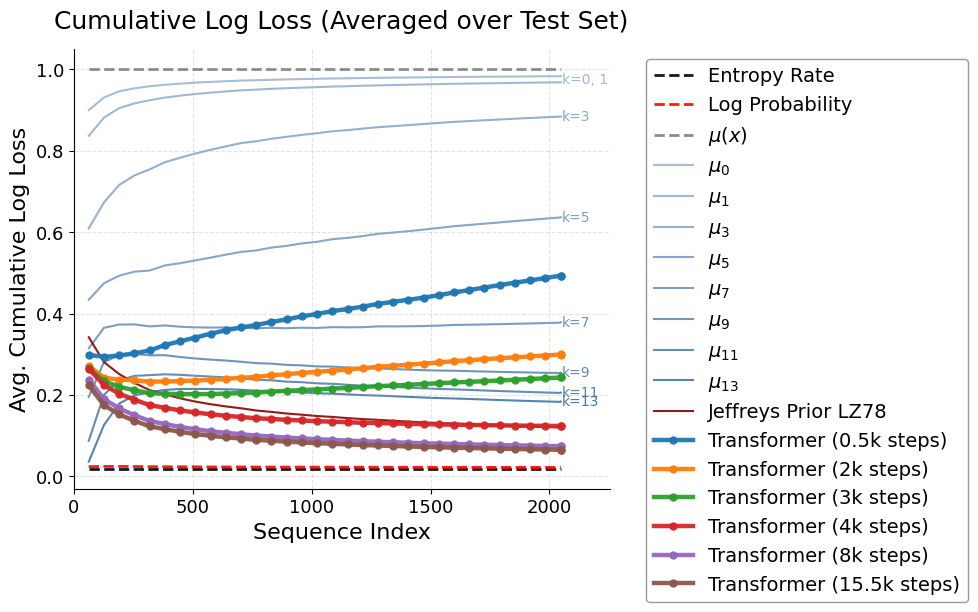}
        \subcaption{\inlineRev $\mu_k$ comparison}
        \label{fig:degenerate_prior_mu_k}
    \end{subfigure}
    \hfill
    \begin{subfigure}{0.48\linewidth}
        \centering
        \includegraphics[width=\linewidth]{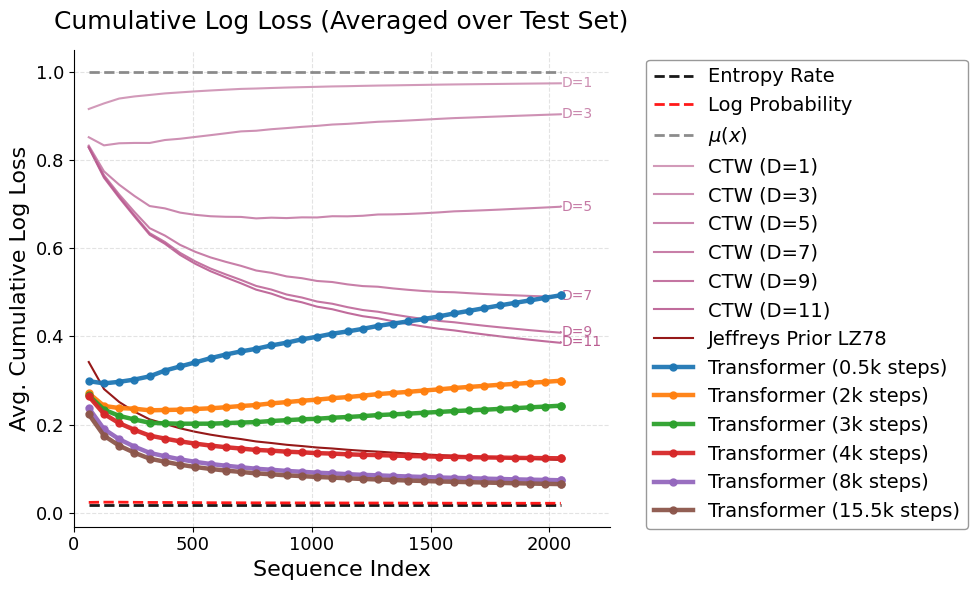}
        \subcaption{\inlineRev CTW comparison}
        \label{fig:degenerate_prior_ctw}
    \end{subfigure}

    \caption{\inlineRev Four-layer transformer compared to $\mu_k$ and CTW, trained on the LZ78 source under the almost-degenerate Dirac--Dirichlet prior.}
    \label{fig:degenerate_prior_comparison}
\end{figure*}

\begin{figure*}[tbp]
    \centering
    \begin{subfigure}{0.48\linewidth}
        \centering
        \includegraphics[width=\linewidth]{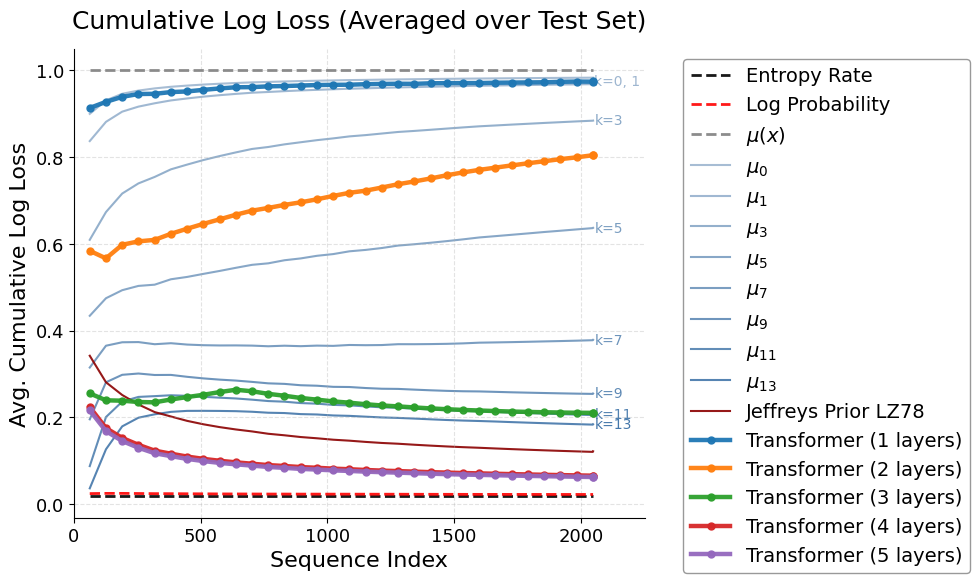}
        \subcaption{\inlineRev$\mu_k$ comparison}
        \label{fig:degenerate_prior_mu_k_depth}
    \end{subfigure}
    \hfill
    \begin{subfigure}{0.48\linewidth}
        \centering
        \includegraphics[width=\linewidth]{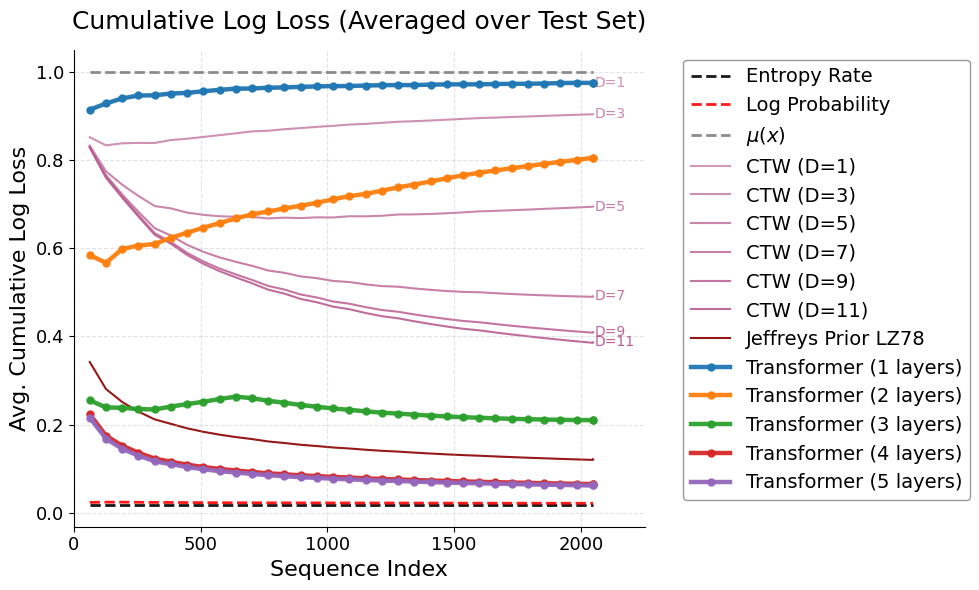}
        \subcaption{\inlineRev CTW comparison}
        \label{fig:degenerate_prior_ctw_depth}
    \end{subfigure}

    \caption{\inlineRev Transformers of different depths compared to $\mu_k$ and CTW, trained on the LZ78 source under the almost-degenerate Dirac--Dirichlet prior.}
    \label{fig:degenerate_prior_comparison_depth}
\end{figure*}

\paragraph{Almost-degenerate prior}
Finally, we consider a prior where 97.5\% of the weight is on equal-height point masses at 0 and 1, with the remainder distributed uniformly.
This source produces an LZ tree that is ``almost degenerate'' in that most nodes only have a single child due to their realized value of $\Theta$ being either 1 or 0.
As a result, the LZ tree grows deeper at a faster rate than the other examples, resulting in more long-range dependencies.
Results, ablating depth and training time, are in Figures \ref{fig:degenerate_prior_comparison} and \ref{fig:degenerate_prior_comparison_depth}, respectively.


The four- and five-layer transformer (at the end of training) do very well, getting relatively close to the true log probability by the end of the sequence (note that this is only possible because we consider a finite-sample regime).
Their log loss surpasses even $\mu_{13}$ and the LZ78 SPA under the Jeffreys prior.
For smaller models (and the four-layer transformer over the course of training), the source is much harder to learn due to the prevalence of long-range dependencies.
For the four-layer transformer over the course of training, the log loss increases towards the end of the context up until 4000 training steps.
For the model depth ablations, there is a clear distinction between transformer depths up until four layers (the one-layer transformer is around $\mu_1$, the two-layer transformer between $\mu_3$ and $\mu_5$, the three-layer transformer around $\mu_{11}$).
This may indicate that number of layers is key to learning long-range dependencies.

\section{Conclusion}\label{sec:conclusion}
We considered the  probability source induced by the celebrated LZ78 universal compressor and studied some of its key properties.
We established the source's entropy rate and almost-sure convergence of a realization's normalized log probability to the entropy rate.
Additionally, we studied the minimum log loss achievable via a finite-state probability model on a realization of the LZ78 source,\footnote{As proven in \cite{sagan2024familylz78baseduniversalsequential}, the finite-state predictability, optimal Markov predictability, and finite-state compressibility (scaled by $\log |\Acal|$) of any individual sequence are all the same quantity.} finding that it almost surely converges to a value larger than the entropy rate by a ``Jensen gap.''
We have also characterized  the relative entropy rate between the LZ78 source and any Markovian law.
We then performed numerical simulations on the log probability of sequences drawn from the source, and the corresponding log loss of the best Markov probability model on the realizations.
These experiments illustrate some of our main theoretical results.
\rev{Finally, we applied the LZ78 source to the study of in-context learning in transformers by training small transformer models on data sampled from the LZ78 source and comparing log loss against the true log probabilities, CTW, and the best possible Markov SPAs. These experiments 
provide further insight into the potential and limitations of transformer models for ICL.   }

% In future work, the LZ source would be a useful probability source for gauging the performance of sequential probability models (such as those based on deep learning) on non-Markovian  non-stationary data.
In future work, it may be interesting to study the properties of sources based on other universal compressors or probability models. 
%, \eg, context tree weighting \cite{willems1995CTW}.
% With respect to theoretical results, we suspect that the condition of $\supp(\Pi) = [0, 1]$ for the entropic results can be relaxed.
% In fact, we suspect that our main results hold without any stipulations on $\Pi$.  
Another potential direction is the development of more refined finite-sample results. 
%for the normalized log probability and Markov model log loss for realizations of this source.
It will also be worthwhile to understand whether the LZ78 sources satisfy some version of Kac's lemma  \cite{kac1947recurrence} that might allow to revisit and combine the settings of \cite{merhav2022universalrandomcodingensemble} and  \cite{8ea2ff2d-04be-3396-ad29-ada206f92b98} toward a sliding window based  universal lossy compressor.  
\rev{Finally, the study of in-context learning presented here is only an initial exploration; further experimentation and analysis can be done, \eg, performing further ablations on the transformer architecture or the training data.}

\appendices
\section{Proofs of \prettyref{thm:zero-order-empirical-distribution} and \prettyref{thm:emp-dist-of-r-tuples} for $\Ebb[H(\Theta)] = 0$}\label{app:proofs-for-supp-pi-01}
In proving \prettyref{thm:zero-order-empirical-distribution} and \prettyref{thm:emp-dist-of-r-tuples} for a general law $\Pi$, there are components that require $h \triangleq \Ebb_{\Theta \sim \Pi}[H(\Theta)] > 0$.
As such, we consider the case $\Ebb[H(\Theta)] = 0$ (\ie, $\supp(\Pi)$ only includes the corners of the simplex, or distributions that place all weight on a single symbol $a \in \Acal$), separately.

\begin{lemma}[Zero-Order Empirical Distribution for $\Ebb \lbrack H(\Theta)\rbrack = 0$]\label{lem:zero-order-dist-pi-01}
    For the LZ78 source with $\Ebb[H(\Theta)] = 0$, $M_n$ converges to $\Pi$ weakly, almost surely.

    \begin{proof}
        Specifically, we show that, for all measurable $A \subseteq \Mcal(\Acal)$, $M_n(A) \convas \Pi(A) \triangleq p$.

        For the case $p=0$, $|A \cap \supp(\Pi)| = 0$, so both $M_n(A)$ and $\Pi(A)$ are deterministically $0$.
        % For $p=1$, $A \in \supp(\Pi)$, so both sides are deterministically $1$.

        Now, consider $p > 0$.
        As $\supp(\Pi)$ is contained within the corners of the simplex, $X_t$ is a deterministic function of $B_t$, for every time point.
        So, the LZ78 prefix tree is a single branch and the $\ell$\textsuperscript{th} phrase has length $\ell$.
        Define $\hat{M}_\ell$ as the empirical measure over the $\ell$\textsuperscript{th} phrase.
        Within a phrase, $\Theta$ values are $\simiid \Pi$, so $\hat{M}_\ell(A) \triangleq \tfrac{1}{\ell} M$, where $M\sim \text{Bin}(p, \ell)$.

        By the Chernoff bound for the binomial distribution,$\forall \epsilon > 0$,
        \[\Pr{\left|\hat{M}_\ell(A) - p\right| > \epsilon} \leq 2\exp{-\frac{\epsilon^2 \ell}{3p},}\]
        where $\exp{\cdot}$ is taken as the base-$e$ exponential function.

        Then, as
        \begin{align*} \sum_{\ell=1}^\infty \Pr{\left|\hat{M}_\ell(A) - p\right| > \epsilon} &\leq 2\sum_{\ell=1}^\infty \left(\exp{-\frac{\epsilon^2}{3p}}\right)^\ell \\ &< \infty,\end{align*}
        by the Borel-Cantelli lemma  $\forall \epsilon > 0$, almost surely,
        \[\limsup_{\ell \to\infty} \indic{\left|\hat{M}_\ell(A) - p\right| > \epsilon} = 0.\]
        We want to show that $\forall \xi > 0$, $\exists N > 0$ s.t. $|M_n(A) - p| < \xi$, $\forall n > N$ (with probability $1$).
        By the Borel-Cantelli result, almost surely $\exists L$ s.t., $\forall \ell > L$, $|\hat{M}_\ell(A) - p| < \frac\xi2$.
        Choose $N$ such that, $\forall n > N$, the first $L$ phrases (plus the potential final, unfinished phrase, which is $O(\sqrt{n})$) are less than a $\frac\xi2$ fraction of the total sequence.
        Then, $\forall n > N$, 
        \[|M_n(A) - p| < \tfrac{\xi}{2}\cdot 1 + \left(1 - \tfrac{\xi}{2}\right) \cdot \tfrac{\xi}{2} \leq \xi\quad\text{(a.s.)}.\]
        $\therefore$, $M_n(A) \convas p$.
    \end{proof}
\end{lemma}

\begin{lemma}[Empirical Distribution of $r$-Tuples for $\Ebb\lbrack H(\Theta)\rbrack = 0$]
Let $A\in \Scal_r$ be any single-sequence event.
Then, for the LZ78 source with $\Ebb[H(\Theta)] = 0$,
\[L_n^{(r)}(A) \convas \etaStarRA\quad \text{as }n\to\infty.\]

\begin{proof}
    The proof is largely the same as that of \prettyref{lem:zero-order-dist-pi-01}.
    Let $A_i$, $1\leq i \leq r$ and $X_*^r$ be as in \prettyref{def:single-sequence-event}.
    For $\etaStarRA \triangleq p = 0$, either $|A_i \cap \mathrm{supp}(\Pi)| = 0$ or $X_{*,i} \not\in A_i$ for some $i \in [r]$.
    In that case, $L_n^{(r)}(A)$ is deterministically $0$.

    Now, assume $p > 0$, and consider the subsequence consisting of every $r$\textsuperscript{th} element of the $\ell$\textsuperscript{th} phrase, starting at element $k$ (for $0 \leq k < r$).
    Define $\hat{M}_\ell^{(k)}$ to be the empirical measure of that subsequence.
    As the elements of the subsequence are in a single phrase and spaced out such that corresponding $r$-tuples do not overlap, all but perhaps the last element of each subsequence (which will include a return to the root) are independent Ber$(p)$ random variables.
    So, $\hat{M}_\ell^{(k)}(A) \triangleq \frac{M + W}{\lfloor\ell/r\rfloor - 1}$ for $M \sim \text{Bin}(p, \lfloor\ell/r\rfloor - 1)$, and $|W| \leq 2$ (covering the last element of the subsequence, as well as edge effects).

    Fix $\epsilon > 0$ and assume $\lfloor\ell/r\rfloor - 1 > \frac{4}{\epsilon}$ (for smaller $\ell$, the trivial bound {\small$\Pr{\left| \hat{M}_\ell^{(k)} -p \right| < \epsilon} \leq 1$} can be used).
    The Chernoff bound on the Binomial distribution produces
    \begin{align*}
        \Pr{\left| \hat{M}_\ell^{(k)} -p \right| < \epsilon} &= \Pr{\left| \tfrac{M}{\lfloor\ell/r\rfloor - 1} - p + \tfrac{W}{\lfloor\ell/r\rfloor - 1} \right| < \epsilon} \\ &\leq \Pr{\left| \tfrac{M}{\ell} - p \right| < \epsilon - \tfrac{2}{\lfloor\ell/r\rfloor - 1}} \\
        &\leq 2\exp{\frac{(\epsilon - \tfrac{2}{\lfloor\ell/r\rfloor - 1})^2(1 - \lfloor\ell/r\rfloor)}{3p}} \\ 
        &\leq 2\exp{\frac{\epsilon^2(1 - \lfloor\ell/r\rfloor)}{12p}}.
    \end{align*}
    Consider the sequence of events 
    % $\left(E_1 = |\hat{M}_1^{(0)} -p| < \epsilon,E_2 = |\hat{M}_1^{(1)} -p| < \epsilon, \dots, E_r = |\hat{M}_2^{(0)} -p| < \epsilon, \dots\right)$.
    \[
        \left(
        \begin{aligned}
        E_1 &\equiv \bigl|\hat{M}_1^{(0)}-p\bigr|<\epsilon,\\
        E_2 &\equiv \bigl|\hat{M}_1^{(1)}-p\bigr|<\epsilon,\\
        &\ \vdots\\
        E_r &\equiv \bigl|\hat{M}_2^{(0)}-p\bigr|<\epsilon,\\
        &\ \vdots
        \end{aligned}
        \right).
    \]
    By the Chernoff bound we derived,
    \[\sum_{i=1}^\infty \Pr{E_i} \leq \frac{4r^2}{\epsilon} + 2r^2 + 2r\sum_{i=1}^\infty \left(\exp{-\frac{\epsilon^2}{12p}}\right)^i < \infty.\]
    Then, by the Borel-Cantelli lemma, $\forall \epsilon > 0$, almost surely $\limsup_{i \to\infty} \indic{E_i} = 0$.
    As a direct result, $\forall 0 \leq k < r$, almost surely $\limsup_{\ell\to\infty} \indic{\left| \hat{M}_\ell^{(k)} -p \right| < \epsilon} =0$ and therefore
    \[\limsup_{\ell \to\infty} \indic{\left|\hat{M}_\ell(A) - p\right| > \epsilon} = 0\quad\text{(a.s.)},\]
    where $\hat{M}_\ell$ is the empirical measure of phrase $\ell$.

    Then, by the argument at the end of \prettyref{lem:zero-order-dist-pi-01} (which we will not repeat here), $L_n^{(r)}(A) \convas p$.
\end{proof}    
\end{lemma}

\section{Moment Definitions}\label{app:moment-definitions}
The following notation for moments of $N^A(\ell)$ and $N(\ell)$ will be used throughout subsequent proofs.
\begin{definition}[First Moments]
    $m_1(\ell) \triangleq \Ebb[N(\ell)]$ and $m_1^A(\ell) \triangleq \Ebb[N^A(\ell)]$.
\end{definition}
\begin{definition}[Second Moment of $N(\ell)$]
    $\hat{m}_2(\ell) \triangleq \Ebb \left[(N(\ell) - m_1(\ell))^2\right]$.
\end{definition}
\begin{definition}[Expression Related to Second Moment of $N^A(\ell)$]
    $\hat{m}_2^A(\ell) \triangleq \Ebb\left[(N^A(\ell) - N(\ell) \Pi(A))^2\right]$.
    
    As we see in \prettyref{app:second-moment-bound-on-NA-proof}, $m_1^A(\ell) = \Pi(A) m_1(\ell)$, so $\hat{m}_2^A(\ell)$ is closely tied to the second moment of $N^A(\ell)$.
\end{definition}

\section{Proof of \prettyref{thm:zero-order-empirical-distribution} Given Claims \ref{claim:NA-to-N-ratio-converges-to-measure} and \ref{claim:N-ratio-converges-to-one}}\label{app:end-of-zero-order-proof}

If we assume \prettyref{claim:NA-to-N-ratio-converges-to-measure} and \prettyref{claim:N-ratio-converges-to-one}, \prettyref{thm:zero-order-empirical-distribution} can be proven directly.
The claims are proven in \prettyref{app:NA-to-N-ratio-converges-to-measure-proof} and \prettyref{app:N-ratio-converges-to-one-proof}, respectively.

\textbf{Theorem \ref{thm:zero-order-empirical-distribution}} (\zeroordertitle)\textbf{.}
    \zeroordercontents
    \begin{proof}
Assuming the results of Claims \ref{claim:NA-to-N-ratio-converges-to-measure} and \ref{claim:N-ratio-converges-to-one} hold, we show that $M_n(A) \convas \Pi(A)$, for any $A \in \Mcal(\Acal)$.

Both $N$ and $N^A$ are monotonically increasing, so
% \begin{equation}\label{eqn:NA-over-N-sandwich-equation}
%     \tfrac{N^A(T_n)}{N(T_n+1)} \leq M_n(A) \leq \tfrac{N^A(T_n + 1)}{N(T_n)} \stackrel{(*)}{\iff} \tfrac{N^A(T_n)}{N(T_n)} \cdot \tfrac{N(T_n)}{N(T_n+1)} \leq M_n(A) \leq \tfrac{N^A(T_n+1)}{N(T_n+1)} \cdot \tfrac{N(T_n+1)}{N(T_n)},
% \end{equation}
\begin{IEEEeqnarray}{c}
    \frac{N^A(T_n)}{N(T_n+1)} \le M_n(A) \le \frac{N^A(T_n+1)}{N(T_n)} \overset{(*)}{\Longleftrightarrow}\label{eqn:NA-over-N-sandwich-equation}\\
    \frac{N^A(T_n)}{N(T_n)}\frac{N(T_n)}{N(T_n+1)}
    \le M_n(A) \le
    \frac{N^A(T_n+1)}{N(T_n+1)}\frac{N(T_n+1)}{N(T_n)} \nonumber
\end{IEEEeqnarray}
where we multiply by $\frac{N(T_n)}{N(T_n)}$ on the left and $\frac{N(T_n+1)}{N(T_n+1)}$ on the right to get $(*)$.

By Claims \ref{claim:NA-to-N-ratio-converges-to-measure} and \ref{claim:N-ratio-converges-to-one}, $\frac{N^A(\ell_k)}{N(\ell_k)} \convas \Pi(A)$ and $\frac{N(\ell_k)}{N(\ell_{k+1})} \convas 1$ (\ie, convergence over a specific sequence $\ell_k$); now, we leverage the monotonicity of $N(\ell)$ and $N^A(\ell)$ to show that $\frac{N^A(T_n)}{N(T_n)} \convas \Pi(A)$ and $\frac{N(T_n)}{N(T_n+1)} \convas 1$.

As $\ell_k$ is a deterministic, strictly increasing sequence, $\forall n > 0$, $\exists k$ s.t. $\ell_k \leq T_n < T_n + 1 \leq \ell_{k+1}$, and $k\to\infty$ with $n$.
By the monotonicity of $N(\ell)$, $\frac{N(\ell_k)}{N(\ell_{k+1})} \leq \frac{N(T_n)}{N(T_n+1)} \leq 1$.
By \prettyref{claim:N-ratio-converges-to-one}, $\frac{N(\ell_k)}{N(\ell_{k+1})} \convas 1$ as $k\to\infty$, so $\frac{N(T_n)}{N(T_n+1)}$ does as well via the squeeze theorem.

Likewise, by monotonicity, $\frac{N^A(\ell_k)}{N(\ell_{k+1})} \leq \frac{N^A(T_n)}{N(T_n)} \leq \frac{N^A(\ell_{k+1})}{N(\ell_k)}$.
Multiplying by $\frac{N(\ell_k)}{N(\ell_k)}$ on the left and $\frac{N(\ell_{k+1})}{N(\ell_{k+1})}$ on the right,
\[\tfrac{N(\ell_k)}{N(\ell_{k+1})}  \cdot \tfrac{N^A(\ell_k)}{N(\ell_k)} \leq \tfrac{N^A(T_n)}{N(T_n)} \leq \tfrac{N^A(\ell_{k+1})}{N(\ell_{k+1})} \cdot \tfrac{N(\ell_{k+1})}{N(\ell_k)}.\]
By Claims \ref{claim:NA-to-N-ratio-converges-to-measure} and \ref{claim:N-ratio-converges-to-one}, both sides almost surely converge to $\Pi(A)$, so, by the squeeze theorem, $\frac{N^A(T_n)}{N(T_n)} \convas \Pi(A)$.

Via an identical argument, applied to \eqref{eqn:NA-over-N-sandwich-equation}, $M_n(A) \convas \Pi(A)$ as $n\to\infty$.
\end{proof}

\section{Zero-Order Empirical Distribution: Proof of \prettyref{claim:NA-to-N-ratio-converges-to-measure}} \label{app:NA-to-N-ratio-converges-to-measure-proof}

\textbf{Claim \ref{claim:NA-to-N-ratio-converges-to-measure}.}
   \NAratioclaimcontents
\begin{proof}
    The proof relies on the following claim, which we will assume true for now and prove in \prettyref{app:second-moment-bound-on-NA-proof}.
    \begin{claim}\label{claim:second-moment-bound-on-NA}
        For $d \triangleq \Ebb[\lVert \Theta \lVert_2^2] < 1$,
        $\hat{m}_2^A(\ell)$ satisfies
        \[\hat{m}_2^A(\ell) \triangleq \Ebb\left[(N^A(\ell) - N(\ell) \Pi(A))^2\right] \leq C_d \ell^2,\]
        where $C_d$ is a finite constant that depends only on $d$.\footnote{$d < 1$ due to the condition $\Ebb_{\Theta \sim \Pi} H(\Theta) > 0$.
        See \prettyref{app:proofs-for-supp-pi-01} for the $\Ebb_{\Theta \sim \Pi} H(\Theta) = 0$ case.}
    \end{claim}

    For arbitrary $\epsilon > 0$, the probability that the ratio $\frac{N^A(\ell)}{N(\ell)}$ is more than $\epsilon$ away from the true measure is
    \begin{align*}
    &\Pr{\left|\tfrac{N^A(\ell)}{N(\ell)} - \Pi(A)\right| > \epsilon} \\
    &\qquad\qquad\qquad= \Pr{\left|N^A(\ell) - \Pi(A)\,N(\ell)\right| > \epsilon\,N(\ell)} .
    \end{align*}

    By \prettyref{rem:extreme-values-of-N-ell}, $N(\ell) \geq C \ell \log(\ell)$, for universal constant $C$.
    Applying this fact, Chebyshev's inequality, and \prettyref{claim:second-moment-bound-on-NA},
    \begin{align*}
    \Pr{\left|\tfrac{N^A(\ell)}{N(\ell)} - \Pi(A)\right| > \epsilon} &\leq \frac{\Ebb\left[(N^A(\ell) - N(\ell) \Pi(A))^2\right]}{C^2\epsilon^2 \ell^2 (\log \ell)^2}  \\
    &= \frac{\hat{m}_2^A(\ell)}{C^2\epsilon^2 \ell^2 (\log \ell)^2} \leq \frac{C_d/C^2}{\epsilon^2 (\log \ell)^2}.
    \end{align*}
    We are now in a position to finish the proof using the Borel-Cantelli lemma:
    \[\sum_{k=1}^\infty \Pr{\left|\tfrac{N^A(\ell_k)}{N(\ell_k)} - \Pi(A)\right| > \epsilon} \leq \frac{C_a}{C^2 \epsilon^2} \sum_{k=1}^\infty (\log \ell_k)^{-2}.\]
    Plugging in $\ell_k \cong \exp{\sqrt{k} (\log k)^\alpha}$ for any $\frac{1}{2} < \alpha < 1$,
    \[\sum_{k=1}^\infty \Pr{\left|\tfrac{N^A(\ell_k)}{N(\ell_k)} - \Pi(A)\right| > \epsilon} \leq \frac{O(1)}{\epsilon^2} \sum_{k=1}^\infty \frac{1}{k(\log k)^{2\alpha}},\]
    which is finitely summable, $\forall \alpha > \frac{1}{2}$.
    
    Thus, by the Borel-Cantelli lemma, $\frac{N^A(\ell_k)}{N(\ell_k)} \convas \Pi(A)$ as $k\to\infty$.
\end{proof}

\subsection{Proof of \prettyref{claim:second-moment-bound-on-NA}}\label{app:second-moment-bound-on-NA-proof}
\textbf{Claim \ref{claim:second-moment-bound-on-NA}.}
        Assume $d \triangleq \Ebb[\lVert \Theta \rVert_2^2] < 1$.
        Then, $\hat{m}_2^A(\ell)$ satisfies
        \[\hat{m}_2^A(\ell) \triangleq \Ebb\left[(N^A(\ell) - N(\ell) \Pi(A))^2\right] \leq C_d \ell^2,\]
        where $C_a$ is a finite constant that depends only on $d$.
        
\textit{Proof}.    
Based on the tree structure of the LZ78 source, we write $N^A(\ell)$, for arbitrary $\ell$, via the following recursion:
\begin{construction}[Tree Recursion of $N^A(\ell)$]\label{con:tree-recursion}
    Let the value of $\Theta$ corresponding to the root of the LZ78 tree be $\theta_0 \sim \Pi$.
    Conditioned on $\theta_0$, let 
    \[R \triangleq \{R_a\}_{a \in \Acal} \sim \mathrm{Multinomial}(\ell - 1, \theta_0)\] be a set of random variables representing the number of times we traverse from the root to the sub-tree corresponding to symbol $a \in \Acal$.
    Then, we have the recursion
    \begin{equation}
        N^A(\ell) = \ell \delta_{\theta_0}(A) + \sum_{a \in \Acal} N_a^A(R_a), \label{eqn:recursion}
    \end{equation}
    where $N_a^A$ is the number of times that $B_t \in A$ for steps in the sub-tree corresponding to symbol $a$.
    $\{N_a^A\}_{a\in \Acal}$, conditioned on their arguments, are independent of each other and have the same distribution  as $N^A$.

    This recursion holds because, for each of the $\ell$ times we visited the root except the first one, we traversed to the node corresponding to drawing symbol $a$ with probability $\theta_{0, a}$ ($R_a$ times in total), for every $a  \in \Acal$.
    This same relation holds for every node of the tree, producing the recursion of \eqref{eqn:recursion}.
\end{construction}

Taking the expectation of \eqref{eqn:recursion} for $N^A(\ell)$ and $N(\ell)$ (plugging in the full unit interval for $A$ to get $N(\ell)$),
\begin{alignat*}{2}
    m_1^A(\ell) &= \Ebb[N^A(\ell)] &&= \ell \Pi(A) + \sum_{a \in \Acal} \Ebb[m_1^A(R_a)], \\
    m_1(\ell) &= \Ebb[N(\ell)] &&= \ell + \sum_{a \in \Acal}  \Ebb[m_1(R_a)],
\end{alignat*}
$m_1^A(0) = m_1(0) = 0$, so this recursion has a unique solution.
By inspection, it is evident that this unique solution satisfies $m_1^A(\ell) = \Pi(A) m_1(\ell)$, which will be helpful in deriving a recursion for $\hat{m}_2^A(\ell)$.

Using \eqref{eqn:recursion} and the law of total expectation,
\begin{align*}
    \hat{m}_2^A(\ell) &= \Ebb \left[ (N^A(\ell) - \Pi(A)N(\ell))^2 \right] \\
    &= \Ebb \Bigg[ \bigg(\ell\left(\delta_{\theta_0}(A) - \Pi(A)\right) \\
    &\quad+ \sum_{a\in\Acal}\left(N_a^A(R_a) - \Pi(A) N_a(R_a)\right)\bigg)^2 \Bigg] \\
    &= \Ebb_{\theta_0, R} \Bigg[\Ebb \bigg[ \bigg(\ell\left(\delta_{\theta_0}(A) - \Pi(A)\right) \\
    &\quad+ \sum_{a\in\Acal} \left(N_a^A(R_a) - \Pi(A) N_a(R_a)\right)\bigg)^2\,\,\bigg\rvert\,\, \theta_0, R\bigg] \Bigg].
\end{align*}

Conditioned on $R$ and $\theta_0$, $\{N_a^A(R_a)\}_{a\in\Acal}$ are independent and $\Ebb[N_a^A(R_a) - \Pi(A) N_a(R_a)] = m_1^A(R_a) - \Pi(A)m_1(R) = 0$, $\forall a \in \Acal$.
So, $\hat{m}_2^A(\ell)$ evaluates to
\begin{align*}
    \hat{m}_2^A(\ell) &=\Ebb_{\theta_0, R}\bigg[ \ell^2 \left( \delta_{\theta_0}(A) - \Pi(A) \right)^2 + \sum_{a\in\Acal} \hat{m}_2^A(R_a) \bigg] \\
    &= \ell^2 \text{var}(\delta_{\theta_0}(A)) + \Ebb_{\theta_0, R} \left[ \sum_{a\in\Acal} \hat{m}_2^A(R_a)\right] \\
    &\leq \ell^2 + \Ebb_{\theta_0, R} \left[ \sum_{a\in\Acal}\hat{m}_2^A(R_a) \right],
\end{align*}
as the variance of an indicator is $\leq 1$.

It now remains to bound this quantity by $O(\ell^2)$.
As will become useful later (in proving the $r$-tuple result), we prove the following, more general statement:
\newcommand{\Ca}{\widetilde{C}_{\hat{d}}}
\begin{lemma}[$\ell^2$ Inductive Bound]\label{lem:inductive-ell-squared-bound}
    Suppose, for some function $f: \mathbb{N} \to \mathbb{R}$ and random variable $\theta$ over $\Mcal(\Acal)$,
    \begin{align*}
        f(\ell) &\leq C_0 \ell^2 + (1+\epsilon) \Ebb\left[{\sum}_{a\in\Acal} f(R_a^{(\ell)})\right],\\
        \quad f(0) &= 0,\\
        \quad R^{(\ell)}\big|\theta &\sim\mathrm{Multinomial}(\ell-1, \theta),
    \end{align*}
    where $0 \leq \epsilon < \frac{1-d}{d}$ (with $d=\Ebb[\lVert \theta \rVert_2^2]$).

    Define $\delta$ such that $\rho \triangleq (1 + \epsilon)(d + \delta) < 1$.\footnote{Such $\delta$ is possible to define, as we require $d < 1$ and $(1 + \epsilon) d < 1$.}
    Then, 
    \[f(\ell) \leq \tilde{C}_\rho \ell^2,\quad \forall \ell > 0\]
    where $\tilde{C}_\rho$ is a constant that only depends on $\rho$ and $C_0$.

    \begin{proof}
        Consider $g(\ell) \triangleq f(\ell) / \ell^2$, for any $\ell > 0$, and define $G_\ell \triangleq \max_{k\leq \ell} g(k)$.
        It suffices to show that $G_\ell \leq \tilde{C}_\rho$, $\forall \ell > 0$.

        By assumption,
        \begin{align*}
            g(\ell) &\leq C_0 + \frac{1+\epsilon}{\ell^2} \Ebb\left[{\sum}_{a\in\Acal} g(R_a^{(\ell)}) (R_a^{(\ell)})^2\right] \\
            &\leq C_0 + \frac{(1+\epsilon) G_{\ell-1}}{\ell^2} \sum_{a\in\Acal} \Ebb\left[ (R_a^{(\ell)})^2 \right],
        \end{align*}
        as $R^{(\ell)}_a \leq \ell - 1$, $\forall a \in \Acal$.

        $R_a^{(\ell)}\big|\theta\sim\text{Bin}(\ell-1, \theta_a)$, where $\theta_a \triangleq \Pr{(\theta \sim \Pi) = a}$, so
        \begin{align*}
            \Ebb \left[\left(R_a^{(\ell)}\right)^2\right] &= (\ell - 1)^2 \Ebb [\theta_a^2] + (\ell-1)\Ebb[\theta_a(1-\theta_a)]; \\
            \sum_{a\in\Acal} \Ebb \left[\left(R_a^{(\ell)}\right)^2\right] &= d(\ell-1)^2 + (1-d)(\ell - 1).
        \end{align*}

        Plugging this in,
        \[g(\ell) \leq C_0 + (1+\epsilon) G_{\ell-1} \underbrace{\frac{d(\ell-1)^2 + (1-d)(\ell - 1)}{\ell^2}}_{\triangleq D_\ell}.\]
        $D_\ell \to d$ as $\ell\to\infty$, so there exists $L > 0$ such that $\forall \ell \geq L$, $D_\ell \leq d + \delta$
        (where $\delta > 0$ is as defined in the Lemma statement).
        So, $\forall \ell \geq L$,
        \begin{equation}
            g(\ell) \leq C_0 + (1+\epsilon)(d + \delta) G_{\ell - 1} = C_0 + \rho G_{\ell - 1},\label{eqn:g-ell-bound-simple}
        \end{equation}
        where, by assumption, $0 < \rho < 1$.

        Hence, by backward recursion, for any $\ell > L$, 
        \[G_\ell \le C_0 [ 1 + \rho + \rho^2 + \rho^{\ell-(L+1)}] + \rho^{\ell-L} G_L
        \le C_0/(1-\rho) + G_L,\]
        with $L$ depending only on $\rho$, and
        with \eqref{eqn:g-ell-bound-simple} holding at any $\ell$, except for having there $D_\ell$ instead of $\rho$.
        So, we conclude that $G_L$ is some explicit function of $C_0$, 
        $d+\delta$ and $D_2,\dots,D_L$ which in turn are determined by $\rho$.
        % $\forall \ell \geq L$, $G_{\ell}$ is bounded by
        % \[G_{\ell} \leq \max(G_{\ell-1}, C_0 + \rho G_{\ell-1}).\]
        % Consider two cases.
        % \textit{Case 1}: $G_{\ell - 1} > \frac{C_0}{1 - \rho}$, then
        % \[G_{\ell-1} - (C_0 + \rho G_{\ell-1}) = G_{\ell-1} (1 - \rho) - C_0 > 0,\]
        % so $G_\ell = G_{\ell - 1}$.
        % \textit{Case 2}: $G_{\ell - 1} \leq \frac{C_0}{1 - \rho}$.
        % Then,
        % \[G_{\ell-1} - (C_0 + \rho G_{\ell-1}) = G_{\ell-1} (1 - \rho) - C_0 \leq 0,\]
        % so we have the bound
        % \[G_\ell = C_0 + \rho G_{\ell-1} \leq C_0 + \rho \frac{C_0}{1-\rho} = \frac{C_0}{1-\rho}.\]
        % This holds for all $\ell \geq L$, so we have 
        % \[G_\ell \leq \max(G_L, \tfrac{C_0}{1-\rho}) \triangleq \tilde{C}_p,\quad \forall \ell > 0.\]
    \end{proof}
\end{lemma}

\section{Zero-Order Empirical Distribution: Proof of \prettyref{claim:N-ratio-converges-to-one}} \label{app:N-ratio-converges-to-one-proof}

First, using the recursive structure of the LZ78 tree, we prove that the expected phrase length grow logarithmically, which is then used to prove that the gap between consecutive $m_1(\ell_k)$ (the expected number of symbols in the first $\ell_k$ phrases) grow slower than $\ell_k \log \ell_k$.
We then prove that the second moment of $N(\ell)$ grows at most quadratically in $\ell$.
Together, these can be used to bound the squared distance between consecutive $N(\ell_k)$, such that Borel-Cantelli can be applied to prove \prettyref{claim:N-ratio-converges-to-one}.

\subsection{Growth of Expected LZ78 Source Phrase Length}
First, we prove that, up to an additive constant gap, the expected length of the $\ell$\textsuperscript{th} phrase grows logarithmically with $\ell$.
This order of growh result forms the basis for several components of \prettyref{claim:N-ratio-converges-to-one}.
\begin{lemma}\label{lem:delta-bound}
    Let $\Delta(\ell) \triangleq m_1(\ell+1) - m_1(\ell)$, \ie, the expected length of the $\ell$\textsuperscript{th} phrase, and 
    \[h \triangleq \Ebb_{\Theta \sim \Pi} H(\Theta) = \Ebb_{\Theta \sim \Pi} \left[{\sum}_{a\in\Acal}\Theta_a \log \tfrac{1}{\Theta_a}\right],\]
    where $\Theta_a \triangleq \Pr{(\Theta \sim \Pi) = a}$.

    If $h > 0$, then $\exists$ finite constant $C_*$ (which only depends on $\Pi$) such that
    \[\left| \Delta(\ell) - \tfrac{1}{h} \log \ell\right| \leq C_*,\quad \forall \ell \geq 1.\]
    \begin{proof}
Applying a similar tree recursion to \prettyref{con:tree-recursion},
\begin{align*}
    \Delta(\ell) = 1 + \Ebb[\Delta(\Gamma_\ell)],\forall \ell \geq 1;\quad \Delta(0) = 1,
\end{align*}
where, for any fixed value of $\Theta$,
\[\Gamma_\ell \stackrel{d}{=} \mathrm{Bin}(\ell-1, \Theta_a)\quad \text{w.p.}\quad\Theta_a.\]
In the expression for $\Delta(\ell)$, the term of $1$ comes from traversing the root, $\Gamma_\ell$ is a random variable representing the number of times the second node in the phrase has been traversed, and $\Delta(r)$ is the expected phrase length, starting from a node that has been previously traversed $r$ times.

Define quantity $\hat{\Delta}(\ell)$ as
\[\hat{\Delta}(\ell) \triangleq h\Delta(\ell) - \log (\ell+\gamma) = h + h\Ebb[\Delta(\Gamma_\ell)] - \log(\ell + \gamma),\]
where $\gamma \triangleq \frac{1}{|\Acal| - 1}$.

We want to show that $-C_* \leq \hat{\Delta}(\ell) \leq C_*$, $\forall \ell \geq 1$, for finite, positive constant $C_*$ that may depend on $\Pi$.
This suffices to prove the claim, as the difference between $\log (\ell+\gamma)$ and $\log (\ell)$ is $O(\frac{1}{\ell})$ (via Taylor series approximation).

Consider $\hat{\Delta}(\ell) - \Ebb[\hat{\Delta}(\Gamma_\ell)]$,
and let 
\[R = \{R_a\}_{a\in\Acal} \sim \mathrm{Multinomial}(\ell - 1, \Theta)\]
be the number of times that each node at the first level of the tree (below the root) was traversed.
Then,
\begin{equation}
    \begin{aligned}
        \hat{\Delta}(\ell) - \Ebb[\hat{\Delta}(\Gamma_\ell)] &= h + \Ebb\left[ \log(\Gamma_\ell + \gamma)\right] - \log(\ell + \gamma) \\\
        &\hspace{-3em}= \Ebb_{R,\Theta} \left[\sum_{a\in\Acal} \Theta_a \log \frac{1}{\Theta_a} + \Theta_a \log \frac{R_a + \gamma}{\ell + \gamma}\right] \\
        &\hspace{-3em}= -\Ebb_{R,\Theta} \left[ D\left(\Theta \lVert \tfrac{R+\gamma}{\ell+\gamma}\right)\right],
    \end{aligned}\label{eqn:delta-hat-and-relative-entropy}
\end{equation}
where the arithmetic operations $\frac{R+\gamma}{\ell + \gamma}$ are taken elementwise over the vector $R$.
$\frac{R+\gamma}{\ell + \gamma}$ is a valid probability distribution, as
\[\sum_{a\in\Acal} \frac{R_a+\gamma}{\ell + \gamma} = \frac{\ell - 1 + \frac{|\Acal|}{|\Acal| - 1}}{\ell + \gamma} = \frac{\ell + \gamma}{\ell + \gamma} = 1.\]

We now use this relationship to prove the upper bound $\hat{\Delta}(\ell) \leq C_U$, followed by the lower bound $-C_L \leq \hat{\Delta}(\ell)$, where $C_*$ from the statement of the claim can be taken to be $\max(C_L, C_U)$.

\textbf{Upper bound}.
By the non-negativity of relative entropy, \eqref{eqn:delta-hat-and-relative-entropy} implies $\hat{\Delta}(\ell) \leq \Ebb[\hat{\Delta}(\Gamma_\ell)]$.
We can use this to prove $\hat{\Delta}(\ell) \leq h$, $\forall \ell \geq 1$ via induction.

For the base case, the first phrase always has length $1$, so
\[\hat{\Delta}(1) = h \Delta(1) - 1 = h - 1 < h.\]
Now, assume that, $\forall 1 \leq k < \ell$, $\hat{\Delta}(k) \leq h$.
Then, by \eqref{eqn:delta-hat-and-relative-entropy} and the inductive hypothesis,
\[\hat{\Delta}(\ell) \leq \Ebb[\hat{\Delta}(\Gamma_\ell)] \leq \Ebb[h] = h.\]

\textbf{Lower bound}.
We first prove the following bound on $\Ebb_{R} \left[ D\left(\Theta \lVert \tfrac{R+\gamma}{\ell+\gamma}\right)\right]$ for fixed $\Theta$.
\begin{lemma}\label{lem:upper-bound-relative-entropy}
    For any fixed $\Theta \in \Mcal(\Acal)$ and $\ell \geq 1$,
    \[\Ebb_{R} \left[ D\left(\Theta \lVert \tfrac{R+\gamma}{\ell+\gamma}\right)\right] \leq \frac{C'}{\ell},\]
    where $C'$ is a finite constant that only depends on $\gamma$ (i.e., on the alphabet size).

    \begin{proof}
        By \cite{nishiyama2020relations}, the relative entropy is bounded by the $\chi^2$ divergence, resulting in
        \[\Ebb_R \left[D\left(\Theta \lVert \tfrac{R+\gamma}{\ell+\gamma}\right)\right] \leq \Ebb_R \sum_{a\in\Acal} \frac{\Theta_a^2 (\ell + \gamma)}{R_a + \gamma} - 1.\]
We show in the sequel that for 
$X \sim \mathrm{Bin}(n,p)$, any $n \ge 0$, $p \in [0,1)$ and $\gamma \in (0,1]$,
\begin{equation}\label{bd-bin}
\Ebb \Big[\frac{(n+1)p}{X+\gamma}\Big] \le 1 + \frac{7}{\gamma (n+1)p} \,.
\end{equation}
Applying \eqref{bd-bin} for $X=R_a$, where $p=\Theta_a$ and $n=\ell-1$, then 
summing over $a \in \Acal$, results with 
\begin{align*}
\Ebb_R \sum_{a\in \Acal} \frac{\Theta_a^2 (\ell + \gamma)}{R_a + \gamma} 
&\le \frac{\ell+\gamma}{\ell} \sum_{a \in \Acal} \Theta_a \Big(1+\frac{7}{\gamma \ell \Theta_a}\Big) \\
&= \Big(1 + \frac{\gamma}{\ell} \Big) \Big(1 + \frac{7 |\Acal|}{\gamma \ell} \Big)
= 1 + \frac{C'}{\ell} 
\end{align*}
for some finite $C'$ that only depends on $\gamma$, as stated. Turning to show 
\eqref{bd-bin}, the following easy to verify identity
\[
\Ebb \Big[\frac{(n+1)p}{X+\gamma}\Big] = \Ebb\Big[Z {\bf 1}_{Y \ge 1} \Big]\,,
\]
with $Y \sim \mathrm{Bin}(n+1,p)$ and $Z:=\frac{Y}{Y-1+\gamma} \le \frac{1}{\gamma}$ 
whenever $Y \ge 1$, implies that 
for $A=\{Y \ge 1,  Z \ge 1 + \frac{3}{\gamma (n+1) p}\}$,
\[
\Ebb \Big[\frac{(n+1)p}{X+\gamma}\Big] \le 1 + \frac{3}{\gamma (n+1)p} +
\frac{1}{\gamma} \Pr\{A\}\,.
\]
Further, with $X \ge 0$ we consider only $(n+1) p \ge \sqrt{7}$,
in which case elementary algebra shows that $Y \le \frac{1}{2} (n+1) p$
throughout $A$. Recall that $\Ebb Y = (n+1) p$ and Var$(Y)
%=(1-p) p (n+1) 
\le \Ebb Y$, so by Chebyshev's inequality we then have
\begin{align*}
\Pr\{A\} \le
\Pr \{ Y \le \frac{1}{2} (n+1) p \} 
& \le \Pr \{ |Y - \Ebb Y| \ge \frac{1}{2} \Ebb Y \} \\
& \le \frac{4 \mathrm{Var(Y)}}{(\Ebb Y)^2} \le \frac{4}{(n+1)p} \,,
\end{align*}
out of which \eqref{bd-bin} immediately follows.
    \end{proof}
\end{lemma}
Applying \prettyref{lem:upper-bound-relative-entropy} and \eqref{eqn:delta-hat-and-relative-entropy},
\[\hat{\Delta}(\ell) \geq \Ebb[\hat{\Delta}(\Gamma_\ell)] - \frac{C'}{\ell}.\]
Define $I(\ell) = \inf_{k\leq \ell} \hat{\Delta}(k)$.
For every phrase, $I(\ell)$ is either equal to $I(\ell - 1)$ or $\hat{\Delta}(\ell)$.
Let $\mathcal{I}$ represent the indices where $I(\ell) = \hat{\Delta}(\ell)$, \ie, those corresponding to ``jumps'' in the value of $I(\ell)$.

We wish to bound $I(\ell)$ for $\ell \in \mathcal{I}$ (which, consequently, bounds $I(\ell),\, \forall \ell$).
To that extent, we prove the following lemma:
\begin{lemma}\label{lem:gamma-k-probabilities}
    If $h > 0$, $\exists\, 0 < r < \frac{1}{2}, 0 < b < 1$ and $\ell_0 < \infty$ (all of which may depend on $\Pi$), such that
    \[q_k \triangleq \Pr{\Gamma_k \leq (1 - r)k} > b,\quad \forall k \geq \ell_0.\]
    \begin{proof}
        Since $\Pr{\mathrm{Bin}(k-1,p) > x)}$ is non-increasing in $p$, for any choice of $k$ and $r$:

        \begin{equation}
            \begin{aligned}
                1-q_k &= \Pr{\Gamma_k > (1-r) k} \\
            &\le \Pr{\max_a \Theta_a > 1-2r} \\
            &\qquad+ \Pr{\mathrm{Bin}(k-1,1-2r) > (1-r) k}. \label{eqn:one-minus-qk}
            \end{aligned}
        \end{equation}

        As $h > 0$, $\exists\, 0 < r < \frac{1}{2}$ and $b > 0$ such that
        \[\Pr{\max_{a \in \Acal} \Theta_a \leq (1-2r)} \geq 2b.\]
        Otherwise, we would have $\max_{a\in\Acal}\Theta_a = 1$ with probability $1$, which is only satisfied by zero-entropy distributions.
        So, the first term on the RHS of \eqref{eqn:one-minus-qk} is at most $1-2b$.
                
        Moreover, from law of large numbers, we know that for some finite $\ell_0=\ell_0(r,b)$ the second term on RHS of  \eqref{eqn:one-minus-qk} is at most $b$ for all $k > \ell_0$, in which case
        \[1-q_k \le (1-2b) + b = 1 -b\]
        as claimed.
    \end{proof}
\end{lemma}
Let $r, b, \ell_0$, and $q_\ell$ be as in \prettyref{lem:gamma-k-probabilities}.
Then, for $\ell \in \mathcal{I}$ such that $\ell \geq \ell_0$,
\begin{align*}
    I(\ell) &= \hat{\Delta}(\ell) \geq \Ebb[I(\Gamma_\ell)] - \frac{C'}{\ell} \\
    &\geq (1-q_\ell)I(\ell-1) + q_\ell I\left((1-r)\ell\right) - \frac{C'}{\ell} \\
    &\geq (1-b)I(\ell) + bI\left((1-r)\ell\right) - \frac{C'}{\ell},
\end{align*}
by \prettyref{lem:gamma-k-probabilities} and the monotonicity of $I(\cdot)$.
Solving for $I(\ell)$,
\[I(\ell) \geq I\left((1-r)\ell\right) - \frac{C'}{b\ell} \triangleq I\left((1-r)\ell\right) - \frac{C}{\ell}.\]
Let $g(\ell) \triangleq \max \{k \leq \ell : k \in \mathcal{I}\}$.
Unrolling the recursion of the above $I(\ell)$ bound,
\begin{align*}
    I(\ell) \geq I\left(g((1-r)\ell)\right) - \frac{C}{\ell} 
    \geq & I\left(g\left\{(1-r)g((1-r)\ell)\right\}\right) \\
    &- \frac{C}{g((1-r)\ell)} - \frac{C}{\ell} \geq \cdots.
\end{align*}
For all $\ell$, $\hat{\Delta}(\ell)$ then satisfies
\begin{align*}
    \hat{\Delta}(\ell) &\geq I(\ell)\geq I(\ell_0) - \max_{\substack{\text{sequences }\{k_i\}_{i=1}^n \text{ s.t. } \\k_0 \geq \ell_0, k_i \leq (1-r)k_{i+1}, k_n=\ell}} C\sum_{i=1}^n \frac{1}{k_i} \\
    &\geq I(\ell_0) - C\sum_{i=0}^{\infty} \frac{1}{(1-r)^i} = I(\ell_0) - \frac{C}{r} \\
    &\triangleq -C_L > -\infty.
\end{align*}        
    \end{proof}
\end{lemma}

\subsection{Convergence Rate of Expected Cumulative Length $m_1(\ell)$}
The bound on the phrase length from \prettyref{lem:delta-bound} also induces a bound on $m_1(\ell)$, i.e., the cumulative length of the first $\ell$ phrases (as defined in \prettyref{app:moment-definitions}).
We then use the bound on $m_1(\ell)$ to prove that the distance between $m_1(\ell_{k})$ for consecutive $k$ grows slower than $\ell_k \log \ell_k$ (\prettyref{lem:m1-difference-grows-slower-than-ell-k-log-ell-k}).
This will later be used to bound the distance between consecutive $N(\ell_k)$.
\begin{corollary}\label{cor:m1ell-bound}
    $\exists$ finite constant $C$, which depends on $\Pi$, such that
    \[\left| m_1(\ell) - \tfrac{1}{h} \ell \log \ell\right| \leq C \ell.\]

    \begin{proof}
        Let $(\ast)$ stand for $ \left| m_1(\ell) - \tfrac{1}{h}\ell \log \ell\right|$.
        By the definition of $\Delta(\ell)$ and the triangle inequality,
        \begin{align*}
            (\ast) &= \left|\sum_{j=0}^{\ell-1}\Delta(j) -  \tfrac{1}{h}\ell \log \ell\right| \\
            &\leq \left|\sum_{j=1}^{\ell-1}\Delta(j) - \sum_{j=1}^{\ell-1} \tfrac{1}{h}\log j\right| \\
            &\qquad+ \left|\sum_{j=1}^{\ell-1} \tfrac{1}{h}\log j - \tfrac{1}{h}\ell \log \ell\right| + 1\\
            &\leq \sum_{j=1}^{\ell-1} |\Delta(j) - \tfrac{1}{h}\log j| + \left|\sum_{j=1}^{\ell-1} \tfrac{1}{h}\log j - \tfrac{1}{h}\ell \log \ell\right| + 1.
        \end{align*}
        Applying \prettyref{lem:delta-bound},
        \begin{align*}
            \left| m_1(\ell) - \tfrac{1}{h}\ell \log \ell\right| &\leq \ell C_* + \left|\sum_{j=1}^{\ell-1} \tfrac{1}{h}\log j - \tfrac{1}{h}\ell \log \ell\right| + 1.
        \end{align*}
        Using Stirling's approximation, the second term simplifies to
        \begin{align*}
            \left|\sum_{j=1}^{\ell-1} \tfrac{1}{h}\log j - \tfrac{1}{h}\ell \log \ell\right| &= \tfrac{1}{h}\left|\log ((\ell - 1)!) - \ell \log \ell\right| \\
            &= \tfrac{1}{h}\left|\log (\ell!) - \ell \log \ell - \log \ell\right| \\
            &=\tfrac{1}{h}\left|\ell\log \ell - O(\ell) - \ell \log \ell\right| \\
            &\leq C'\ell,
        \end{align*}
        for some constant $C'$, so the overall $\left| m_1(\ell) - \tfrac{1}{h}\ell \log \ell\right|$ is bounded by a constant times $\ell$.
    \end{proof}
\end{corollary}

To extend this corollary from $m(\ell)$ to $m(\ell_k)$, we need the following result:
\begin{fact}[$\ell_{k+1}$ with respect to $\ell_k$]\label{fact:ell-k-plus-1}
    Recall $\ell_k = f(k) \exp{\sqrt{k}(\log k)^\alpha}$ for any fixed $\frac{1}{2} < \alpha < 1$ and $f(k) > 0$ such that $f(k+1) = f(k) + o(1)$.
    Then,
    \[\ell_{k+1} = \ell_k (1 + o(1))\]
    and
    \[\quad \ell_{k+1}\log \ell_{k+1} = \ell_k \log \ell_k (1 + o(1).\]

    \begin{proof}
        Applying Taylor's theorem to the two factors in argument of the exponent,
        \begin{align*}
            &\sqrt{k+1}(\log (k+1))^\alpha \\
            &\qquad= \left(\sqrt{k} + O\left(k^{-1/2}\right)\right)\left((\log k)^\alpha + O\left(\tfrac{(\log k)^{\alpha - 1}}{k}\right)\right) \\
            &\qquad= \sqrt{k}(\log k)^\alpha + o(1).
        \end{align*}
        Plugging this into $\ell_{k+1}$,
        \[\ell_{k+1} = f(k+1) \exp{\sqrt{k}(\log k)^\alpha + o(1)} = \ell_k (1 + o(1)).\]
        In addition, $\ell_{k+1}\log \ell_{k+1}$ evaluates to
        \begin{align*}
            \ell_{k+1}\log \ell_{k+1} &= f(k+1) \exp{\sqrt{k}(\log k)^\alpha + o(1)}  \\ &\qquad\cdot\left(\sqrt{k}(\log k)^\alpha + \log f(k+1) + o(1)\right) \\
            &= \ell_k (1 + o(1)) (\log \ell_k + o(1)) \\
            &= (1 + o(1))\ell_k \log \ell_k .
        \end{align*}
    \end{proof}
\end{fact}

Now, we prove the desired bound on the difference between consecutive $m(\ell_k)$.
\begin{lemma}\label{lem:m1-difference-grows-slower-than-ell-k-log-ell-k}
    Assume $\Ebb_{\Theta\sim\Pi}H(\Theta) > 0$.
    Then, for the same deterministic $\ell_k$ as in \prettyref{claim:NA-to-N-ratio-converges-to-measure},
    \[\frac{m_1(\ell_{k+1}) - m_1(\ell_k)}{\ell_k \log \ell_k} \to 0.\]
\end{lemma}
\begin{proof}
    Applying \prettyref{cor:m1ell-bound}, 
    \begin{equation}
        \begin{aligned}
            0 &\leq \frac{m_1(\ell_{k+1}) - m_1(\ell_k)}{\ell_k\log\ell_k} \\
            &\leq  \frac{\ell_{k+1} \log \ell_{k+1} - \ell_k \log \ell_k}{h\ell_k \log \ell_k} + \frac{2C \ell_{k+1}}{h\ell_k\log\ell_k}.
        \end{aligned}\label{eqn:m1lk-sandwich}
    \end{equation}
    % The second term on the right-hand side is trivially $o(1)$, and we can show the first term is $o(1)$ as well.
    By \prettyref{fact:ell-k-plus-1}, the first term on the right-hand side of \eqref{eqn:m1lk-sandwich} becomes
    \begin{align*}
        \frac{\ell_{k+1} \log \ell_{k+1} - \ell_k \log \ell_k}{h\ell_k \log \ell_k} = \frac{\ell_k \log \ell_k (1 + o(1)) - \ell_k \log \ell_k}{h\ell_k \log \ell_k},
    \end{align*}
    which is $o(1)$.
    The second term is also
    \[\frac{O(1) \ell_{k+1}}{\ell_k \log \ell_k} = \frac{O(1)(1+o(1)) \ell_{k}}{\ell_k \log \ell_k} = o(1).\]
    Therefore, by the squeeze theorem, because
    \[0 \leq \frac{m_1(\ell_{k+1}) - m_1(\ell_k)}{\ell_k\log\ell_k} \leq o(1),\]
    it then holds that, as $k\to\infty$
    \[
    \frac{m_1(\ell_{k+1}) - m_1(\ell_k)}{\ell_k\log\ell_k} \to 0.
    \]
\end{proof}

\subsection{Second Moment Bound}
\prettyref{claim:N-ratio-converges-to-one} also requires a bound on the second moment of $N(\ell)$, $\hat{m}_2(\ell)$ (as defined in \prettyref{app:moment-definitions}).

First, we bound a related quantity, as defined below, and then prove the full bound on $\hat{m}_2(\ell)$ in \prettyref{lem:second-moment-bound-on-N-ell}.

\begin{claim}\label{claim:gamma-terms-bound}
    Suppose $\Ebb_{\Theta\sim\Pi} H(\Theta) > 0$.
    For 
    \[R = \{R_a\}_{a\in\Acal} \sim \text{Multinomial}(\ell - 1, \theta)\]
    and $\theta \sim \Pi$, $\exists$ finite constant $C_0$ that depends only on $\Pi$ s.t.
    \[\Gamma \triangleq \Ebb\left[ \left({\sum}_{a\in\Acal} m_1(R_a) + \ell - m_1(\ell)\right)^2 \right] \leq C_0 \ell^2.\]
    \begin{proof}
    Let $\mathscr{M}$ be shorthand for the quantity we are trying to bound in expectation, i.e., 
    \[\mathscr{M} \triangleq {\sum}_{a\in\Acal} m_1(R_a) + \ell - m_1(\ell).\]
    
        We will show that, for any $R$, $|\mathscr{M}| = O(\ell)$, where the constant underlying the big-$O$ term does not depend on $R$.
        % WLOG, we take $R \leq \ell/2$.
        % If $R \geq \ell/2$, the proof is identical, but with the roles of $R$ and $\ell-1-R$ reversed.
        
        Applying \prettyref{cor:m1ell-bound}, $m_1(k) = \frac{1}{h}k \log k + O(k)$, $\forall k > 0$.
        \begin{align*}
            h\,\left|\mathscr{M}\right| &= O(\ell) + \left|{\sum}_{a\in\Acal} R_a \log R_a - \ell\log  \ell\right|.
        \end{align*}
        Define probability mass function $P$ such that $\Pr{a \in \Acal} = P_a = \frac{R_a}{\ell-1}$.
        \begin{align*}
            \sum_{a\in\Acal} R_a \log R_a &= \sum_a (\ell-1) P_a (\log (\ell-1) + \log P_a) \\
            &= \ell \log (\ell-1) - \log(\ell-1) - (\ell-1) H(P) \\
            &= \ell \log \ell + \ell O(\ell^{-1}) + O(\log \ell) - \ell H(P) \\
            &= \ell \log \ell + O(\log \ell) - \ell H(P).
        \end{align*}
        As $|H(P)| \leq \log |\Acal|$,
        \begin{align*}
            h\,\left|\mathscr{M}\right| &\leq O(\ell) + \ell \log |\Acal| = O(\ell),
        \end{align*}
        where the big-$O$ term is independent of $R$.
        Then, $\Gamma \leq \Ebb_R[(O(\ell) + \ell)^2] \leq C_0\ell^2,$ for some finite constant $C_0$.
    \end{proof}
\end{claim}

\begin{lemma}\label{lem:second-moment-bound-on-N-ell}
    If $\Ebb_{\Theta \sim\Pi}H(\Theta) > 0$, then $\exists C' < \infty$ s.t. $\hat{m}_2(\ell) \leq C'\ell^2$, where $C'$ is a constant depending only on $\Pi$.
    \begin{proof}
        Via the recursion of \prettyref{con:tree-recursion},
        \begin{align*}
            \hat{m}_2(\ell) &= \Ebb\Big[\bigg({\sum}_{a\in\Acal} \underbrace{N_a(R_a) - m_1(R_a)}_{\triangleq T_a} \\
            &\qquad+ \underbrace{\ell - m_1(\ell) + {\sum}_{a\in\Acal} m_1(R_a)}_{\triangleq T'}\bigg)^2\Big],
        \end{align*}
        where $R = \{R_a\}_{a\in\Acal} \sim \mathrm{Multinomial}(\ell-1, \theta)$ for $\theta \sim \Pi$.
        Conditioned on $R$ and $\theta$, $\{T_a\}_{a\in\Acal}$ are independent and zero-mean, and $T'$ is a measurable function of $R$, so
        \begin{align*}
            \hat{m}_2(\ell) &= \Ebb\left[\left(T' + {\sum}_{a\in\Acal} T_a\right)^2\right] \\
            &= \Ebb_{R, \theta} \left[ \Ebb\left[\left(T' + {\sum}_{a\in\Acal} T_a\right)^2\,\bigg|\, R,\theta\right] \right] \\
            &= \Ebb_{R, \theta}\left[T'^2 + {\sum}_{a\in\Acal} T_a^2 \right] \\
            &= \Ebb\left[{\sum}_{a\in\Acal} \hat{m}_2(R_a)\right] + \Gamma,
        \end{align*}
        where $\Gamma$ is as defined in \prettyref{claim:gamma-terms-bound}.
        
        Applying \prettyref{claim:gamma-terms-bound} ($\Gamma$ grows at most quadratically), the setting of \prettyref{lem:inductive-ell-squared-bound} holds (with $\epsilon=0$), and thus $\hat{m}_2(\ell) \leq C' \ell^2$, where $C' < \infty$ depends on $\Pi$.
    \end{proof}
\end{lemma}

\subsection{Main Proof of \prettyref{claim:N-ratio-converges-to-one}}
\textbf{Claim \ref{claim:N-ratio-converges-to-one}.}
    \Nratioclaimcontents
    \begin{proof}
        We first reduce the statement of the claim to one where we can apply the bound on consecutive $m_1(\ell_k)$ from \prettyref{lem:m1-difference-grows-slower-than-ell-k-log-ell-k}, and the bound on $\hat{m}_2(\ell)$ from \prettyref{lem:second-moment-bound-on-N-ell}.
        
        Equivalent to this claim is the statement
        \[\frac{N(\ell_{k+1}) - N(\ell_k)}{N(\ell_k)} \convas 0.\]
        By \prettyref{rem:extreme-values-of-N-ell}, $N(\ell) \geq C \ell \log \ell$, for universal constant $C$, so it suffices to prove
        \[\frac{N(\ell_{k+1}) - N(\ell_k)}{\ell_k \log \ell_k} \convas 0.\]
        By \prettyref{lem:m1-difference-grows-slower-than-ell-k-log-ell-k}, this is equivalent to proving
        \[\frac{N(\ell_{k+1}) - N(\ell_k)}{\ell_k \log \ell_k} - \frac{m_1(\ell_{k+1}) - m_1(\ell_k)}{\ell_k \log \ell_k} \convas 0.\]
        The square of the numerator is bounded by
        \begin{align*}
            &\Ebb \left[\left( (N(\ell_{k+1}) - m_1(\ell_{k+1})) - (N(\ell_{k}) - m_1(\ell_{k})) \right)^2\right] \\
            &\leq \Ebb \left[\left( (N(\ell_{k+1}) - m_1(\ell_{k+1})) - (N(\ell_{k}) - m_1(\ell_{k})) \right)^2\right] \\
            &+ \Ebb \left[\left( (N(\ell_{k+1}) - m_1(\ell_{k+1})) + (N(\ell_{k}) - m_1(\ell_{k})) \right)^2\right] \\
            &=2\hat{m}_2(\ell_{k+1}) + 2\hat{m}_2(\ell_k) \leq 2C'(\ell_{k+1}^2 + \ell_k^2),
        \end{align*}
        where the final inequality is achieved by \prettyref{lem:second-moment-bound-on-N-ell}.

        As a result of \prettyref{fact:ell-k-plus-1}, $\frac{\ell_{k+1}}{\ell_k} = O(1)$, and thus
        \[\Ebb \left[\left( (N(\ell_{k+1}) - m_1(\ell_{k+1})) - (N(\ell_{k}) - m_1(\ell_{k})) \right)^2\right]  \leq C\ell_k^2,\]
        where $C$ is a constant that depends on $\Pi$.

        By a Chebyshev and Borel-Cantelli argument identical to \prettyref{app:NA-to-N-ratio-converges-to-measure-proof}, this bound implies the almost-sure convergence of $\frac{N(\ell_{k+1}) - N(\ell_k)}{\ell_k \log \ell_k} - \frac{m_1(\ell_{k+1}) - m_1(\ell_k)}{\ell_k \log \ell_k}$ to $0$:
        \begin{align*}
            &\Pr{\left|\frac{N(\ell_{k+1}) - N(\ell_k)}{\ell_k \log \ell_k} - \frac{m_1(\ell_{k+1}) - m_1(\ell_k)}{\ell_k \log \ell_k}\right| > \epsilon} \\
            % &= \Pr{\left|N(\ell_{k+1}) - N(\ell_k) - (m_1(\ell_{k+1}) - m_1(\ell_k))\right| > \epsilon \ell_k \log \ell_k} \\
            &\quad\quad\stackrel{(*)}{\leq} \frac{\Ebb\left[\left(N(\ell_{k+1}) - N(\ell_k) - (m_1(\ell_{k+1}) - m_1(\ell_k))\right)^2\right]}{\epsilon^2 \ell_k^2 (\log\ell_k)^2} \\
            &\qquad\qquad\qquad\leq \frac{O(1)}{\epsilon^2 (\log \ell_k)^2},
        \end{align*}
        where $(*)$ follows from Chebyshev's inequality.
        Then, for $\ell_k \cong \exp{\sqrt{k} (\log k)^\alpha}$ with $\frac{1}{2} < \alpha < 1$,
    \begin{align*}
        \sum_{k=1}^\infty &\Pr{\left|\frac{N(\ell_{k+1}) - N(\ell_k)}{\ell_k \log \ell_k} - \frac{m_1(\ell_{k+1}) - m_1(\ell_k)}{\ell_k \log \ell_k}\right| > \epsilon} \\ &\leq \frac{O(1)}{\epsilon^2} \sum_{k=1}^\infty \frac{1}{k(\log k)^{2\alpha}} < \infty,
    \end{align*}
    so an application of the Borel-Cantelli lemma completes the proof.
    \end{proof}

\section{Extension of \prettyref{thm:zero-order-empirical-distribution} to $r$-Tuples}\label{app:r-tuple-proofs}
\subsection{Proof of \prettyref{lem:LnrA-lower-bound}: Prerequisite Bound}

\newcommand{\mTwoRLower}{\underline{\hat{m}}_2^{(r),A}}
\newcommand{\mOneRLower}{\underline{m}_1^{(r),A}}
\newcommand{\GammaRLower}{\underline{\Gamma}^{(r),A}}
\newcommand{\NRALower}{\underline{N}^{(r),A}}
\newcommand{\mOneTilde}{\widetilde{\underline{m}}_1^{(r),A}}

\begin{definition}
    $\NRALower(\ell)$ is defined as the number of terms of $\underline{L}_n^{(r)}(A)$ that are equal to $1$, up to the $\ell$\textsuperscript{th} visit to the root.
\end{definition}
\begin{definition}
    For this section, we define the following ``moment-like quantities'':
    \begin{align*}
        \mTwoRLower(\ell) &\triangleq \Ebb\left[\left( \NRALower(\ell) - \etaStarRA N(\ell) \right)^2\right], \\
        \mOneRLower(\ell) &\triangleq \Ebb\left[\NRALower(\ell)\right].
    \end{align*}
\end{definition}
Also recall the definitions of $N(\ell)$ and $T_n$ from Definitions \ref{def:step-counts} and \ref{def:root-visits}.

The following result comprises the bulk of \prettyref{lem:LnrA-lower-bound}; subsequently, the rest of the proof is the same as \prettyref{thm:zero-order-empirical-distribution}.
\begin{claim}\label{claim:moment-bound-NALower}
    If $d \triangleq \Ebb[\lVert\Theta\rVert_2^2] < 1$, then, for finite constant $C_d$ that depends on $\Pi$,
    \[\mTwoRLower(\ell) \leq C_d \ell^2,\quad \forall \ell > 0.\]

    \begin{proof}
        As in \prettyref{con:tree-recursion}, $R = \{R_a\}_{a\in\Acal}$, or the distribution of symbols drawn from the root node, is multinomial with parameters $\ell - 1$, $\Theta$ for $\Theta \sim \Pi$.
        
        First, define sequence $\underline{p}_\ell$ such that $\mOneRLower(\ell) = \underline{p}_\ell m_1(\ell)$.
        We will derive bounds on $\underline{p}_\ell$, which will subsequently be used to bound $\mTwoRLower$.

        Developing a recursion relation similar to \prettyref{con:tree-recursion},
        \begin{equation}
            \NRALower(\ell) = \GammaRLower(\ell) + {\sum}_{a\in\Acal}\NRALower_a(R_a), \label{eqn:NrAell-recursion}
        \end{equation}
        where $\GammaRLower$ is the contribution to $\NRALower$ from $r$-tuples starting at the root, and the summation is the contribution from each of the $|\Acal|$ possible branches from the root.

        Taking the expectation,
        \begin{align*}
            \mOneRLower(\ell) &= \Ebb\left[\GammaRLower(\ell)\right] + \sum_{a\in\Acal} \Ebb_{R,\theta_0}\left[\mOneRLower(R_a) \right].
        \end{align*}
        Let event $X_*^r \in \{0, 1\}^r$ and $A_1, \dots, A_r$ be as defined in \prettyref{def:single-sequence-event}, the definition of single-sequence events.
        Then, $\GammaRLower(\ell)$ counts the number of paths traversed from the root, that follow the path $X_*^r$ and encounter an $r$-tuple of $\Theta$ values in $A_1 \times \cdots \times A_r$, without returning to the root within the $r$-tuple.

        The first time that each node in the desired path, $X_*^r$, is encountered, we return to the root directly after.
        This results in $r$ returns to the root for paths that trace $X_*^r$.
        All subsequent paths that trace $X_*^r$ will not return to the root in the duration of the $r$-tuple.
        As a result,
        \begin{equation}
            \ell \etaStarRA - r \leq \Ebb[\GammaRLower(\ell)]  \leq \ell \etaStarRA. \label{eqn:expec-gamma-sandwich}
        \end{equation}
        As such, we can bound $\mOneRLower(\ell)$ by
        \begin{equation}
            \mOneTilde(\ell) \leq \mOneRLower(\ell) \leq \etaStarRA m_1(\ell), \label{eqn:m1-rAell-sandwich}
        \end{equation}
        where $\mOneTilde(\ell)$ takes the lower bound from \eqref{eqn:expec-gamma-sandwich} in each step, producing the following recursion:
        \begin{align*}
            \mOneTilde(\ell) =& \ell \etaStarRA - r + \sum_{a\in\Acal} \Ebb\left[ \mOneTilde(R_a) \right],
        \end{align*}
        with base case $\mOneTilde(0) = 0$.
        To evaluate $\mOneTilde$, define
        \begin{align*}
            \underline{\epsilon}(\ell) &\triangleq \etaStarRA m_1(\ell) - \mOneTilde \\
            &= r + \sum_{a\in\Acal} \Ebb\left[\underline{\epsilon}(R_a)\right],\quad \underline{\epsilon}(0) = 0.
        \end{align*}
        This recursion admits a unique solution, which can be found via inspection to be $\underline{\epsilon}(\ell) = r\ell$.
        So,
        \[\etaStarRA m_1(\ell) - r\ell \leq \mOneRLower(\ell) \leq \etaStarRA m_1(\ell).\]
        By \prettyref{cor:m1ell-bound}, $\exists$ positive constant $C$ (only depending on $\Pi$) such that $m_1(\ell) \geq C\ell \log \ell$.
        Using this fact,
        \begin{equation}
            \etaStarRA - \tfrac{r}{C\log \ell} \leq \underline{p}_\ell \leq \etaStarRA  . \label{eqn:pell-sandwich}
        \end{equation}
        Now bounding $\underline{\hat{m}_2}^{(r),A}(\ell)$ by a process analogous to \prettyref{app:second-moment-bound-on-NA-proof}, we first write
        \begin{align*}
            \NRALower(\ell) - \etaStarRA N(\ell)& \\
            &\hspace{-3em}= \GammaRLower(\ell) - \etaStarRA \ell \\
            &\hspace{-2em}+ \sum_{a\in\Acal} \left( \NRALower_a(R_a) - \etaStarRA N_a(R_a) \right).
        \end{align*}
        As shorthand, let 
        \[\underline{m}_\text{root}(\ell) \triangleq \left( \NRALower(\ell) - \etaStarRA N(\ell) \right).\]
        Taking the expectation and rearranging terms,
        \begin{align*}
            \mTwoRLower(\ell) &= \Ebb\left[\left( \NRALower(\ell) - \etaStarRA N(\ell) \right)^2\right] \\
            &= \sum_{a\in\Acal} \Ebb\left[ \mTwoRLower(R_a)\right] \\
            &\qquad+2\Ebb\bigg[ \left( \GammaRLower(\ell) - \etaStarRA\ell \right) \underline{m}_\text{root}(\ell)\bigg] \\
            &\qquad- \Ebb\left[ \left( \GammaRLower(\ell) - \etaStarRA\ell \right)^2 \right] \\
            &\qquad+ \sum_{a\in\Acal} \sum_{b\in\Acal\backslash a} \Ebb\bigg[\underline{m}_\text{root}(R_a) \underline{m}_\text{root}(R_b) \bigg] \\
            &\leq \sum_{a\in\Acal} \Ebb\left[ \mTwoRLower(R_a)\right] + \Ebb[V_\ell],
        \end{align*}
        where $V_\ell$ is defined such that
        \begin{align*}
            \Ebb[V_\ell] &= 2\Ebb\left[ \left( \GammaRLower(\ell) - \etaStarRA\ell \right)\underline{m}_\text{root}(\ell)\right] \\
            &\qquad+\sum_{a\in\Acal} \sum_{b\in\Acal\backslash a}  \Ebb\bigg[\underline{m}_\text{root}(R_a) \underline{m}_\text{root}(R_b)\bigg].
        \end{align*}
        To bound the double summation expression, we recognize that, conditioned on $(R, \theta_0)$, the tuples $(\NRALower_a, N_a)$ for $a\in\Acal$ are independent of each other.
        This is because they involve disjoint branches of the LZ78 tree, and, if a return to the root occurs within an $r$-tuple, it is not counted in any $\NRALower_a$, so no dependence arises from returns to the root.

        Then, by the definition of $\underline{p}_\ell$, for any $a, b \in \Acal \times \Acal$ where $a \neq b$,
        \begin{align*}
            &\Ebb\bigg[\underline{m}_\text{root}(R_a) \underline{m}_\text{root}(R_b)\bigg] \\
            &\qquad=\Ebb\bigg[ \left( \NRALower_a(R_a) - \etaStarRA N_a(R_a) \right) \\
            &\qquad\qquad\qquad \cdot \left( \NRALower_b(R_b) - \etaStarRA N_b(R_b) \right)\bigg] \\
            &\qquad=\Ebb_{R,\theta_0}\bigg[\left(\underline{p}_{R_a} - \etaStarRA\right)m_1(R_a)\\
            &\qquad\qquad\qquad\qquad \cdot \left(\underline{p}_{R_b} - \etaStarRA\right)m_1(R_b)\bigg],
        \end{align*}
        which we define as $\Ebb_{R, \Theta}[\alpha_\ell]$.
        From \eqref{eqn:pell-sandwich}, there is some constant $\beta$ such that $|\underline{p}_k - \etaStarRA| \leq \frac{\beta}{\log k}$, $\forall k > 0$.
        By \prettyref{cor:m1ell-bound}, $m_1(k) \leq \frac{1}{h}k\log k + O(k)$, so $\exists$ constant $C' < \infty$ such that 
        \[|\underline{p}_k - \etaStarRA| m_1(k) \leq C'k,\quad \forall k > 0.\]
        Thus, $\alpha_\ell$ can be bounded by
        \[\alpha_\ell \leq C'^2R_aR_b \leq C'^2\ell^2,\]
        and the double summation can be bounded by 
        \[|\Acal| (|\Acal| - 1) C'^2\ell^2 \triangleq C'' \ell^2.\]
        
        We next bound the first term of $\Ebb[V_\ell]$.
        As $2xy \leq \frac{1}{\delta}x^2 + \delta y^2$, for any $\delta > 0$,\footnote{This is achieved by rearranging terms of the inequality $0 \leq \left( \frac{1}{\sqrt{\delta} x} - \sqrt{\delta} y\right)^2 = \frac{1}{\delta}x^2 + \delta y^2 - 2xy$.}
        \begin{align*}
            &2\Ebb\left[ \left( \GammaRLower(\ell) - \etaStarRA\ell \right)\left( \NRALower(\ell) - \etaStarRA N(\ell) \right)\right] \\
            &\qquad\leq \frac{1}{\delta} \Ebb\left[ \left(\GammaRLower(\ell) - \etaStarRA\ell\right)^2 \right] \\
            &\qquad \qquad + \delta \Ebb\left[ \left( \NRALower(\ell) - \etaStarRA N(\ell) \right)^2 \right] \\
            &\qquad\leq \tfrac{1}{\delta}\ell^2 + \delta \mTwoRLower(\ell),
        \end{align*}
        applying $\GammaRLower \leq \ell$ on the first term and the definition of $\mTwoRLower$ on the second.
        
        Thus, for all $\ell > 0$ and $\delta > 0$,
        \[\Ebb[V_\ell] \leq \left(C'' + \tfrac{1}{\delta}\right)\ell^2 + \delta \mTwoRLower(\ell).\]
        Plugging this bound into the bound on $\mTwoRLower(\ell)$ defined above and solving for $\mTwoRLower$ (with $0 < \delta < 1$),
        \begin{align*}
            \mTwoRLower(\ell) &\leq \tfrac{1}{1-\delta} \left(C'' + \tfrac{1}{\delta}\right) \ell^2 + \tfrac{1}{1-\delta} \sum_{a\in\Acal} \Ebb\left[\mTwoRLower(R_a)\right],\\
            \mTwoRLower(0) &= 0.
        \end{align*}
        Choose any $\delta \in (0, 1-d)$, where $d = \Ebb[\lVert\Theta\rVert_2^2 ]$ (we know that, as $\Ebb_{\Theta\sim\Pi} H(\Theta) > 0$, $d < 1$).
        Then, $\frac{d}{1-\delta} < 1$, so we can directly apply \prettyref{lem:inductive-ell-squared-bound} with $\epsilon = \frac{1}{1-\delta} - 1$ to get
        \[\mTwoRLower(\ell) \leq C_d\ell^2,\]
        where $C_d$ is a constant that depends on $C''$ and $d$ (and the choice of $\delta$).
    \end{proof}
\end{claim}

\subsection{Proof of \prettyref{lem:LnrA-lower-bound}}\label{app:LnrA-lower-bound-proof}
\textbf{Lemma \ref{lem:LnrA-lower-bound}.}
Assume $\Ebb_{\Theta\sim\Pi} H(\Theta) > 0$.
Then, for any $A \in \Scal_r$, $\underline{L}_n^{(r)}(A) \convas \etaStarRA$ as $n\to\infty$.
\begin{proof}
    Given \prettyref{claim:moment-bound-NALower}, by the exact same process as in \prettyref{app:NA-to-N-ratio-converges-to-measure-proof}, we can apply Chebyshev's inequality and the Borel-Cantelli lemma to conclude that
    \footnote{As this component and the subsequent monotonicity argument are identical to those in the proof of \prettyref{thm:zero-order-empirical-distribution}, they are not re-stated here.}
    \[\lim_{k\to\infty} \frac{\NRALower(\ell_k)}{N(\ell_k)} = \etaStarRA\quad\text{(a.s.)}.\]
    Then, using the same monotonicity arguments as in \prettyref{app:end-of-zero-order-proof},
    \[\tfrac{\NRALower(T_n)}{N(T_n + 1)} \leq \underline{L}^{(r)}(A) \leq \tfrac{\NRALower(T_n +1)}{N(T_n)},\]
    implying that
    \[\underline{L}^{(r)}(A) \convas \etaStarRA.\]
\end{proof}

\subsection{Proof of \prettyref{lem:LnrA-upper-bound}}\label{app:LnrA-upper-bound-proof}
\textbf{Lemma \ref{lem:LnrA-upper-bound}.}
Assume $\Ebb_{\Theta\sim\Pi} H(\Theta) > 0$.
Then, for any $A \in \Scal_r$, $\overline{L}_n^{(r)}(A) \convas \etaStarRA$ as $n\to\infty$.

\begin{proof}
    The proof is almost identical to that of \prettyref{lem:LnrA-lower-bound}, so only the parts that differ are recorded here.
    In terms of notation, all variables that are underlined in the proof \prettyref{lem:LnrA-lower-bound} will be replaced by versions with an overline.
    
    Let $\overline{N}^{(r),A}(\ell)$, the number of terms in $\overline{L}_n^{(r)}(A)$ that are $1$ up to the $\ell$\textsuperscript{th} visit to the root, replace $\underline{N}^{(r),A}(\ell)$.
    As returns to the root are counted as automatic ``successes'' rather than ``failures,'' \eqref{eqn:expec-gamma-sandwich} becomes
    \[\ell \etaStarRA \leq \Ebb\left[ \overline{\Gamma}^{(R),A}(\ell)\right] \leq \ell \etaStarRA + r,\]
    where $ \overline{\Gamma}^{(R),A}$ is defined by counting $r$-tuples that start at the root and are $\in A$ \textit{or} include a return to the root.
    Subsequently,
    \[\etaStarRA m_1(\ell) \leq \overline{m}_1^{(r),A}(\ell) \leq {\widetilde{\overline{m}}_1}^{(r),A}(\ell),\]
    where $\widetilde{\overline{m}}_1^{(r),A}(\ell)$ is now defined by taking the upper bound in each step:
    \[
        \begin{gathered}
        \widetilde{\overline{m}}_1^{(r),A}(\ell)
        = \ell \etaStarRA + r
        + \sum_{a\in\Acal} \Ebb\!\left[\widetilde{\overline{m}}_1^{(r),A}(R_a)\right], \\
        \widetilde{\overline{m}}_1^{(r),A}(0) = 0 .
        \end{gathered}
    \]
    The solution to this recursion is $\widetilde{\overline{m}}_1^{(r),A} = \etaStarRA m_1(\ell) + r\ell$, so
    \[\etaStarRA m_1(\ell) \leq \overline{m}_1^{(r),A}(\ell) \leq \etaStarRA m_1(\ell) + r\ell.\]
    Then, \eqref{eqn:pell-sandwich} becomes
    \[\etaStarRA \leq \overline{p}_\ell \leq \etaStarRA + \tfrac{r}{C\log \ell}.\]
    The remainder of the proof requires no changes (other than replacing the appropriate notation).
\end{proof}

\section{Entropy Rate and Shannon-Mcmillan-Breiman Result: Proof of \prettyref{thm:shannon-mcmillan-breiman}}\label{app:smb-proof}
By \prettyref{rem:lz78-codelength-convergence}, the result of \prettyref{thm:shannon-mcmillan-breiman} reduces to showing that $\frac{T_n \log T_n}{n} \convas \Ebb[H(\Theta)]$.
Equivalently, we aim to prove $\frac{\ell \log \ell}{N(\ell)} \convas \Ebb[H(\Theta)]$.
From \prettyref{cor:m1ell-bound}, $\frac{\ell \log \ell}{m_1(\ell)} \to \Ebb[H(\Theta)]$, so we only need the following result:
\begin{lemma}\label{lem:N-ell-expectation-ratio-conv-one}
    Assume we have an LZ78 probability source with $\Ebb_{\Theta\sim\Pi} H(\Theta) > 0$.
    For $N(\ell)$ from \prettyref{def:step-counts},
    \[\frac{N(\ell)}{m_1(\ell)} \triangleq \frac{N(\ell)}{\Ebb [N(\ell)]} \convas 1,\quad \text{as } \ell\to\infty.\]
    \begin{proof}
        By \prettyref{lem:second-moment-bound-on-N-ell}, $\exists$ finite constant $C$ that depends only on $\Pi$ such that
        \[\Ebb[(N(\ell) - m_1(\ell))^2] \leq C\ell^2,\quad \forall \ell \geq 1.\]
        For any $\epsilon > 0$, the probability that $\frac{N(\ell)}{m_1(\ell)}$ is more than $\epsilon$ away from $1$ has upper bound
        \begin{align*}
            \Pr{\left| \frac{N(\ell)}{m_1(\ell)} - 1 \right| > \epsilon} &= \Pr{\left|N(\ell) - m_1(\ell)\right| > \epsilon m_1(\ell)} \\
            &\stackrel{(a)}{\leq} \frac{\Ebb[(N(\ell) - m_1(\ell))^2]}{\epsilon^2 m_1(\ell)^2} \\
            &\stackrel{(b)}{\leq} \frac{C\ell^2}{\epsilon^2 m_1(\ell)^2} \\
            &\stackrel{(c)}{\leq} \frac{C'\ell^2}{\epsilon^2 \ell^2 (\log(\ell))^2} = \frac{C'}{\epsilon^2(\log(\ell))^2},
        \end{align*}
        where $C'$ only depends on $\Pi$.
        $(a)$ is by Chebyshev's inequality, $(b)$ follows from \prettyref{lem:second-moment-bound-on-N-ell}, and $(c)$ follows from \prettyref{cor:m1ell-bound}.
        Using an identical Borel-Cantelli lemma argument to the one presented in \prettyref{app:NA-to-N-ratio-converges-to-measure-proof} (which we will not repeat here), we can conclude that $\frac{N(\ell_k)}{m_1(\ell_k)} \convas 1$, where $\ell_k$ is the sequence defined in \prettyref{claim:NA-to-N-ratio-converges-to-measure}.

        To extend this result beyond convergence along the sequence $\ell_k$, we leverage the monotonicity of $N(\ell)$ and $m_1(\ell)$.
        For any $\ell$, there is some $\ell_k$ such that $\ell_k \leq \ell \leq \ell_{k+1}$.
        Then, as both $N(\ell)$ and $m_1(\ell)$ are monotonically increasing,
        \begin{IEEEeqnarray}{c}
            \tfrac{N(\ell_k)}{m_1(\ell_{k+1})} \leq \tfrac{N(\ell)}{m_1(\ell)} \leq \tfrac{N(\ell_{k+1})}{m_1(\ell_k)} \iff \label{eqn:N-by-m1-sandwich}\\
            \tfrac{m_1(\ell_k)}{m_1(\ell_{k+1})} \cdot \tfrac{N(\ell_k)}{m_1(\ell_k)} \leq \tfrac{N(\ell)}{m_1(\ell)} \leq \tfrac{m_1(\ell_{k+1})}{m_1(\ell_k)} \cdot \tfrac{N(\ell_{k+1})}{m_1(\ell_{k+1})},\nonumber
        \end{IEEEeqnarray}
        with $\ell_k$ as defined above.

        We just proved that $\frac{N(\ell_k)}{m_1(\ell_k)} \convas 1$.
        Also, by \prettyref{lem:m1-difference-grows-slower-than-ell-k-log-ell-k}, $m_1(\ell_{k+1}) = m_1(\ell_k) + o(\ell_k \log \ell_k)$.
        Applying the fact from \prettyref{cor:m1ell-bound} that $m_1(\ell) = \frac{1}{h}\ell \log \ell + O(\ell)$, where $h = \Ebb[H(\Theta)]$,
        \begin{align*}
            \frac{m_1(\ell_{k+1})}{m_1(\ell_k)} &= 1 + \frac{o(\ell_k \log \ell_k)}{\frac{1}{h}\ell_k \log \ell_k + O(\ell_k)} \\
            &= 1 + \frac{o(\ell_k \log \ell_k)}{\Omega(\ell_k \log \ell_k)} = 1 + o(1),
        \end{align*}
        which approaches $1$ as $k\to\infty$.
        Then, both the lower and upper bound on $\frac{N(\ell)}{m_1(\ell)}$ from \eqref{eqn:N-by-m1-sandwich} converge a.s. to $1$, so $\frac{N(\ell)}{m_1(\ell)} \convas 1$ by the squeeze theorem.
    \end{proof}
\end{lemma}

\textbf{Theorem \ref{thm:shannon-mcmillan-breiman}} (\smbtheoremtitle)\textbf{.}
    \smbtheoremcontents

\begin{proof}
    Taking the $\log$ of \prettyref{rem:probability-of-lz78-sequence},
    \[\frac{1}{n}\log \frac{1}{P_{X^n}(X^n)} = \frac{1}{n}\sum_{z \in \Zcal(X^n)} \log \frac{1}{q^\Pi(Y_z^{m_z})},\]
    where $Y_z^{m_z}$ is the subsequence of $X^n$ generated from node $z$ of the LZ78 tree, and $q^\Pi$ is an i.i.d. Bayesian mixture distribution under prior $\Pi$.
    By \prettyref{rem:lz78-codelength-convergence},
    \[\lim_{n\to\infty} \max_{X^n \in \{0,1\}^n} \left| \frac{1}{n}\sum_{z \in \Zcal(X^n)} \log \frac{1}{q(Y_z^{m_z})} - \ell_\text{LZ}(X^n) \right| = 0,\]
    where $\ell_\text{LZ}(X^n) \triangleq \frac{T_n(X^n) \log T_n(X^n)}{n}$.
    It then suffices to show that $\frac{T_n \log T_n}{n} \convas h \triangleq \Ebb[H(\Theta)]$ as $n\to\infty$.
    By \prettyref{cor:m1ell-bound},
    \[m_1(\ell) = \tfrac{1}{h}\ell \log \ell + O(\ell).\]
    Additionally, by \prettyref{lem:N-ell-expectation-ratio-conv-one}, $\frac{N(\ell)}{m_1(\ell)} \convas 1$ as $\ell\to\infty$.
    Combining the two results,
    \[\frac{\ell \log \ell}{h N(\ell)} + \frac{O(\ell)}{N(\ell)} \convas 1 \implies \frac{\ell \log \ell}{N(\ell)} \convas h,\]
    as $\frac{O(\ell)}{N(\ell)} = \frac{O(\ell)}{m_1(\ell)} \cdot \frac{m_1(\ell)}{N(\ell)} \convas 0 \cdot 1 = 0.$
    Then, as $T_n$ is an increasing sequence that goes $\to\infty$ with $n$,
    \[\frac{T_n \log T_n}{N(T_n)} \convas \Ebb[H(\Theta)]\quad\text{as }n\to\infty.\]
\end{proof}

\section{Entropy Rate Without Support Conditions}
    \subsection{Convergence to Sums of Empirical Entropies}

    The goal of this subsection is to prove~\lemref{lem:log_probs_to_entropy_in_prob}, which is a modified version of Lemma D.6 in~\cite{sagan2024familylz78baseduniversalsequential}. Without the assumption that $\supp(\Pi) = \Mcal(\Acal)$, we do not have the same uniform convergence, and so instead we show convergence in expectation. The proof of~\lemref{lem:log_probs_to_entropy_in_prob} breaks the summands into terms which are bounded via~\lemref{lem:suff_stat_theta} and~\lemref{lem:plugin-stopping-time-bd}. Finally,~\lemref{lem:sum_nodes_bound} yields the convergence to $0$ in expectation.

    We let $\theta_{z,i} \triangleq \Pr{\Theta_z = i}$ be the probability the random variable $\Theta_z$ over $\Acal$ selected at node $z$ takes a value of $i$. We also define $n_i \triangleq \sum_{k=1}^{m-1} \mathds{1}[y_{z,k}=i]$.
    
    \begin{lemma}\label{lem:suff_stat_theta}
        We have that $\left(m_z, (n_i)_{i \in \Acal}, y_{z,m_z}\right)$ is a sufficient statistic of $Y_z^{m_z}$ for inferring $\Theta_z$.
    \end{lemma}
    \begin{remark}
        The probability distribution of what the next phrase will be when we are at the root only depends on the current structure of the tree and not the order in which the nodes in the tree were initialized. Therefore, we do not need the entire history $Y_z^{m_z}$ at a node when predicting $\Theta_z$, it suffices to consider what the recent history has looked like and the structure of the tree that was formed up until the last time the node was visited.
    \end{remark}
    \begin{proof}
        Define for brevity of notation 
        \[P_{m,y_z^m} \triangleq \Pr{m_z=m|Y_z^{m-1} = y_z^{m-1},Y_{z,m}=y_{z,m}}. \]
        Then we have that
        \begin{align*}
            p_{\theta_z}\left(Y_z^{m_z}=y_z^m \right) &= p_{\theta_z}\left(Y_z^{m}=y_z^m\right) P_{m,y_z^m} \\
            &= \prod_{i \in \Acal} \theta_{z,i}^{n_i + \mathds{1}[y_{z,m}=i]} P_{m,y_z^m}.
        \end{align*}
        By the Fisher-Neyman theorem, it suffices to show that $P_{m,y_z^m}$ only depends on $(m, (n_i)_{i \in \Acal}, y_{z,m})$. In other words, it suffices to show that reordering $y_z^{m-1}$ does not change this conditional probability. We will show that the number of sequences $x^n$ consistent with any such reordering is the same by constructing a bijection via rearranging phrases, and then conclude by noting that all sequences resulting in the same final tree and ending at the same final node have the same likelihood.
        % To do this, we will construct a bijection between sequences $x^n$ reaching node $z$ $m$ times and with the same sum at node $z$ over the first $m-1$ times reached, but with different orderings of $y_z^{m-1}$. 
    
        For ease of notation, let $T_z(k)$ be the index in $x^n$ corresponding to the $k$th visit to node $z$, and define $y_{z,k}$ to be the bit outputted at that index in $x^n$.
        Let $T_{i,z}(k)$ be the index of the $k$th visit to node $z$ having an $i$.
    
        For any string $x^n$ satisfying the properties that the number of visits to node $z$ is $m$, we partition $x^n$ into the following blocks of characters:
        \begin{itemize}
            \item Substring $a_k$ is defined as the substring beginning at the start and ending at index $T_z(1)-1$ when $k=1$, and otherwise the first time the root is reached after index $T_z(k-1)$ when $k > 2$. and  and ending at index $T_z(k)-1$ (inclusive).
            \item Substring $b_{j,i}$ with $i \in \Acal$ is defined as the substring beginning at index $T_{i,z}(j)$ and ending at the index before we next visit the root node.
            \item Substring $c$ is defined to be the substring beginning at index $T_z(m)$ and continuing to the end of $x^n$.
        \end{itemize}
        Each $x^n$ is therefore a concatenation of blocks of type $a_i$ with blocks of type $b_{j,i}$ separating adjacent blocks of type $a_i$, terminated by a block of type $c$. We now consider operations that swap two blocks in this decomposition
        and then relabel the blocks $b_{j,i}$, for each fixed $i\in\mathcal A$, according
        to their new left-to-right order. To go from one $y_z^{m_z}$ to a reordering, we can always choose a set of such operations which forms a bijection of sequences and by construction preserves the values $(n_i)_{i \in \Acal}$.
        
        Explicitly, we now have
        % \begin{align*} 
        % &\Pr{m_z=m, Y_{z}^{m-1} = y_z^{m-1} \mid Y_{z,m}=y_{z,m}} \\
        % &= \frac{1}{\binom{m-1}{(n_i)_{i \in \Acal}}} \Pr{m_z=m, \forall i \in \Acal,\, \sum_{k=1}^{m-1} \mathds{1}[Y_{z,k}=i] = n_i \mid Y_{z,m}=y_{z,m}}
        % \end{align*}
        \begin{multline*}
        \Pr\{m_z=m, Y_{z}^{m-1} = y_z^{m-1} \mid Y_{z,m}=y_{z,m}\} \\
        = \frac{1}{\binom{m-1}{(n_i)_{i \in \mathcal{A}}}} \Pr\biggl\{m_z=m, \forall i \in \mathcal{A}, \\
        \sum_{k=1}^{m-1} \mathds{1}[Y_{z,k}=i] = n_i \mid Y_{z,m}=y_{z,m}\biggr\}
        \end{multline*}
        and
        {\small
        \begin{align*} &\Pr{Y_{z}^{m-1} = y_z^{m-1} \mid Y_{z,m}=y_{z,m}} \\
        &= \frac{1}{\binom{m-1}{(n_i)_{i \in \mathcal{A}}}} \Pr{\forall i \in \Acal, \sum_{k=1}^{m-1} \mathds{1}[Y_{z,k} = i] = n_i \mid Y_{z,m}=y_{z,m}},
        \end{align*}}
        which then implies
        {\small
        \begin{align*} 
            &P_{m,y_z^m} \\
            &= \frac{\Pr{m_z=m, Y_z^{m-1}= y_z^{m-1} \mid Y_{z,m}=y_{z,m}}}{\Pr{Y_z^{m-1}=y_z^{m-1} \mid Y_{z,m}=y_{z,m}}} \\
            &= \frac{\Pr{m_z=m, \forall i \in \Acal, \sum_{k=1}^{m-1} \mathds{1}[Y_{z,k}=i]= n_i \mid Y_{z,m}=y_{z,m}}}{\Pr{\forall i \in \Acal, \sum_{k=1}^{m-1} \mathds{1}[Y_{z,k}=i]= n_i \mid Y_{z,m}=y_{z,m}}} \\
            &= \Pr{m_z=m \, \Bigg|\,\, \forall i \in \Acal, \sum_{k=1}^{m-1} \mathds{1}[Y_{z,k}=i]= n_i, Y_{z,m}=y_{z,m}}
        \end{align*}}
        which only depends on $m, (n_i)_{i \in \Acal}, y_{z,m}$.
    \end{proof}
    \begin{lemma}\label{lem:plugin-stopping-time-bd}
    Let $T$ be a bounded stopping time satisfying $1 \le T \le n$ with probability $1$. Let $\theta_z$ be a random variable over $\Acal$ with distribution $P_{\theta_z}$. We have that for $y_{z} \overset{iid}{\sim} P_{\theta_z}$,
        \[\Ebb_{T,y_z^T}[T \left(H(\theta_z) - H_0(y_z^{T})
        \right)] \le 2\abs{\Acal}\Ebb_T[\log (T+1)].\]
    \end{lemma}
    \begin{proof}
        Denote $\theta_i \triangleq \Pr{\theta_z=i}$, for $i \in \Acal$.
        As shown in~\cite{10.1162/089976603321780272}, if $n_{T, i} \triangleq \abs{\set{k: y_{z,k} = i}}$ for $i \in \Acal$, and if we define $P_{\theta_0}$ to be the empirical distribution, then we have that 
        \begin{align*}
            &H(\theta_z) - H_0(y_z^{T}) \\ 
            &\qquad= \sum_{i \in \Acal} \log \theta_i \left(-\frac{n_{T, i}}{T} + \theta_i\right) + D\left(P_{\theta_0} \mid\mid P_{\theta_z} \right).
        \end{align*}
        Using boundedness of $T$ we may compute via the optional stopping theorem that
        \begin{align*}
            &\Ebb[T(H(\theta_z) - H_0(y_z^{T}))] \\
            &\qquad= \Ebb\left[TD\left(P_{\theta_0} \mid\mid P_{\theta_z} \right) + \sum_{i \in \Acal} \log \theta_i \left(-\frac{n_{T, i}}{T} + \theta_i\right) \right] \\
            &\qquad= \Ebb\left[TD\left(P_{\theta_0} \mid\mid P_{\theta_z} \right)\right] \\
            &\qquad\le \Ebb\left[T\chi^2\left(P_{\theta_0} \mid\mid P_{\theta_z} \right)\right]\\
            &\qquad= \Ebb\left[T\sum_{i \in \Acal} (n_{T,i}/T)^2/\theta_i - T
            \right].
        \end{align*}
        % Letting $S_T := n_{T,1}$ for ease of notation, we then can write that
        % \begin{align*}
        %     &\Ebb[T\chi^2(n_{T,1}/T\mid\mid\theta)] \\
        %     &\qquad= \Ebb\left[T(S_T/T)^2/\theta + T((T-S_T)/T)^2/(1-\theta) - T\right].
        % \end{align*}
        Let $M_{T,i} := n_{T,i} - \theta_i T$ (so $M_{T,i}$ is a martingale stopped at time $T$). We can then write
        \begin{align*}
            \Ebb\left[T(n_{T,i}/T)^2/\theta\right] &= \Ebb\left[T((M_{T,i} + \theta_i T)/T)^2/\theta_i\right] \\
            &= \frac{1}{\theta_i}\left(\Ebb[M_{T,i}^2/T] +2\theta_i \Ebb[M_{T,i}] + \theta_i^2 \Ebb[T]\right) \\
            &= \frac{1}{\theta_i} \Ebb[M_{T,i}^2/T]+\theta_i \Ebb[T],
        \end{align*}
        where again by boundedness of $T$ we can apply the optional stopping theorem.  
        Combining tells us that 
        \[ \Ebb\left[T\chi^2\left(P_{\theta_0} \mid\mid P_{\theta_z} \right)\right] = \sum_{i \in \Acal} \frac{1}{\theta_i} \EE[M_{T,i}^2/T].\]
    
        It therefore suffices to prove the following claim.
    
        \begin{claim}
        We have that for any $i \in \Acal$,
            \[ \frac{1}{\theta_i}\Ebb[M_{T,i}^2/T] \le 2\Ebb[\log (T+1)]\]
            for any positive stopping time $T$ satisfying $1 \le T \le n$ almost surely.
        \end{claim}
        For brevity of notation, because we are proving this for all $i \in \Acal$, we drop the subscript $i$ throughout the proof of this claim.
    
        We induct on $n$. In the case when $n = 1 $, the claim is easy to verify. We now assume that for all stopping times $1 \le T \le k$ the claim is true. Consider any stopping time $T$ satisfying $1 \le T \le k+1$. We can construct a stopping time $T' \triangleq T \wedge k$, and so $\Pr{T' \ge k+1} = 0$.
    
        % By our hypothesis, we know that 
        % \[ \frac{1}{\theta} \Ebb[M_{T'}^2/T'] \le 2\Ebb[\log(T'+1)].\]
        % We also have by definition of $M_T$,
        % \[ M_{T} = M_{T'} + \mathds{1}[T=k+1](\mathds{1}[Y_{z,{k+1}} = i] - \theta).\]
        Letting $S(y_z^{t}) \triangleq \abs{\set{k: 1 \le k \le t, y_{z,k} = i}} - t\theta,$
        we may write that:
        \begin{align*}
            &\Ebb[M_{T}^2/T] - \Ebb[M_{T'}^2/T'] \\
            &\qquad= \sum_{y_z^{k+1}}  \Pr{Y_z^{k+1}=y_z^{k+1}, T=k+1} \\
            &\qquad\qquad\cdot \left(\frac{S\left(y_z^{k+1}\right)^2}{k+1} - \frac{S\left(y_z^{k}\right)^2}{k}\right) \\
            &\qquad= \sum_{y_z^{k}}  \Pr{Y_z^{k}=y_z^{k}, T=k+1} \\
            &\qquad\qquad\cdot \Biggl(\theta\frac{\left(S\left(y_z^{k}\right) + 1 - \theta\right)^2}{k+1} \\
            &\qquad\qquad+ (1-\theta)\frac{\left(S\left(y_z^{k}\right) - \theta\right)^2}{k+1} - \frac{S\left(y_z^{k}\right)^2}{k}\Biggr) \\
            &\qquad= \sum_{y_z^{k}}  \Pr{Y_z^{k}=y_z^{k}, T=k+1} \\
            &\qquad\qquad\cdot \left(\frac{S\left(y_z^{k}\right)^2 + \theta(1-\theta)}{k+1} - \frac{S\left(y_z^{k}\right)^2}{k}\right) \\
            &\qquad\le \sum_{y_z^k} \Pr{T=k+1, Y_z^{k}=y_z^{k}} \left(\frac{\theta(1-\theta)}{k+1}\right) \\
            &\qquad\le \Pr{T=k+1} \left(\frac{\theta(1-\theta)}{k+1}\right).
        \end{align*}
        Because we have
        \begin{align*} 
            &\Ebb[\log(T+1)] - \Ebb[\log(T'+1)] \\
            &\qquad= \Pr{T=k+1}(\log (k+2)-\log(k+1)) \\
            &\qquad\ge \frac{\Pr{T=k+1}}{2(k+1)},
        \end{align*}
        we can compute
        \begin{align*} 
            \frac{1}{\theta}\Ebb[M_{T}^2/T] &\le \frac{1}{\theta}\Ebb[M_{T}^2/T'] + \Pr{T=k+1}\frac{1-\theta}{k+1} \\
            &\le 2\Ebb[\log(T'+1)] + \Pr{T=k+1}\frac{1-\theta}{k+1} \\
            &\le 2\Ebb[\log(T+1)].
        \end{align*}
        By induction, the claim follows.
    \end{proof}
    
    \begin{lemma} \label{lem:sum_nodes_bound}
        Let $\mathcal{Z}_n$ be the (infinite) set of all nodes. When $\Ebb[H(\Theta)] > 0$,
            \[ \sum_{z \in \Zcal_n} \Ebb[\log(m_z+1)] = o(n). \]
    \end{lemma}
    \begin{proof}
        We have that
        \[ \sum_{z \in \mathcal{Z}_n} \Ebb[\log(m_z+1)] = \Ebb_{x^n}\left[\abs{Z(x^n)} \sum_{z \in Z(x^n)} \frac{\log (m_z+1)}{\abs{Z(x^n)}} \right]. \]
        We can interpret the inner sum as
        \begin{align*} \sum_{z \in Z(x^n)} \frac{\log (m_z+1)}{\abs{Z(x^n)}} &= \Ebb_{u \sim \text{Unif}(Z(x^n))} [\log(m_u+1)] \\
        &\le \log\left(\Ebb_{u \sim \text{Unif}(Z(x^n))} [m_u] + 1]\right) \\
        &= \log \left(\Ebb[n/\abs{Z(x^n)} + 1]\right).
        \end{align*}
        Then \[\abs{Z(x^n)}\log \Ebb\left[n/\abs{\set{Z(x^n)}} + 1\right] \lesssim \abs{Z(x^n)} \log\log n,\]
        noting that the number of nodes in the LZ tree is on the order of $\frac{n}{\log n}$ when $\Ebb[H(\Theta)] > 0$ (e.g. by Corollary~\ref{cor:m1ell-bound}). Taking the outer expectation over $x^n$ we have that
        \[ \sum_{z \in \mathcal{Z}_n} \Ebb[\log(m_z+1)] \lesssim \frac{n\log\log n}{\log n}. \]
    \end{proof}
    \begin{lemma}\label{lem:log_probs_to_entropy_in_prob}
        Let $q^\Pi(x^n)$ be a Bayesian mixture distribution (\prettyref{def:bayesian-mixture-dist}) and such that $\Ebb[H(\Theta)] > 0$. Then we have that
        \begin{align*} 
        &\lim_{n \to \infty} \Ebb\left[ \abs{\frac{1}{n} \sum_{z \in \mathcal{Z}(x^n)} \log \frac{1}{q\left(y_z^{m_z}\right)} - \frac{1}{n} \sum_{z \in \mathcal{Z}(x^n)} m_z H_0\left(y_z^{m_z}\right)}\right]  \\&\qquad\qquad= 0.
        \end{align*}
    \end{lemma}
    \noindent\textit{Proof Sketch:}
        Let $n_i := \abs{\set{1 \le k \le n: y_{z,k} = i}}$ as before. We first show that the quantity inside the absolute value is always nonnegative, and then decompose the quantity inside the sum as follows:
        \begin{align*}
            &\Ebb\left[\log \frac{1}{q\left(y_z^{m_z}\right)} - m_z H_0\left(y_z^{m_z}\right)\right] \\
            &\qquad= \MI{m_z, (n_i)_{i \in \Acal}, Y_{z,m_z}}{\Theta} \\
            &\qquad\quad+ \Ebb\left[m_z H(\Theta) - m_z H_0(y_z^{m_z})\right],
        \end{align*} 
        where $y_{z,i}$ is the output when we reach node $z$ on the $i$th time. Viewing $m_z$ as a stopping time of the sequence $y_z^t$ produced at node $z$, we bound each term by $\Ebb[\log(m_z +1)]$. We then use \lemref{lem:sum_nodes_bound} to show that \[\Ebb[\log(m_z +1)] \lesssim \frac{n \log\log n}{\log n}.\]
    \begin{proof}
    For any $z \in \mathcal{Z}(x^n)$, it is the case that the term inside the absolute value is always nonnegative, as
    \begin{align*}
        &\log \frac{1}{q(y_z^{m_z})}\\
        &= -\log\int \prod_{i \in \Acal} \theta_i^{n_i} \, {\Pi(d\theta_z)}\\
        % &\qquad= -\log\int \exp\left(m_z\left(\frac{n_0}{m_z} \log(1-\theta)+\frac{n_1}{m_z}\log(\theta)\right)\right)\, {\Pi(d\theta)} \\
        &= -\log\int \exp\left(m_z\left(-D(P_{\theta_0}\mid\mid P_{\theta_z}) - H_0(y_z^{m_z})\right)\right)\, {\Pi(d\theta_z)} \\
        &= m_zH_0(y_z^{m_z}) - \log \int 2^{-m_zD(P_{\theta_0}\mid\mid P_{\theta_z})}\, {\Pi(d\theta_z)}
    \end{align*}
    where $P_{\theta_0}$ is the empirical distribution. We proceed to analyze terms of the form
    \[ S_z := \Ebb\left[\log \frac{1}{q(y_z^{m_z})} - m_z H_0(y_z^{m_z})\right].\]
        Fix $n$, and recall that $m_z$ is the number of times we reach node $z$ in the first $n$ symbols of the LZ source. We therefore can think of $m_z$ as a stopping time of the process $(y_{z,i})_{i \in \mathbb{Z}^+}$ where $y_{z,i}$ is the $i$th element of $y_z^{m_z}$. Then taking the expectation over $\Theta \sim \Pi, \,Y_z^{\infty} \overset{\text{iid}}{\sim} P_{\theta_z}, m_z \sim P_{m_z|Y_z^{\infty}}$,
        \begin{align*}
            \Ebb\left[\log \frac{1}{q(y_z^{m_z})}\right] &= \Ebb\left[\log \frac{p_\theta(y_z^{m_z})}{q(y_z^{m_z})}\right] - \Ebb\left[\log p_\theta(y_z^{m_z})\right] \\
            &= \MI{Y_z^{m_z}}{\Theta} + \Ebb\left[m_z H(\Theta)\right] \\
            &= \MI{(n_i)_{i \in \Acal}, Y_{z,m_z}, m_z}{\Theta} + \Ebb\left[m_z H(\Theta)\right]
        \end{align*}
        where we use the optional stopping theorem on the martingale $\log(p_\theta(Y_z^{m_z})) - m_zH(\Theta)$ to compute the term on the right (noting that $m_z < n$ so our stopping time is bounded) and where we use~\lemref{lem:suff_stat_theta} to rewrite the mutual information term.
        We then note that 
        \begin{align*}
            &\MI{(n_i)_{i \in \Acal}, Y_{z,m_z}, m_z}{\Theta} \\
            &\qquad= \CMI{(n_i)_{i \in \Acal}}{\Theta}{m_z, Y_{z, m_z}} +\CMI{Y_{z,m_z}}{\Theta}{m_z} \\
            &\qquad\quad\quad+ \MI{m_z}{\Theta} \\
            &\qquad\le \Ebb_{m_z}\left[H\left((n_i)_{i \in \Acal}\right)\right] + 1 + \MI{m_z}{\Theta} \\
            &\qquad\le \Ebb\left[\,\abs{\Acal}\log (m_z+1)\right] + \MI{m_z}{\Theta} + 1 \\
            &\qquad\le \left(\,\abs{\Acal}+2\right)\Ebb[\log(m_z + 1)] + O(1),
        \end{align*}
        where we apply the well-known result that any random variable $X$ supported on $\mathbb{Z}^+$ satisfies $H(X) \le 2\Ebb[\log(X+1)] + O(1)$.
        Applying~\lemref{lem:plugin-stopping-time-bd}, we find that 
            \[\Ebb[m_z \left(H(\Theta) - H_0(y_z^{m_z})\right)] \lesssim \Ebb[\log (m_z+1)],\]
        implying that
        \[ S_z \lesssim \Ebb[\log(m_z+1)],\]
        and finally~\lemref{lem:sum_nodes_bound} gives the desired bound.
    \end{proof}
    \subsection{Convergence of Log-Likelihoods and Uniform Convergence}

    In this subsection we prove~\lemref{lem:smb-in-prob-no-support}, an analog of the Shannon-McMillan-Breiman result in Theorem~\ref{thm:shannon-mcmillan-breiman} but without requiring the support condition and with convergence in probability replacing almost sure convergence. We also prove uniform integrability in~\lemref{lem:ui_for_L} which we use in the calculation of the entropy rate in Theorem~\ref{thm:entropy-rate}.
    
    \begin{lemma}\label{lem:log-probability-converges-to-lz-compression-ratio-in-prob}
        Let $q$ be a mixture of distributions over $\Acal$ satisfying $H(q) > 0$. Then we have that 
        \[ \sum_{z \in Z(x^n)} \log \frac{1}{q(y_z^{m_z})} - \frac{T_n(x^n) \log T_n(x^n)}{n} \overset{p}{\to} 0.\]
    \end{lemma}
    \begin{proof}
        By Theorem III.10 of~\cite{sagan2024familylz78baseduniversalsequential}, the log-loss of probability families with density bounded away from $0$ uniformly converges to $\frac{T_n(x^n) \log T_n(x^n)}{n}$. Then taking $\hat{q}(\cdot)$ to be a density satisfying these properties,
        \begin{align*} 
            &\Ebb\left[\abs{\frac{1}{n} \sum_{z \in Z(x^n)} \log \frac{1}{q(y_z^{m_z})} - \frac{T_n(x^n)\log T_n(x^n)}{n}}\right] \\
            &\quad\le \, \Ebb\left[\abs{\frac{1}{n} \sum_{z \in Z(x^n)} \log \frac{1}{q(y_z^{m_z})} - \frac{1}{n} \sum_{z \in Z(x^n)} m_z H_0(y_z^{m_z})}\right] \\
            &\quad+ \, \Ebb\left[\abs{\frac{1}{n} \sum_{z \in Z(x^n)} \log \frac{1}{\hat{q}(y_z^{m_z})} - \frac{1}{n} \sum_{z \in Z(x^n)} m_z H_0(y_z^{m_z})}\right] \\
            &\quad+ \, \Ebb\left[\abs{\frac{1}{n} \sum_{z \in Z(x^n)} \log \frac{1}{\hat{q}(y_z^{m_z})} - \frac{T_n(x^n)\log T_n(x^n)}{n}}\right]
        \end{align*}
        by the triangle inequality. Taking $n \to \infty$, we have that the first term converges to $0$ by~\lemref{lem:log_probs_to_entropy_in_prob}, the second term converges to $0$ by Lemma D.6 in~\cite{sagan2024familylz78baseduniversalsequential}, and the third term converges to $0$ by Theorem III.10 of~\cite{sagan2024familylz78baseduniversalsequential}, implying convergence in probability of the term inside the expectation on the left hand side.
    \end{proof}
    \begin{lemma}\label{lem:smb-in-prob-no-support}
        Let $\Xv$ be generated from the LZ78 source $Q^{\text{LZ}, \Pi}$ and assume that $\Ebb[H(\Theta)] > 0$.
        Then
        \[ \frac{1}{n}\log \frac{1}{Q^{\text{LZ}, \Pi}_{X^n}(X^n)} \overset{p}{\rightarrow}  \Ebb[H(\Theta)]. \]
    \end{lemma}
    \begin{proof}
        This proof is identical to the proof of Theorem~\ref{thm:shannon-mcmillan-breiman} but using~\lemref{lem:log-probability-converges-to-lz-compression-ratio-in-prob} in place of Remark \ref{rem:lz78-codelength-convergence}.
    \end{proof}

    \begin{lemma}\label{lem:ui_for_L}
        Define the set \[L_n := \set{\frac{1}{n} \log \frac{1}{P_{X^n}(X^n)}}_{n \ge 1}.\]
        The set $\set{L_n}$ is uniformly integrable.
    \end{lemma}
    \begin{proof}
        It suffices to show that
        \[ \lim_{K \to \infty} \sup_n \Ebb\left[L_n \mathds{1}[L_n > K]\right] = 0.\]
        Note that 
        \[ L_n \mathds{1}[L_n > K] = (L_n - K)_+ + K\mathds{1}[L_n > K]\]
        and 
        \[ \Ebb[(L_n - K)_+] = \int_{K}^\infty \Pr{L_n > t} dt.\]
        We can therefore write
        \begin{align*}
            &\sup_n \Ebb\left[L_n \mathds{1}[L_n > K]\right] \\
            &\qquad\le \sup_{n} \Ebb[(L_n - K)_+] + K\sup_n \Ebb[\mathds{1}[L_n > K]] \\
            &\qquad\le \sup_n \int_{K}^\infty \Pr{L_n > t} dt + K\sup_n \Pr{L_n > K} \\
            &\qquad\le \int_{K}^\infty \sup_n \Pr{L_n > t} dt + K\sup_n \Pr{L_n > K} \\
            &\qquad\le \abs{\Acal}(K+1)2^{-K},
        \end{align*}
        using Chernoff's inequality to conclude that 
        \[ \Pr{L_n > t} \le \frac{\Ebb\left[2^{nL_n}\right]}{2^{nt}} = \abs{\Acal}^n 2^{-nt}.\]
    \end{proof}

\section*{Acknowledgment}
Discussions with Andrea Montanari at an early stage of this work are acknowledged with thanks.

We gratefully acknowledge support from the SystemX Alliance at Stanford University. Naomi Sagan and Matthew Ho acknowledge support from the National Science Foundation Graduate Research Fellowship Program under Grant No. DGE-2146755. Naomi also acknowledges support from the Cisco Systems Fellowship, part of the Stanford Graduate Fellowship (SGF) program.
Amir Dembo acknowledges support from NSF grant DMS-2348142.

\ifCLASSOPTIONcaptionsoff
  \newpage
\fi

\bibliographystyle{IEEEtran}
\bibliography{main.bib}

\begin{IEEEbiographynophoto}{Naomi Sagan}
received her B.S. and M.S. in Electrical Engineering and Computer Science from University of California, Berkeley, and is currently a third-year PhD student in Stanford's Department of Electrical Engineering. She is a Hertz Fellowship finalist and a recipient of the NSF Graduate Research Fellowship and Stanford Graduate Fellowship. Her research spans information theory, statistics, optimization, and natural language processing.
\end{IEEEbiographynophoto}

\begin{IEEEbiographynophoto}{Matthew Ho}
received his B.S. in Mathematics from the Massachusetts Institute of Technology and is currently a first-year PhD student in Stanford's Department of Electrical Engineering. He is a recipient of the NSF Graduate Research Fellowship. His research interests include information theory, statistics, and machine learning.
\end{IEEEbiographynophoto}

\begin{IEEEbiographynophoto}{Amir Dembo}
obtained his PhD in Electrical Engineering from Technion, Israel. Since 1990 he has been on the faculty of Stanford University,
Statistics and Mathematics departments, where since 2012 he is a Marjorie Mhoon Fair Professor in Quantitative Science and since 2013
he is also a Courtesy Professor of Electrical Engineering. His areas of specialization are probability theory and stochastic processes, information theory, large deviations, and their applications. Together with Ofer Zeitouni, he has authored a book on the theory of large deviations which is now a classical reference
in the field. He has served as editor of Probability Theory and Related Fields and of the Annals of Probability. He is a fellow of the Institute of Mathematical Statistics, a member of the National Academy of Sciences and a follow of the American Academy of Arts and Sciences.
\end{IEEEbiographynophoto}

\begin{IEEEbiographynophoto}{Tsachy Weissman} is the Robert and Barbara Kleist Professor of Electrical Engineering at Stanford, where he has been since 2003, researching and teaching the science of information, with applications spanning genomics, neuroscience, and technology.

He has been serving on editorial boards for scientific journals, technical advisory boards in industry, and as founding director of the Stanford Compression Forum. His recent projects include the SHTEM science and humanities summer internship program for high schoolers, Stagecast, a low-latency video platform allowing actors and musicians to perform together in real-time while geographically distributed, and the Starling Lab for authentication of digital content.

An IEEE Fellow, he has received multiple awards for his research and teaching, including best paper awards from the IEEE Information Theory and Communications societies, while his students received best student-authored paper awards at the top conferences of their areas of scholarship.

His students have become faculty at top institutions, industry leaders, and serial entrepreneurs. He has played key roles in the formation of companies and the development of technology powering Guardant Health's blood tests for early detection of cancer, Amazon's storage and machine learning, Google's search, HP's printing, Ford's streaming of self-driving footage, Siemens' streaming of sensor data, Apple's image and video compression, and Yahoo's bidding platform, among others. His favorite gig to date was advising the HBO show ``Silicon Valley''.
\end{IEEEbiographynophoto}


%\vfill


\end{document}